%% file: 0main.tex
\documentclass[10pt,screen]{acmart}
\usepackage{lineno}
\usepackage{xcolor,colortbl}
\usepackage{tabularx}
\usepackage{multirow}
\usepackage{color}
\usepackage{listings}
\usepackage{lstautogobble}
\usepackage{xspace}
\usepackage{amsmath}
\usepackage{hyperref}
\usepackage{soul}
\usepackage{pifont}

\newcommand{\name}{StarPlat\xspace} 
\usepackage{subcaption}
\usepackage{caption}

\definecolor{light}{rgb}{0.5, 0.5, 0.5}
\def\light#1{{\color{light}#1}}
\newcommand{\keywordSP}[1]{{\color{blue}{#1}}}

\newcommand{\todo}[1]{{\color{blue}{#1}}}
\newcommand{\REM}[1]{}

\makeatletter
\AtBeginDocument{%
  \let\c@figure\c@lstlisting
  
  \let\ftype@lstlisting\ftype@figure 
}

\AtBeginDocument{%
  \providecommand\BibTeX{{%
    \normalfont B\kern-0.5em{\scshape i\kern-0.25em b}\kern-0.8em\TeX}}}

\setcopyright{acmcopyright}
\copyrightyear{2025}
\acmYear{2025}
\acmDOI{XXXXXXX.XXXXXXX}

\acmConference[]{}{}{}
\acmBooktitle{} 
\acmPrice{}
\acmISBN{}
\usepackage{color}
\definecolor{backcolour}{rgb}{0.95,0.95,0.92}

\lstdefinelanguage{NEAR}
{
  morekeywords= [1]{ COPY,
  ENTRYPOINT, function, fixedPoint, filter, Min, forall, Dynamic, Batch, OnAdd, OnDelete, Static, Incremental, Decremental},
  morekeywords = [2]{propEdge, updates,propNode, Graph, node, edge, int, bool, in, until, for, if, from, do, while, void, _global_, unsigned, long, auto, return,},
  morekeywords = [3]{updateCSRDel,updateCSRAdd,currentBatch,attachNodeProperty, nodes, neighbors, get_edge, staticSSSP},
  morekeywords = [4]{True, False},
  morecomment=[l]{\//},
  morestring=[b]",
}

\lstdefinestyle{mystyle}{
  basicstyle=\small,
  numbers=left,
  stepnumber=1,                   
  numbersep=8pt,                  
   showspaces=false,
  rulecolor=\color{black},
  stringstyle=\color{green},
  numberstyle=\tiny\color{black},
  commentstyle=\color{gray},
  identifierstyle=\color{black},
  keywordstyle=[3]\color{purple}\bfseries,
  keywordstyle=\color{blue},
  showstringspaces=false,        
   showtabs=false,                
  tabsize=1,                      
  captionpos=b,                   
  breaklines=true,                
  breakatwhitespace=true,         
  title=\lstname 
}

\begin{document}

\title{Generating Dynamic Graph Algorithms for Multiple Backends for a Graph DSL}

\author{Nibedita Behera}
\orcid{}
\email{}
\affiliation{%
  \institution{IIT Madras}
  \city{Chennai}
  \country{India}
}

\author{Ashwina Kumar}
\orcid{0000-0001-6425-7479}
\email{cs20d016@cse.iitm.ac.in}
\affiliation{%
 \institution{IIT Madras}
 \city{Chennai}
 \country{India}}

\author{Atharva Chougule}
\orcid{0009-0006-6369-6668}
\email{chouguleatharva@gmail.com}
\affiliation{%
  \institution{IIT Madras}
  \city{Chennai}
  \country{India}}

\author{Mohammed Shan P S}
\orcid{0009-0005-9836-3233}
\email{mohammedshan.p.s@gmail.com}
\affiliation{
  \institution{IIT Madras}
  \city{Chennai}
  \country{India}}

\author{Rushabh Nirdosh Lalwani}
\orcid{0009-0007-3135-3035}
\email{rushilal2411@gmail.com}
\affiliation{%
  \institution{IIT Madras}
  \city{Chennai}
  \country{India}}

\author{Rupesh Nasre}
\orcid{0ss000-0001-7490-625X}
\email{rupesh@cse.iitm.ac.in}
\affiliation{%
  \institution{IIT Madras}
  \city{Chennai}
  \country{India}}

\renewcommand{\shortauthors}{Kumar et al.}
\newcommand{\MLE}{\color{red}{OOM}}
\newcommand{\OOT}{\color{red}{OOT}}

\begin{abstract}
 With the growth of unstructured and semistructured data, parallelization of graph algorithms is inevitable for efficiency. Unfortunately, due to the inherent irregularity of computation, memory access, and communication, graph algorithms are traditionally challenging to parallelize. To tame this challenge, several libraries, frameworks, and domain-specific languages (DSLs) have been proposed to reduce the parallel programming burden of users, who are often domain experts.  Several frameworks exist today that partially or fully hide the parallelism intricacies, provide mnemonics for scheduling strategies, and perform program analysis to identify races to generate synchronization code. However, existing frameworks are limited by their support in the form of abstractions and run-time optimizations, primarily for static graphs. However, most real-world graphs are dynamic in nature; that is, their structure changes over time. There has been a growing emphasis on understanding dynamic graphs (also called as \textit{morph} graph algorithms), which evolve over time via insertion, deletion, and modification of vertices, edges, and the associated attributes. However, generating code for morph algorithms that is not only correctly synchronized but also efficient is a challenging task.

In this work, we present an abstraction scheme and runtime optimizations for efficient processing of morph
algorithms. In particular, given an initial graph \textit{G} and updates $\Delta$\textit{G} involving edge additions and deletions, the work entails specifying its dynamic processing as part of a domain-specific language (DSL), and generating parallel code corresponding to the specification  
for the multicore, distributed,  and many-core environments.
 We illustrate the effectiveness of our approach by running the DSL-generated parallel code on ten large graphs with varied characteristics and three popular algorithms: Shortest Paths computation, Page Rank processing, and Triangle Counting.
 
 Our dynamic processing paradigm demonstrates marked and consistent performance enhancements across a comprehensive suite of large-scale, real-world graphs when juxtaposed with conventional static graph algorithms. Specifically, the CUDA backend yields an average acceleration ranging from 1.4× to 5.1× for Page Rank, 1.5× to 34.1x for Single-Source Shortest Path (SSSP), and 1.2× to 4.5x for Triangle Counting (TC) compared to a recent static baseline. The OpenMP backend exhibits analogous performance gains, attaining 1.4x - 5.19x for Page Rank, 1.3x - 15.9x for SSSP, and 2.4x - 65.6x for TC. Similarly, the MPI backend achieves average speedups ranging from 1.1× to 1.8× for Page Rank, 1.2x to 6.6× for SSSP, and 0.8× to 11× for TC depending on the percentage of updates. These empirical results underscore the robustness, adaptability, and computational efficacy of our dynamic processing framework across diverse parallel execution environments and heterogeneous architectures.

To the best of our knowledge, this is the first work on code generation for parallel dynamic graph algorithms for multiple CPU backends (both OpenMP and MPI) and a GPU backend (CUDA).
\end{abstract}




\maketitle

\section{Introduction}
Dynamic graph algorithms~\cite{dynamicgraphRECENT,bladyg} can outperform their static counterparts as they update only the affected portions of the graph when changes occur, rather than recomputing the solution for the entire graph as a static algorithm would.
Further, since only parts of the graph are processed, dynamic algorithms save on computational resources such as time and memory.
For real-time systems, such as social networks or traffic monitoring, dynamic algorithms can provide near-instant updates, whereas static algorithms require recomputation, leading to delays. Dynamic graphs, by nature, evolve frequently. For example, nearly 20,000 transactions are processed per second during sales on Alibaba's e-commerce site~\cite{alibaba}. Similarly, on Twitter, we have 500 million tweets~\cite{tweet} every second tweeted by users.
Static algorithms may struggle with such frequent recomputations for large graphs, while
a dynamic algorithm can keep pace with these changes.
The growing adoption of real-time analytical processing~\cite{iyerdynamic,braundynamic,andrewdynamic} across industries has made optimizing computational efficiency for large-scale dynamic graph operations an imperative requirement today. 

A fundamental challenge in the domain of dynamic graph algorithms lies in the efficient preservation and real-time adaptation of graph properties amidst continuous structural modifications~\cite{McGregordynamic}. To mitigate the substantial computational burden typically incurred by full recomputation, dynamic variants employ strategies such as \textit{incremental updates}, wherein changes are processed selectively. 
Efficient processing of the updates often demands identifying the portion of the graph for which adopting to the changes suffices. This is done either by explicit computation, or by keeping track of auxiliary information otherwise not required for its static counterpart, leading to computation-memory trade-off. For instance, a dynamic breadth-first search (BFS) may maintain the underlying BFS DAG in addition to the BFS level number information.

\textit{Partially dynamic} graph algorithms process changes to the graph properties for specific types of updates, such as insertions (incremental) or deletions (decremental). 
\textit{Fully dynamic} algorithms~\cite{dynamicgraphRECENT}, on the other hand, handle changes to graph properties for both insertions and deletions. Prior art also varies in terms of the support for vertex vs. edge updates. Certain methods support only edge updates with the assumption that the vertex updates can be simulated as multiple edge updates on an originally disconnected vertex. 

Parallelization of static graph algorithms has traditionally been viewed as challenging due to irregular data access patterns, the inherent sequential nature of some algorithms, memory access patterns, synchronization overhead, and data dependencies. The challenges are exacerbated when structural modifications are allowed. Therefore, it is only in the recent past that the community has started looking for efficient parallel solutions for dynamic graph algorithms. Maintaining the graph attributes consistently under structural modifications happening simultaneously demands clever synchronization mechanisms, in the absence of which, the processing becomes sluggish. As a side effect though, this makes the code complex, and often the logic gets lost in the nitty-gritty details of handling parallelism and synchronization. It would be ideal, of course, if the parallel code could be \textit{generated} rather than written. 
In this work, we address this concern.

We utilize and build upon a static graph processing DSL, StarPlat~\cite{starplat}, to create StarPlat Dynamic -- a DSL for specifying dynamic graph algorithms. This new DSL provides constructs (and the corresponding parser) for easy and intuitive specification of dynamic updates on graphs. We add support for generating parallel code for these dynamic algorithms for three different backends: OpenMP, CUDA, and MPI.

Using a DSL-based approach for dynamic graph processing has several reasons and advantages:
 \begin{itemize}
    \item     Manually optimizing for parallelism on different kinds of hardware is cumbersome and error-prone.
    The programmer needs to be an expert in the application domain and HPC to exploit algorithmic properties as well as to suit the processing to the target hardware. Further, the strategies used across these parallel programming paradigms (multicore, distributed, manycore) are quite different.
    \item Users can focus on their algorithms rather than the nuances of parallel hardware and synchronization.

    \item StarPlat has high-level constructs for graphs and properties that expose the inherent parallelism in the algorithm and can apply a series of high-level optimizations and parallelization, finally producing an optimized parallel implementation of the given algorithm for the three backends.
\end{itemize}


A range of interesting frameworks have emerged to address the complexities of dynamic graph algorithms, such as cuSTINGER~\cite{cutstinger}, faimGraph~\cite{famigraph}, aimGraph~\cite{autonomous}, Hornet~\cite{hornet}, and Meerkat~\cite{concessao2024meerkat}. Each of them provides a carefully crafted dynamic graph data structure, the associated primitives, and the algorithms built with the primitive operations. 
Despite their technical merits, a key weakness of these existing frameworks or data structures is their confinement to a single hardware paradigm. In particular, a majority of these systems are exclusively optimized for or support only the CUDA backend. In contrast, our proposed StarPlat Dynamic framework is designed with cross-platform versatility, offering native support for CUDA, OpenMP, and MPI backends, thereby enhancing adaptability across various parallel computing environments.


This work makes the following contributions.
\begin{itemize}
\item We propose the first code generation scheme for dynamic graph algorithms for multiple types of parallel hardware. The language, named StarPlat Dynamic, is designed to express the intended aggregate operations while remaining independent of the underlying hardware architecture.
\item We propose language abstractions embedded into a graph DSL (named StarPlat), leading to custom code generation in OpenMP for multi-core backend, in MPI for distributed systems backend, and in CUDA for many-core architecture. For efficient code generation, the StarPlat Dynamic compiler also performs backend-specific optimizations from a common intermediate representation.
\item Using a variety of graphs with various characteristics and three popular graph algorithms (Shortest Paths computation, Triangle Counting, and Page Rank), we illustrate the effectiveness of the generated dynamic processing against the static ones (StarPlat Static~\cite{starplat}, Galois~\cite{Galois}, Ligra~\cite{Ligra}, Green-Marl~\cite{GreenMarl}). We observe that the dynamic processing outperforms its static counterpart for up to a certain percentage of updates, beyond which the static version outperforms.
\end{itemize}

The rest of the article is organized as follows. Section~\ref{background} presents the background of StarPlat Static. 
Section~\ref{dynamicmodel} presents the language specification and the intermediate representation. It also deals with diff-CSR representation, which is crucial for dynamic graph representation. Section~\ref{sec code gen} describes the code-generation scheme, followed by the overall flow of StarPlat Dynamic. Section~\ref{sec optimizations} provides an overview of the backend-specific optimizations \name Dynamic employs for efficient code generation. 
The experimental evaluation of the generated code for each backend is discussed in Section~\ref{sec exp evaluation}. Section~\ref{sec relatedwork} discusses the relevant related work in the context of graph analytics. We summarize our experience and conclude the article in Section~\ref{sec conclusion}.

\begin{figure}
\centering
\includegraphics[scale=0.5]{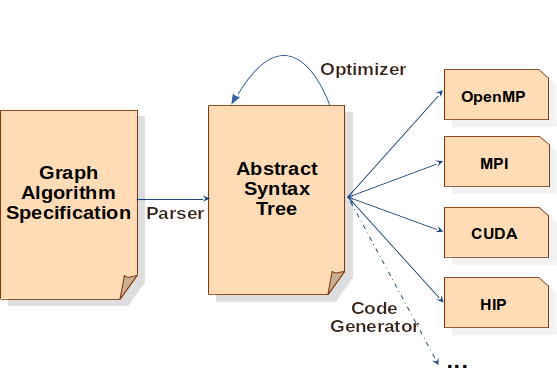}
\caption{Process Flow of StarPlat Static}
\label{fig:starplatflow}
\end{figure}

\section{Background: StarPlat Static}\label{background}
\textit{StarPlat Dynamic} is built upon \textit{StarPlat Static}~\cite{starplat}, whose overview we present in this section. 
StarPlat is a domain-specific language for static graph algorithms. It supports three backends: OpenMP, MPI, and CUDA. An example code (\texttt{staticSSSP}) is shown in Listing~\ref{ssspdyn-dsl} in Appendix~\ref{starplat:codes} and its process flow is depicted in Figure~\ref{fig:starplatflow}. StarPlat supports primitive data types: \texttt{int, bool, long, float}, and \texttt{double}. It also supports \texttt{Graph, node, edge}, as first-class types. Node and edge attributes can be added readily. \texttt{forall} is an aggregate construct that can process a set of elements in parallel. Depending upon the backend chosen by the user, it gets translated into an OpenMP pragma, or an MPI process, or a CUDA kernel.
Currently, StarPlat supports vertex-based processing. Several solutions to graph algorithms are iterative and converge based on the
conditions on the node attributes. StarPlat provides a \texttt{fixedPoint} construct to specify this succinctly. Its syntax involves
a boolean variable and a boolean expression on node-properties forming the fixed-point convergence condition. 
StarPlat provides constructs Min and Max, which perform multiple assignments atomically based on a comparison criterion.

To cater to the need of being versatile to generate efficient code for multiple backends, StarPlat relies on an intermediate representation common across backends. Its parser converts the source program into an abstract syntax tree (AST) recursively while parsing constructs. Since the source is not explicitly parallel but the generated code is, StarPlat needs to insert proper synchronization to avoid datarace. This demands program analysis on the AST. Various kinds of analyses are performed by StarPlat: to identify read-write sets in \texttt{forall} to ensure proper \texttt{cudaMemcpy}s generated in the right direction; to identify datarace within \texttt{forall}'s statements to insert correct synchronization (e.g., OpenMP \texttt{critical}); to optimize CPU-GPU data transfer (e.g., to retain graph properties on GPU across kernel calls); etc. Program transformations on the AST also modify it towards optimized code generation (e.g., removing dead code).

StarPlat is designed for multiple backends. It currently supports OpenMP, MPI, CUDA (while HIP, DPC++, multi-GPU are in development). A user can choose a backend via a command-line argument, based on which an appropriate code generator gets invoked. The code generator takes the generically-optimized AST as the input, applies backend-specific optimizations, and produces the further-optimized C++ code. The generated code can then be integrated by the users with their applications.

\section{StarPlat Dynamic: Programming Model}\label{dynamicmodel}
StarPlat is designed to ease the programming burden by introducing high-level language constructs to achieve performance competitive to hand-tuned codes. It is challenging to generate efficient code for different backends from the same algorithmic specification. For instance, when a DSL construct demands lock-based synchronization in OpenMP, it also expects to have inter-process communication in MPI, and a lock-free synchronization in CUDA! Generating the correct code for each backend poses difficulties.

\begin{figure}
    \includegraphics[scale=0.5]{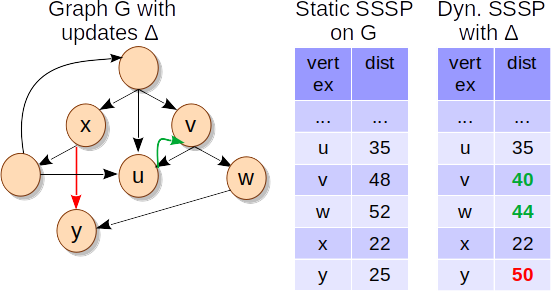} 
    \caption{Example graph to illustrate dynamic SSSP}
    \label{fig:dynsssp}
\end{figure}
 
\subsection{Example to Illustrate Dynamic SSSP}
Consider an example directed graph $G$ illustrated in Figure~\ref{fig:dynsssp}. It consists of vertices $u$, $v$, $w$, $x$, $y$, among others. Red edge $x {\color{red}{\rightarrow}} y$ exists in $G$, but gets deleted. Green edge $u {\color{green}{\rightarrow}} v$ is not originally present in $G$, but gets added. Thus, the two colored edges are part of the dynamic updates $\Delta$. If we execute a static SSSP computation (Dijkstra's algorithm or Bellman-Ford), we can find the shortest distance of each vertex from a designated source (not shown). Let us assume that the computed distances are as shown in the first table.

Dynamic SSSP begins with this computed distance table, applies $\Delta$ updates to the graph (edges $x \rightarrow y$ and $u \rightarrow v$) and needs to update the distances of the affected subgraph. In the context of this example, we need to check if the (incremental) edge $u \rightarrow v$ reduces $v$'s distance, and if yes (e.g., from value 48 to 40), the new distance needs to be propagated further to the subgraph reachable from $v$ until distance updates do not happen (e.g., $w$'s distance changes from 52 to 44, but does not change $y$'s distance yet). Similarly, we need to check that by deleting edge $x \rightarrow y$ (decremental update), whether $y$'s distance increases (to 50), and if yes, the new distance needs to be propagated further, similar to the incremental edge (not needed in this case since $y$ does not have an outgoing edge). The propagation can be readily implemented with StarPlat's \texttt{fixedPoint} construct, which processes the graph till any change happens in the attributes (see Appendix~\ref{ssspdyn-dsl}).
Note that the dynamic processing worked only on a subset of vertices, depending upon $\Delta$.

\subsection{Language Overview with a Motivating Example: Dynamic SSSP}
\begin{lstlisting}[language=NEAR, style=mystyle, label=ssspdyn-dsl-dyn, caption = Dynamic SSSP driver in StarPlat. It makes use of three functions \texttt{staticSSSP, Decremental, and Incremental}, %xleftmargin = 0.01\textwidth
]
Dynamic DynSSSP(Graph g, propNode<int> dist, propNode<int> parent, propEdge<int> weight, updates<g> allUpdates, int batchSize, int source) {  
    // Run Static SSSP on the original graph g.
    staticSSSP(g, dist, parent, weight, source);/*@\label{stat-70}*/  

    // Then apply the updates from the batch and process dynamically. 
    Batch(allUpdates:batchSize) {/*@\label{stat-71}*/  
        propNode<bool> activeOnDelete;/*@\label{stat-100}*/  
        propNode<bool> activeOnAdd;/*@\label{stat-101}*/  
        g.attachNodeProperty(activeOnDelete = False, activeOnAdd = False);

        OnDelete(e in allUpdates.currentBatch()) {  /*@\label{stat-27}*/
            node src = e.source;
            node dest = e.destination;
            if (dest.parent == src) {
                dest.dist = INF;
                dest.activeOnDelete = True;
                dest.parent = -1;
            }
        }  /*@\label{stat-39}*/
        g.updateCSRDel(allUpdates);/*@\label{stat-41}*/  
        Decremental(g, dist, parent, weight, activeOnDelete);  /*@\label{stat-42}*/  
       
        OnAdd(e in allUpdates.currentBatch()) { /*@\label{stat-103}*/ 
            node src = e.source;
            node dest = e.destination;
            int edge = g.getEdge(src, dest);
            if (dest.dist > src.dist + edge.weight) {
                dest.activeOnAdd = True;
                src.activeOnAdd = True;
            }
        }/*@\label{stat-104}*/  
        g.updateCSRAdd(allUpdates);/*@\label{stat-58}*/     
        Incremental(g, dist, parent, weight, activeOnAdd);/*@\label{stat-59}*/
    }
}
\end{lstlisting}


Figure~\ref{ssspdyn-dsl-dyn} shows a snippet of dynamic SSSP written in the StarPlat Dynamic DSL.
It takes various parameters: original graph \texttt{g}, vertex attribute \texttt{dist} for storing the shortest distances, \texttt{parent} for each vertex via which the shortest distance is computed (which can be used to trace the shortest path), edge attribute \texttt{weight}, a sequence of updates (that is, insertions and deletions) processed \texttt{batchsize}-many in parallel, and the source vertex \texttt{src} for running SSSP.
The processing has two parts: static and dynamic. The static part begins with calling static SSSP (Line~\ref{stat-70}) on the original graph \texttt{g}. The static SSSP DSL code is shown in Appendix~\ref{ssspdyn-dsl}, which implements a Bellman-Ford-style processing, which exhibits better parallelism compared to Dijkstra's algorithm. 
It computes the distances of all the vertices in \texttt{g} from vertex \texttt{src}.
The dynamic part involves processing the graph updates in chunks of size \textit{batchsize}. A batched processing is vital to achieve parallelism benefits. Within each batch, the programmer can decide how to process the updates. For instance, the code shown in Figure~\ref{ssspdyn-dsl-dyn} processes all the decremental updates in a batch followed by all the incremental updates. 
StarPlat Dynamic currently supports edge-updates as part of a batch -- that is, edge additions and edge deletions. Vertex additions can be simulated by adding edges to a disconnected vertex, while vertex deletion can be simulated by disconnecting a vertex from the rest of the graph.
Each batch performs some bookkeeping by maintaining \texttt{activeOnDelete} (Line~\ref{stat-100}) and \texttt{activeOnAdd} (Line~\ref{stat-101}) variables to flag the affected vertices from an update.
Note that it is easy in StarPlat to add new vertex properties.

The decremental and the incremental processing often involve three steps: pre-processing, updating the graph, and propagating the update as part of the algorithmic computation.
The pre-processing step uses \texttt{OnDelete} construct and involves setting the properties for the destination nodes of the deleted edges to appropriate values before invoking the \texttt{Decremental} call (Lines~\ref{stat-27}--\ref{stat-39}). In SSSP, it involves resetting the \texttt{dest} node's \texttt{dist} property to $\infty$, \texttt{dest's} \texttt{parent} to \texttt{-1}, and setting its \texttt{activeOnDelete} flag. The next step is a call to the graph library helper function \texttt{updateCSRDel} (Line~\ref{stat-41}) that modifies the graph by updating the original graph representation with the changes after deletion of the edges (mentioned in the batch). StarPlat Static relies on the compressed sparse row (CSR) format for storing the graph, while StarPlat Dynamic uses diff-CSR representation~\cite{diffcsr} (more in Section~\ref{sec diffcsr}). The last step is the \texttt{Decremental} function call (Line~\ref{stat-42}) which updates the distance values of the affected nodes in the modified graph by propagating the information to reachable vertices. The DSL implementation for the \texttt{Decremental} SSSP function is shown in Appendix~\ref{ssspdyn-dsl}.

A similar procedure is followed for the incremental updates, which relies on \texttt{OnAdd} and \texttt{Incremental} constructs. The \texttt{activeOnAdd} flag is set to true for nodes whose distance values have decreased by the edge addition (from the batch). Then a call to the graph library function \texttt{updateCSRAdd} (Line~\ref{stat-58}) modifies the graph and adds the new edges into StarPlat Dynamic's diff-CSR structure which represents the dynamic graph. The \texttt{Incremental} (Line~\ref{stat-59}) function then computes the distance values for the nodes only in the affected subgraph. The DSL implementation for \texttt{Incremental} is also shown in Appendix~\ref{ssspdyn-dsl}.


\subsection{Constructs for Dynamic Processing}
StarPlat Dynamic DSL enables users to write dynamic graph algorithms~\cite{HPECGraph} using various new constructs as discussed in the preceding section. We describe each of those below along with their syntax.

\subsubsection{\textcolor{purple}{Batch}} 
Given a bunch of updates, even the serialized processing of updates using the dynamic algorithm can be time-consuming. The batched update  bears the potential to improve the execution time with parallelism. This construct encloses the preprocessing logic and the order in which the handlers are called for updates in the current batch. The construct sweeps through all the updates in batches of size mentioned as part of the syntax. Thus, by simply changing the value of \texttt{batchsize}, a programmer can tune the value depending on the amount of parallelism available (for instance, the batch size may be smaller on a CPU and larger on a GPU). Further, for applications that do not involve fully-dynamic processing, it is easy to specify the incremental-only or decremental-only functionality, along with any auxiliary processing.

\texttt{\textcolor{purple}{Batch} (updateList: batchsize) \{...\}}

Within the body of the \texttt{batch} construct, the incremental and the decremental processing can be specified.
The currently selected batch can be accessed using the \texttt{currentBatch()} function provided.
 
\subsubsection{\textcolor{purple}{OnAdd} and \textcolor{purple}{OnDelete}}

\texttt{OnAdd} is an iterative construct that goes over all the incremental updates in the selected batch and performs the preprocessing step enclosed in the block for each update. For instance, the step may involve marking the edges or nodes that are part of the edge addition as active. This facilitates detecting the subgraph which requires recomputation of the properties in the Incremental algorithm. Similarly, \texttt{OnDelete} permits specifying the preprocessing logic for detecting the subgraph that may undergo modification in its (node or edge) properties during the decremental processing. The preprocessing logic may vary across algorithms, depending upon the application requirements. For instance, a programmer may decide to process incremental updates prior to the decremental updates, or an application may be written only for incremental updates. Due to these variations, it is left to the user to code up this logic. 

\texttt{\textcolor{purple}{OnAdd} (e in updateList.currentBatch()) \{...\}}

\texttt{\textcolor{purple}{OnDelete} (e in updateList.currentBatch()) \{...\}}

\subsubsection{\textcolor{purple}{Incremental} and \textcolor{purple}{Decremental}}

These are the two special functions the user needs to define. These two functions enclose the actual logic for the dynamic graph algorithm. \textit{Incremental} function defines the dynamic graph algorithm corresponding to the incremental changes or the addition of edges to the graph, while the \textit{Decremental} function defines the decremental changes. The order in which the \textit{Incremental} or the \textit{Decremental} functions are called in the dynamic graph algorithm can be chosen by the programmer in the \textit{Dynamic} version of the algorithm. 
Figure~\ref{ssspdyn-dsl-incremental} shows the incremental processing for SSSP. Complete specifications of incremental and decremental graph updates are presented in Appendix~\ref{TCdyn-dsl} for Triangle Counting, Appendix~\ref{PRdyn-dsl} for Page Rank, and Appendix~\ref{ssspdyn-dsl} for SSSP.

\begin{lstlisting}[language=NEAR, style=mystyle, label=ssspdyn-dsl-incremental, caption = Incremental SSSP computation in StarPlat (called from the Dynamic SSSP driver in Figure~\ref{ssspdyn-dsl-dyn}), %xleftmargin = 0.01\textwidth
]
Incremental(Graph g, propNode<int> dist, propNode<int> parent, propEdge<int> weight, propNode<bool> modified) {
  propNode <bool> modified_nxt;
  g.attachNodeProperty(modified_nxt = False);
  bool finished =False;

  fixedPoint until (finished:!modified) {
    forall (v in g.nodes().filter(modified == True) ) {/*@\label{line:modified}*/  
      forall (nbr in g.neighbors(v)) {          
        edge e = g.get_edge(v, nbr);
        <nbr.dist,nbr.modified_nxt,nbr.parent> = <Min (nbr.dist, v.dist + e.weight), True,v>;
      }
    }
    modified = modified_nxt;
    g.attachNodeProperty(modified_nxt = False);
  }
}
\end{lstlisting}

The incremental procedure in Figure~\ref{ssspdyn-dsl-incremental} runs a fixed-point computation to process the modified (or active) vertices marked by the preprocessing. For each such modified vertex (Line~\ref{line:modified} which uses a \texttt{forall}) in parallel, its neighbors' distances are updated. The \texttt{Min} construct updates a vertex's distance with a critical section (implemented with an atomic instruction), and if the distance is updated, the vertex is marked active for the next iteration. To avoid race condition, it uses \texttt{modified} bits as read-only and writes to the \texttt{modified\_nxt} array. For typical updates, the dynamic fixed-point processing is considerably faster than its static counterpart, proportional to the size of the affected subgraph, dictated by the input batch of updates.

\REM {


The \texttt{Graph} data type encapsulates the operations and properties of a standalone graph. The properties include its nodes, edges, number of nodes, number of edges, etc. The graph representation followed for graph problems in StarPlat is of Compressed Sparse Row (CSR) format and the  Graph object contains fields to expose this representation. 
StarPlat stores the graph in Compressed Sparse Row (CSR) format, which provides the storage benefits of adjacency lists, and also allows seamless transfer across devices, due to the use of offsets.
The data type also facilitates information gathering and manipulation at the node and the edge levels. Since the nodes and edges are tightly bound to the graph, it becomes convenient for Graph to support this through various library functions. For instance, as per the semantics of \texttt{neighbors(u)}, it returns the outgoing neighbors in case of a directed graph and all the neighbors in case of an undirected one. For directed graphs, graph type also exposes a function that returns the incoming neighbours to a node, \texttt{nodesTo()}. This needs Graph to maintain a CSR representation for also its transpose, which becomes handy in algorithms which perform computation on a transposed version of the input graph (e.g., Betweenness Centrality).

\begin{figure}
\begin{lstlisting}[language=NEAR, style=mystyle, label=toy-code,  caption = %, xleftmargin = 0.01\textwidth %\todo{Let's show some useful toy example.} %\rtodo{could not get center align for this code.-DONE}, 
]
function foo(Graph g, propNode<int> modified, propNode<int> data) {
    SetN<g> nodeSet; /*@\label{toy-stat-2}*/
    for(v in g.nodes().filter(modified)) { // Sequential loop /*@\label{toy-stat-3}*/
        for(nbr in g.neighbors(v)) {  /*@\label{toy-stat-4}*/
            if(nbr.data > THRESHOLD) {
                nodeSet.addNode(nbr); /*@\label{toy-stat-8}*/ 	  	     
}   }   }   }
\end{lstlisting} 
    \caption{Example program to illustrate various data types in \name}
    \label{fig:toy-code}
\end{figure}

A node and an edge can in themselves have properties associated with them. In the SSSP problem setting, a vertex's distance can be viewed as a node property, being computed by the corresponding algorithm. Similarly, in the BC computation, the betweenness centrality values of each node can be viewed as a property. The \texttt{propNode} datatype facilitates declaring property for nodes of a graph with the provision of specifying its type. The \texttt{attachNodeProperty} function provided by the Graph type binds this property to the graph’s nodes and initializes the property values if provided. Line~\ref{computeSSSP-stat-3} in our SSSP code from Figure~\ref{fig:staticSSSP} specifies the declaration of a node property dist of type \texttt{int}. The \texttt{attachNodeProperty} binds dist to the graph and optionally, initializes the distance attribute for each vertex (e.g., to infinity in Figure~\ref{fig:staticSSSP}). Similarly, \texttt{propEdge} datatype is associated with edges, and has otherwise the same semantics as that of \texttt{propNode}. The \texttt{attachEdgeProperty} function binds the property to the graph’s edges. 

\name also provides collection types such as \texttt{List}, \texttt{updates}, \texttt{SetN}, and \texttt{SetE}. \texttt{List} allows the presence of duplicates whereas \texttt{SetN} and \texttt{SetE} store unique nodes and edges respectively. Line~\ref{toy-stat-2} in Figure~\ref{fig:toy-code} shows an example usage. The separation of sets between nodes and edges enables choosing the relevant implementation in vertex-based vs. edge-based codes. The \texttt{updates} collection type becomes handy while designing dynamic graph algorithms where they store the updates information on the current graph.

\subsubsection{Parallelization and Iteration Schemes}

\subsection{\texttt{forall }statement}
\texttt{forall} is an aggregate construct in \name which can process a set of elements in parallel. Its sequential counterpart is a simple \light{for} statement. Currently, \name supports vertex-based processing\footnote{Addition of support for edge-based processing is in our plans, which needs changes to the underlying data representation. Compressed Sparse Row (CSR) storage format is suited for vertex-based processing.}.  
The parallel \texttt{forall} supports various ranges it can iterate on (e.g., nodes in the whole graph or neighbors of a node, as shown in Figure~\ref{fig:staticSSSP}, Lines~\ref{computeSSSP-stat-11} and \ref{computeSSSP-stat-13}). 

The function \texttt{g.nodes()} called on a graph \texttt{g} returns a sequence of nodes which can be iterated upon. To iterate over the neighbors of a node \texttt{u}, the functions \texttt{g.neighbors(u)}, \texttt{g.nodesTo(u)} and \texttt{g.nodesFrom(u)} return a similar sequence.
The \texttt{forall} body can be executed selectively for the nodes satisfying a certain boolean expression based on the node label or, node's property by including a \texttt{filter} construct. Line~\ref{toy-stat-3} in Figure~\ref{fig:toy-code} shows its usage using the node property \texttt{modified}.

Considering that several graph algorithms can be well represented using a single outer parallel loop, and that the analysis of nested parallel loops gets complicated, currently, \name supports only outer level parallelism. Hence a nested \texttt{forall} in the DSL results in a parallel outer loop and a sequential inner loop in the target code. 
One may argue that an outer loop over vertices and an inner loop over neighbors can benefit from nested parallelism, and we agree. However, (i) such a processing can be well taken care of by an edge-based parallelism (to be supported), and (ii) since we target large graphs, even a vertex-based processing has enough parallelism to keep the resources busy.

\subsection{\texttt{iterateInBFS/iterateInReverse} statement}
For a graph processing DSL, traversals become the fundamental building blocks.
\name provides breadth-first traversal as a construct, borrowed from Green-Marl~\cite{Green-Marl}. 

\texttt{\keywordSP{iterateInBFS}(v \keywordSP{in} g.nodes() \keywordSP{from} root) \{...\}}

\noindent \texttt{iterateInBFS} performs a BFS traversal of the graph from the given root node. The underlying processing is level-by-level and iterates in parallel over the visited nodes in a specific level. On visiting a node in a level, it executes the body statements and forms the next set of visited nodes. The \texttt{filter()} construct can be utilized to explore the neighborhood of a visited node selectively.  Similarly, \texttt{iterateInReverse} performs a reverse BFS traversal in a level-synchronous manner and extracts parallelism at each level in the computation of the body statements. Note that \texttt{iterateInBFS} is a prerequisite to use \texttt{iterateInReverse}, since the former builds the BFS DAG to be traversed through in the latter. The functions \texttt{neighbors}, \texttt{nodesTo} and \texttt{nodesFrom} have a subtle change in their meaning when used inside \texttt{iterateIn...} constructs: they correspond to the neighbors in the BFS DAG rather than the original graph \texttt{G}.
\subsubsection{Reductions}

Reductions are one of the popular parallel programming primitives, and can be useful in achieving efficient computation. Specifying a reduction in the DSL also helps in conveying a necessity of synchronization. Unfortunately, it does not directly fit into the philosophy of StarPlat design to support reduction as a language construct. Therefore, as a golden-mid, StarPlat permits usage of certain relative C-operators (e.g., \texttt{+=}) to convey reduction. This "trick" allows us to retain the abstraction and still achieve efficiency of the generated code.
As an effort to include the best practices in the language, StarPlat supports \textit{assignment reduction} for commonly known associative operators.
The reduction operators supported by StarPlat are tabulated in Table~\ref{tab:reduction}.

\begin{table}[!htb]
\centering
\captionsetup{justification=centering}
\begin{center}
\begin{tabular}{ |c|c| } 
 \hline
 {\textbf{Operator}} & {\textbf{Reduction Type}}\\
 \hline
 {\texttt{+=}} & {Sum} \\
 \hline 
 {\texttt{*=}} & {Product} \\ 
 \hline 
 {\texttt{++}} & {Count} \\
 \hline 
  {\texttt{\&\&}} & {All } \\
 \hline 
 {\texttt{||}} & {Any} \\
 \hline 
 
\end{tabular}
\caption{Reduction operators in StarPlat}
\label{tab:reduction}
\end{center}
\end{table}

We illustrate the usage of reduction in Figure~\ref{reduction-dsl}. The introduction of reduction (Line~\ref{reduction-accum} in the code makes sure the \texttt{accum} variable has a deterministic result at the end of the parallel region. Note that Line~\ref{reduction-count} involves a thread-local variable \texttt{count} and does not need reduction. On the other hand, if nested parallelism was supported, \texttt{count} would also need a reduction. The reduction operators in \name translate to library based implementations of reduction in the target backend.

\begin{lstlisting}[language=NEAR, style=mystyle, label=reduction-dsl,  caption = Reduction example]
int accum = 0;
forall(v in g.nodes()){   
    int count = 0;
    forall(nbr in g.neighbors(v)){
           count = count + nbr.A;   // regular assignment /*@\label{reduction-count}*/
    }
   accum += count; //sum reduction in assignment form /*@\label{reduction-accum}*/
}
\end{lstlisting}

\section{fixedPoint and Min/Max}
Several solutions to graph algorithms are iterative, and converge based on conditions on node attributes. 
\name provides a fixedPoint construct to specify this succinctly. Its syntax involves a boolean variable and a boolean expression on node-properties forming the convergence condition, as shown below.

\texttt{\keywordSP{fixedPoint until} (var: convergence expression) \{...\}}

\noindent Line~\ref{computeSSSP-stat-9} of Figure~\ref{fig:staticSSSP} in SSSP's specification  uses the \texttt{fixedPoint} construct to define the convergence condition. The loop iterates till at least one node's \texttt{modified} property is set to true. 

\name provides constructs \texttt{Min} and \texttt{Max} which perform multiple assignments based on a comparison criterion. This can be useful in update-based algorithms like SSSP, where an update on node properties is carried out on a  desired condition, while taking care of potential data races.

In the SSSP computation, the \texttt{Min} construct is used to encode the relaxation criteria in Line~\ref{computeSSSP-stat-16}.
The neighboring node's distance is updated if the alternative distance via vertex \texttt{v} is smaller than \texttt{nbr}'s current distance. The update of the \texttt{dist} property based on this comparison specified using \texttt{Min} results in an update of the \texttt{modified} property to \texttt{True}. 

\name also has aggregate functions \texttt{minWt} and \texttt{maxWt} to find the minimum and maximum edge weights.
\subsubsection{fixedPoint and Min/Max Constructs}

Several solutions to graph algorithms are iterative, and converge based on conditions on node attributes. 
\name provides a fixedPoint construct to specify this succinctly. Its syntax involves a boolean variable and a boolean expression on node-properties forming the convergence condition, as shown below.

\texttt{\keywordSP{fixedPoint until} (var: convergence expression) \{...\}}

\noindent Line~\ref{computeSSSP-stat-9} of Figure~\ref{fig:staticSSSP} in SSSP's specification  uses the \texttt{fixedPoint} construct to define the convergence condition. The loop iterates till at least one node's \texttt{modified} property is set to true. 

\name provides constructs \texttt{Min} and \texttt{Max} which perform multiple assignments based on a comparison criterion. This can be useful in update-based algorithms like SSSP, where an update on node properties is carried out on a  desired condition, while taking care of potential data races.

In the SSSP computation, the \texttt{Min} construct is used to encode the relaxation criteria in Line~\ref{computeSSSP-stat-16}.
The neighboring node's distance is updated if the alternative distance via vertex \texttt{v} is smaller than \texttt{nbr}'s current distance. The update of the \texttt{dist} property based on this comparison specified using \texttt{Min} results in an update of the \texttt{modified} property to \texttt{True}. 

\name also has aggregate functions \texttt{minWt} and \texttt{maxWt} to find the minimum and maximum edge weights.
}

\subsection{Abstract Syntax Tree}
StarPlat Dynamic uses Abstract Syntax Tree (AST) as the intermediate representation of the source DSL. An AST serves as a hierarchical representation of a program's syntactic structure, where each node corresponds to a specific programming construct. These constructs include fundamental operations such as variable assignments (e.g., \texttt{x = 5}), declarations (e.g., int y), control flow structures like if-else conditionals, and iterative operations such as for and while loops. It also includes nodes for the Batch Processing which process the sets of input files in the series of batch. Moreover, it contains AST nodes for addition and deletion of edges to the graph which is OnAdd and OnDelete respectively. 

\REM{
The tree structure mirrors the nested relationships between these constructs, enabling precise modeling of code semantics.
The \texttt{ASTNode} class forms the foundational base for all node types within the AST. This parent class is subclassed into specialized categories which reflect distinct programming constructs. For instance, \texttt{StatementNode} encapsulates executable operations, while \texttt{ExpressionNode} represents computational logic.

The StatementNode class branches into granular subtypes, each governing a specific category of executable instructions:

\begin{itemize}
    \item Assignment statements handle variable value assignments.
    \item Declaration statements manage variable or function definitions.
    \item Control flow statements dictate program flow through conditionals (e.g., if, switch) and loops (e.g., while, for).
    \item Procedure calls execute predefined functions or methods.
\end{itemize}
The AST’s design directly reflects the parser’s grammatical specifications. Each node’s composition—such as the nesting of loop bodies within while nodes—is dictated by the language’s syntax rules. This ensures that the tree not only captures surface-level syntax but also enforces semantic validity through its hierarchical constraints.
}

\begin{figure}
\centering
\includegraphics[scale=0.8]{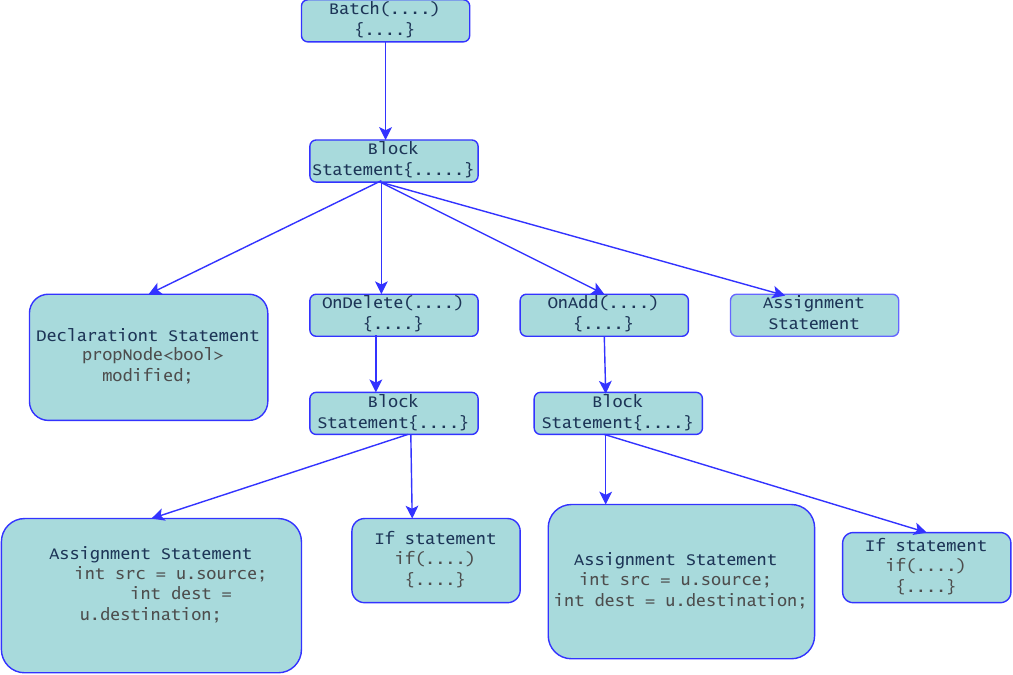}
\caption{AST for Dynamic SSSP in Figure~\ref{ssspdyn-dsl-dyn} } 
\label{fig:ast-sssp}
\end{figure}

Figure~\ref{fig:ast-sssp} shows AST for Dynamic SSSP code from Figure~\ref{ssspdyn-dsl-dyn}.

\subsection{Modified Graph Representation: diff-CSR}
\label{sec diffcsr}
Compressed Sparse Row (CSR) is a way to represent (primarily) sparse matrices, rows of which mostly contain zeroes. The idea is to lay out the non-zero entries in a row into contiguous memory locations, and concatenating these adjacencies. This is unlike the memory representation followed in adjacency lists and adjacency matrices. It reduces space overheads as well as random memory accesses. Another advantage of the CSR representation is that it works with offsets rather than pointers, making it better-suited for cross-device transfers without needing serialization-deserialization (needed in pointer based representations such as adjacency lists). However, the CSR representation is not suited for dynamic graphs.

\REM {
Fig~\ref{fig:CSR} depicts the CSR representation of an example graph. The out-degree of vertex \textit{i} is the difference between the offset values of \textit{i} and \textit{i+1}. 
\begin{figure}
\centering
\includegraphics[scale=0.3]{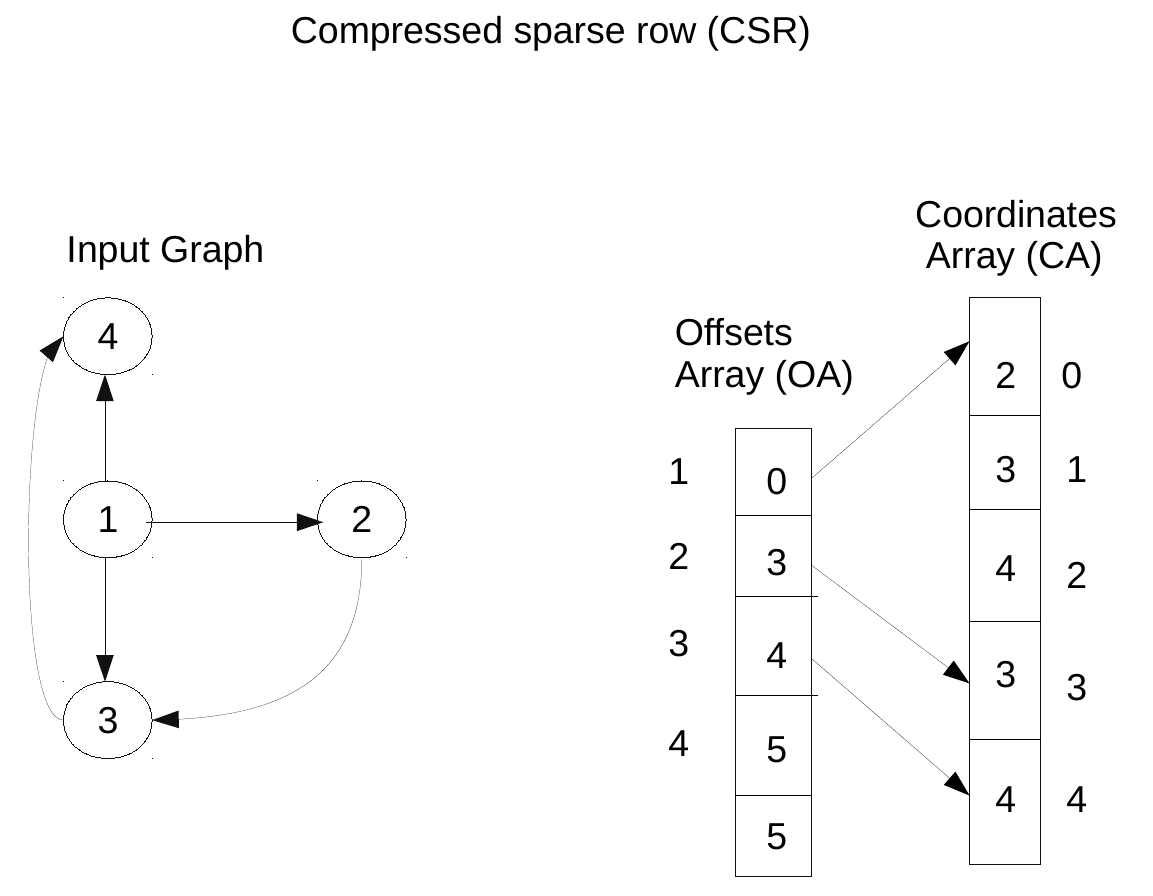}
\caption{CSR Representation}
\label{fig:CSR}
\end{figure}
}

For instance, insertion of a new edge \textit{u → v} necessitates the augmentation of \textit{v} into \textit{u}’s adjacency offset list, thereby necessitating a cascading shift of all the subsequent elements in the coordinate / neighbour array — an operation marked by pronounced data movement and latency. Further, concurrent edge insertions across multiple threads introduces formidable synchronization challenges. 
To ameliorate these inefficiencies, we incorporate an auxiliary diff-CSR~\cite{10.1007/978-3-030-35225-7_17} structure in tandem with the canonical CSR array, thereby facilitating seamless dynamic updates, alleviating synchronization overheads, and preserving structural coherence with minimal perturbation to the underlying CSR structure. We explain this with an example.  

\begin{figure}
\centering
\includegraphics[scale=0.8]{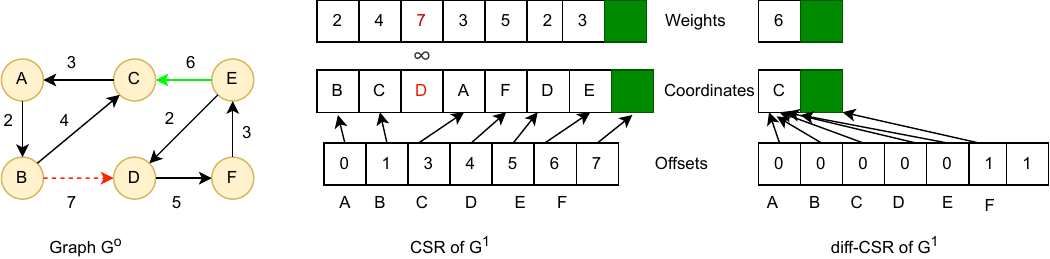}
\caption{Dynamic graph representation. Edge B→D got deleted and edge E→C got
added to G$^0$.} 
\label{fig:graphRep}
\end{figure}

Figure~\ref{fig:graphRep} shows an example initial graph $G^0$ with six vertices and seven edges. It is represented in the CSR format. The CSR representation consists of two arrays: \texttt{offsets} which stores the index where the adjacency list for each vertex begins in the second array, and \texttt{coordinates} which stores the neighbors of each vertex. For instance, \texttt{offsets[C]} is 3, which indicates that vertex \texttt{C}'s adjacency list in the \texttt{coordinates} array begins at offset 3. The number of neighbors of a vertex \texttt{i} is calculated with the help of the next vertex's offset: \texttt{offsets[i+1] - offsets[i]}. Thus, vertex \texttt{C}'s adjacency list is \texttt{\{A\}} and that of vertex \texttt{B} is \texttt{\{C, D\}}. Note that the \texttt{coordinates} array is simply a concatenation of the adjacency lists of vertices. For weighted graphs, a third array \texttt{weights} is maintained, whose length is equal to the number of edges (plus one marking the end of the array). CSR of the static graph $G^0$ is shown in the figure.

Let the modified graph for a batch of updates be $G^1$. Let the batch consist of a decremental update for edge $B \rightarrow D$ and an incremental update for edge $E \rightarrow C$. As can be seen, insertion of a new edge in the \texttt{coordinates} array necessitates shifting of elements and updates to the offsets, which is time-consuming and sequential. The diff-CSR representation marks the deleted edges by a sentinel $\infty$ in the coordinates array. This not only avoids array offset adjustment, but also reduces synchronization requirement with concurrent inserts, deletes, and traversals. A newly added edge will be placed in a vacant position (marked by $\infty$) in CSR, if available. If no vacant position is available, then the new edge gets added to a new diff-CSR structure (as illustrated in Figure~\ref{fig:graphRep}). The diff-CSR structure also consists of the three arrays: \texttt{offsets}, \texttt{coordinates}, and \texttt{weights}, with the size of the latter two arrays dictated by the number of incremental updates in the batch. For instance, for a single edge $E \rightarrow C$, \texttt{offset[E]} is 0, \texttt{offset[F]} is 1, the only entry in the \texttt{offsets} array is \texttt{C} (apart from the end-of-array marker). An additional incremental update $B \rightarrow E$ could be added either in the CSR array or the diff-CSR array. The CSR and diff-CSR arrays together represent graph $G^1$.

Each update batch, having incremental and decremental updates, leads to a new diff-CSR to represent graph snapshots $G^1, G^2, ...$. After a configurable number of batches (which could be 1), this chain of diff-CSRs could be merged into the main CSR structure, similar to a static graph representation.

While originally proposed for the GPU, we exploit diff-CSR for both OpenMP and CUDA backends, and extend it for the MPI backend.

\subsection{The Distributed diff-CSR for MPI}
\begin{figure}
    \begin{center}
\resizebox{150mm}{!}{\includegraphics *{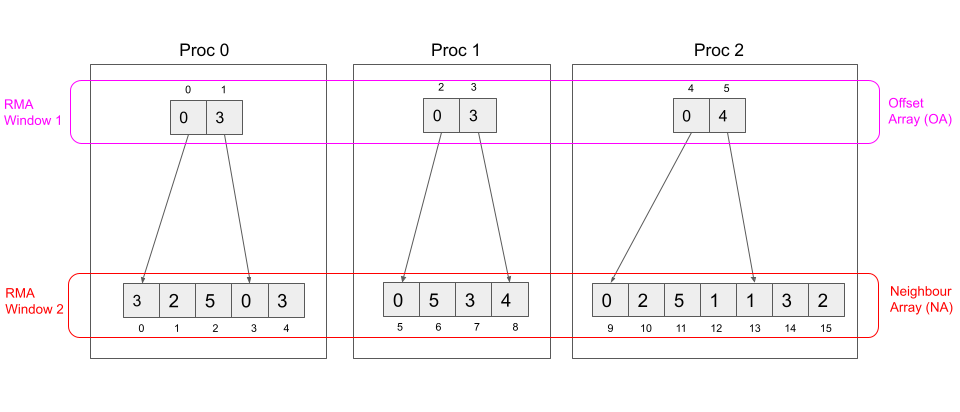}}
    \caption{Distributed CSR for static graphs in the MPI backend of StarPlat Static}
    \label{csr_rma}
    \end{center}
\end{figure}

In MPI, StarPlat stores the graph in a distributed manner across all the processes, as shown in Figure~\ref{csr_rma}, where each node in the graph is owned by a particular process. A process stores only those edges for which the source node is owned by that process. Due to this, we need a way for a process to be able to get the neighbors of nodes stored in some other process. This is readily implementable with \texttt{send-recv} API of MPI. To improve execution, StarPlat uses the Remote-Memory-Access (RMA) feature of MPI, introduced in MPI 3.0. RMA involves only one-sided communication (unlike \texttt{send-recv}) and avoids synchronization requirement across the sender-receiver pair. However, it does not always outperform two-sided communication. We observe that retrieving the adjacency list and other attributes of a vertex does not need to be synchronized, and hence is amenable for RMA. RMA permits exposing data as a \texttt{window}. Thus, each process builds up a CSR for the nodes it owns and all the \texttt{offset} arrays of each CSR resident across all the processes are jointly exposed as an RMA window. Similarly, all the \texttt{coordinate} or neighbor arrays are exposed as another RMA window. Figure~\ref{csr_rma} shows how an example static graph is stored by StarPlat Static in a distributed setup consisting of three processes, with RMA windows exposing offsets and neighbors.

\begin{figure}
    \begin{center}
\resizebox{150mm}{!}{\includegraphics *{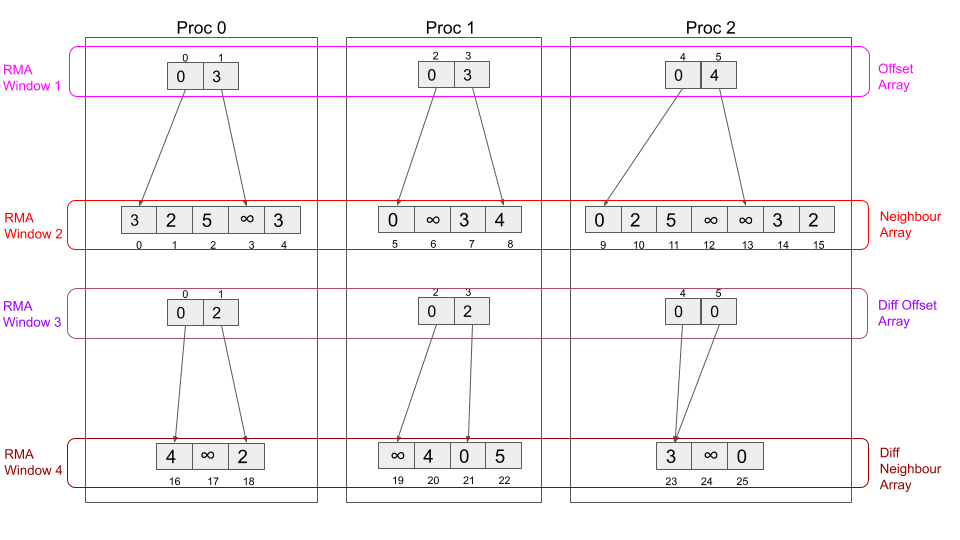}}
    \caption{Distributed diff-CSR for dynamic graphs in the MPI backend of StarPlat Dynamic}
    \label{distributrd_diff_csr}
    \end{center}
\end{figure}

StarPlat Dynamic extends the above technique for storing dynamic graphs too, by distributing the diff-CSR structures across all the processes in a similar way. Every process keeps only that part of diff-CSR which corresponds to the nodes owned by that process. The \texttt{offset} arrays of the diff-CSR structure resident across all the processes are exposed as a single RMA window, while the neighbor arrays are exposed as another RMA window. Thus, the distributed diff-CSR representation consists of four RMA windows, two for the original CSR and two more for the diff-CSR. Figure~\ref{distributrd_diff_csr} shows how the original distributed CSR in Figure~\ref{csr_rma} is extended to also store the diff-CSR in a distributed manner among the MPI processes.

\section{\name Code Generator} \label{sec code gen}
We now explain the the overall flow of StarPlat dynamic.
\subsection{Overall Flow}

The StarPlat Compiler is architecturally bifurcated into two main components: the frontend and the backend, each designed to facilitate a seamless translation from high-level graph abstractions to highly optimized parallel code. The frontend serves as the intellectual gateway to the compiler, ingesting graph algorithms articulated in StarPlat’s expressive domain-specific language (DSL), which gracefully accommodates both static and dynamic variants through a diverse array of syntactic constructs.

This high-level specification undergoes rigorous lexical, syntactic, and semantic analysis, wherein the frontend transmutes the input into a structured Intermediate Representation (IR), embodied as an Abstract Syntax Tree (AST). In tandem, it constructs a richly annotated Symbol Table, encapsulating the semantic essence of program entities—variables, types, scopes, and bindings—thereby laying the semantic groundwork for precise and efficient code synthesis.

\begin{figure}[htpb]
    \begin{center}
\resizebox{130mm}{!}{\includegraphics *{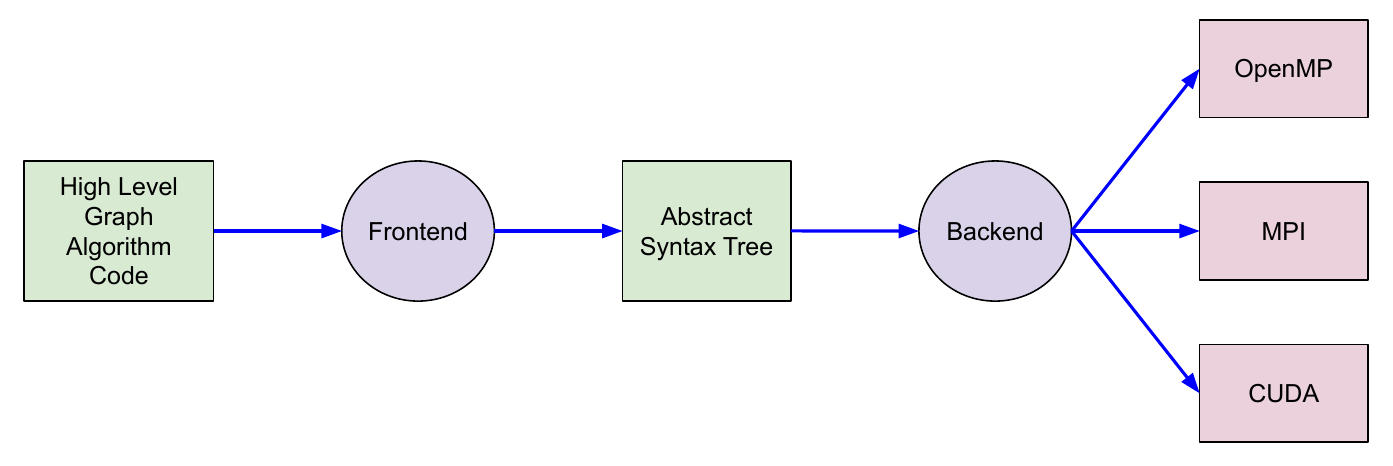}}
    \caption{Overall flow of the StarPlat Compiler}
    \label{dyn-compiler}
    \end{center}
\end{figure}

Subsequently, the AST and Symbol Table are conveyed to the backend, a sophisticated code-generation engine that orchestrates the synthesis of performant, parallelized low-level code tailored for heterogeneous computing environments. The backend emits optimized C/C++ code utilizing OpenMP for shared-memory multi-core architectures, MPI for distributed-memory clusters, and CUDA for many-core, GPU-accelerated systems. This end-to-end compilation workflow, visualized in Fig.~\ref{dyn-compiler}, epitomizes StarPlat’s prowess in bridging high-level declarative algorithm design with the intricacies of low-level parallel execution across diverse architectural paradigms.

\section{Optimizations} \label{sec optimizations}
We explain a few optimizations done for the OpenMP, MPI, and CUDA backends in this section.

\subsection{OpenMP} \label{subsec opt omp}

\textbf{Using built-in atomics.} Atomics have exhibited exceptional efficiency and minimal overhead. Consequently, the generated code harnesses GCC’s intrinsic atomic primitives to ensure thread-exclusive execution of individual statements, thereby obviating the need for conventional locking constructs and enhancing concurrency performance.

\subsection{MPI}
\textbf{Parallel processing of Graph Updates in distributed diff-CSR.}
In MPI, we store the diff-CSR representation of the graph in a distributed manner where each process owns some of the nodes of the graph, and each process stores the CSR and diff-CSR of only the nodes that the process owns. This leads to storing the dynamic graph in an efficient manner and also facilitates faster parallel processing of dynamic graph updates. Each graph processes the updates of only those nodes that it owns and updates diff-CSR accordingly.  \\

\noindent\textbf{RMA Communication Optimization.}
In MPI, we use one-sided communication by using the Remote-Memory-Access (RMA) for a process to be able to get the neighbors of nodes that are stored at some other process. Using MPI RMA over another basic mode of communication allows decoupling of data movement and process synchronization. This decoupling is important for code generation for dynamic graph algorithms as it leads to more generalization of the code generation process, which is rather essential as computation in graph algorithms is inherently irregular. We also experimented with RMA in MPI using passive mode synchronization using the \texttt{MPI\_Put} and \texttt{MPI\_Get} calls, but this required using an \texttt{MPI\_Exclusive} lock on the target process while using \texttt{MPI\_Put}. This initially increased the synchronization overhead in inter-process communication for MPI-generated dynamic graphs as while one process had an \textit{exclusive lock} on one process, other processes could not access the data on the same process even if they were requesting a different part of the data. To optimize this we moved towards the use of MPI atomic operations for inter-process communication like \texttt{MPI\_Accumulate} and \texttt{MPI\_Get\_Accumulate} function which can work with the MPI Shared lock and hence multiple processes can have an MPI Shared lock on the same target process and achieve a higher level of synchronization mechanism~\cite{Dinan2016AnIA}. 

\subsection{CUDA}
\noindent\textbf{Optimized host-device transfer.}
We perform a rudimentary program analysis of the AST to identify variables that need to be transferred across devices. For instance, since the graph is dynamic, it need not be copied back from GPU to CPU at the end of the kernel launch. In contrast, the modified properties need to be transferred back. Similarly, the \texttt{finished} flag is set on CPU, conditionally set on GPU, and read on the CPU again in the fixed-point processing. Therefore, the variable needs to be transferred to-and-fro. The \texttt{forall}-local variables are generated as device-only variables.

\section{Experimental Evaluation} \label{sec exp evaluation}
With StarPlat Dynamic, we generate code for three dynamic graph algorithms in the OpenMP, MPI, and CUDA backends: Single
 Source Shortest Path (SSSP), Page Rank (PR), and Triangle Counting (TC). Their Dynamic DSL codes are presented in Appendix~\ref{starplat:codes}. We use ten
 large graphs to assess the feasibility and effectiveness of our approach. These graphs are a mix of different types:
six large social networks, two road networks, and two are synthetically generated.
One synthetic graph has a uniform random distribution (generated using Green-Marl's graph generator~\cite{GreenMarl}), while the other one has a skewed degree distribution following the recursive-matrix format (generated using SNAP’s RMAT generator with parameters a = 0.57, b = 0.19, c = 0.19, d = 0.05). They are listed in Table~\ref{graph-inputs}, sorted on the number of edges in each category.

\begin{table}
\caption{Input graphs ($\delta$ indicates degree)} 
\centering
\small
\setlength{\tabcolsep}{4pt}
\begin{tabular}{r|c|r|r|r|r}
 \hline
 \textbf{Graph} & \textbf{Short} & \multicolumn{1}{c|}{\textbf{$|$V$|$}} &  \multicolumn{1}{c|}{\textbf{$|$E$|$}}  & \textbf{Avg. $\delta$} & \textbf{Max. $\delta$} \\
    & \textbf{name} & \multicolumn{1}{c|}{$\times 10^6$} & \multicolumn{1}{c|}{$\times 10^6$} & & \\
 \hline
 twitter-2010 & TW   & 21.2  & 265.0 & 
 12.0 & 302,779\\ 
 soc-sinaweibo & SW &   58.6  & 261.0 & 
 4.0 & 4,000 \\
 orkut & OK & 3.0 & 234.3 
 & 76.3  &33,313\\
 wikipedia-ru & WK & 3.3 & 93.3 
 & 55.4 & 283,929\\
 livejournal &LJ &  4.8 & 69.0 
 &28.3 & 22,887\\
 soc-pokec & PK & 1.6 & 30.6 
 & 37.5 & 20,518\\
 \hline
 usaroad & US & 24.0  & 28.9 
 & 2.0 & 9  \\
 germany-osm & GR & 11.5  & 12.4 
 & 2.0  & 13\\ \hline
 rmat876 & RM & 16.7 & 87.6  
 & 5.0 & 128,332\\
 uniform-random & UR & 10.0 & 80.0 
 & 8.0 & 27 \\
 \hline
\end{tabular}
\label{graph-inputs}
\end{table}

All our experiments (with OpenMP, MPI, and CUDA) were run on a compute cluster where the configuration of each compute node as follows: Intel Xeon Gold 6248 CPU with 40 (hyperthreaded) hardware threads spread over two sockets, 2.50~GHz clock, and 192 GB memory. 
All the codes are in C++, compiled with GCC, with optimization flag -O3. Various backends have the following versions: OpenMP version 4.5, OpenMPI version 3.1.6, and CUDA version 10.1.243. The CUDA code was executed on an Nvidia Tesla V100 GPU with 5120 CUDA cores spread uniformly across 80 SMs clocked at 1.38 GHz with 32 GB global memory. The GPUs are installed on the same host nodes used for running the CPU codes.

The execution times for different graphs are plotted against varying percentages of graph updates. This is due to the varying number of edges in the graphs, which differ significantly. Therefore, employing a percentage of the total number of edges instead of a fixed number of updates for all graphs provides a more accurate assessment of the dynamic algorithm’s performance for each individual graph. The results are obtained by running the update as a batch. For a given percentage of updates, we calculate the execution time for both dynamic and static processing for each algorithm. For calculating the static algorithm time, the updates to the graph are performed at the start in a static manner, and the properties are calculated from scratch. The analysis of these plots offers a clear understanding of the efficiency that dynamic computation provides compared to the static computation. We show this analysis for OpenMP, MPI, and CUDA. 

We present a thorough and nuanced comparative evaluation of StarPlat’s statically compiled graph algorithms against a selection of prominent and widely adopted graph processing frameworks, including Galois~\cite{Galois}, Ligra~\cite{Ligra}, Green-Marl~\cite{GreenMarl}, Gemini~\cite{GeminiGraph}, GRAFS~\cite{grafs-ligra}, and Gunrock~\cite{Gunrock}. These results are presented in Tables~\ref{table:openmp-table} through~\ref{table:cuda-table}. 
It is important to highlight that none of these frameworks currently supports automated code generation for dynamic or morph graph workloads. As a result, a direct, apples-to-apples comparison with StarPlat's dynamic compilation capabilities is not feasible. Therefore, to ensure a well-rounded and comprehensive evaluation, we constrain our analysis to the static graph domain, wherein StarPlat’s statically generated code is benchmarked against the best-performing implementations from these existing frameworks.

\subsection{Overall Results: OpenMP, MPI, CUDA}
Tables~\ref{table:all:openmp}, \ref{table:all:mpi}, and \ref{table:all:cuda} present a comparison between StarPlat Static and StarPlat Dynamic across the ten graphs for the three algorithms for three backends. The second column indicates the percentage of updates, which includes both incremental and decremental ones. We evaluate the runtimes of dynamic algorithms for varying percentages of updates from 1\% to 20\%. However, for the SSSP and PR algorithms on the MPI backend, we used updates ranging from 0.1\% to 2\%, as we found this range provided a better comparison with its static counterpart as seen in Figure \ref{fig:sssp_chart} and \ref{fig:pr_chart}.
Next, we provide a detailed description of each backend individually.

\input{10expt-openmp}

\input{11expt-mpi}

\input{12expt-cuda}

\subsection{OpenMP}\label{subsec exp omp}
Table~\ref{table:all:openmp} provides a comprehensive comparison between static and dynamic graph algorithms evaluated over ten representative graphs. It details the execution times for SSSP, TC, and Page Rank algorithms under varying update percentages, thereby highlighting the performance trends and adaptability of each approach.
Now, we will discuss the trends for each algorithm one by one.

\begin{figure}[htpb]
    \begin{center}
    \resizebox{100mm}{!}{\includegraphics *{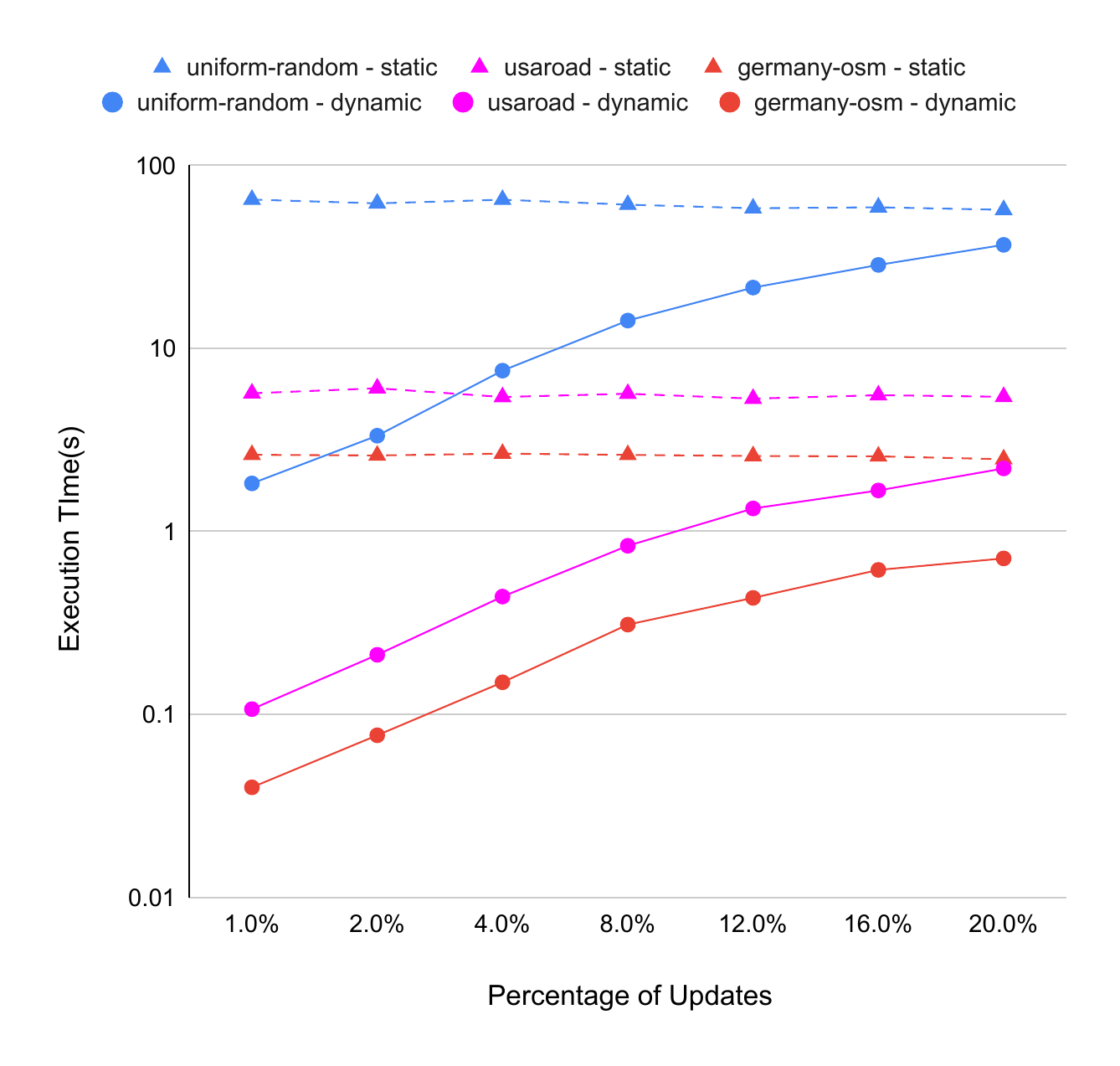}}
    \caption{Dynamic and Static Algorithms Runtime Comparison for Triangle Counting (TC) for OpenMP. (Note different y-axis scales)}
    \label{fig:tcUpdates}
    \end{center}
\end{figure}

\textbf{Triangle Counting.} The results of the plot in Fig.~\ref{fig:tcUpdates} demonstrate that the dynamic triangle counting approach yields a notable improvement in performance compared to the static variant, regardless of the percentage of updates. The execution times for the static implementation remained relatively consistent across different percentages of updates for a given graph. On average, the dynamic variant outperformed the static variant by nearly 30 times for lower percentages of updates.

\begin{figure}[htpb]
    \begin{center}
    \resizebox{100mm}{!}{\includegraphics *{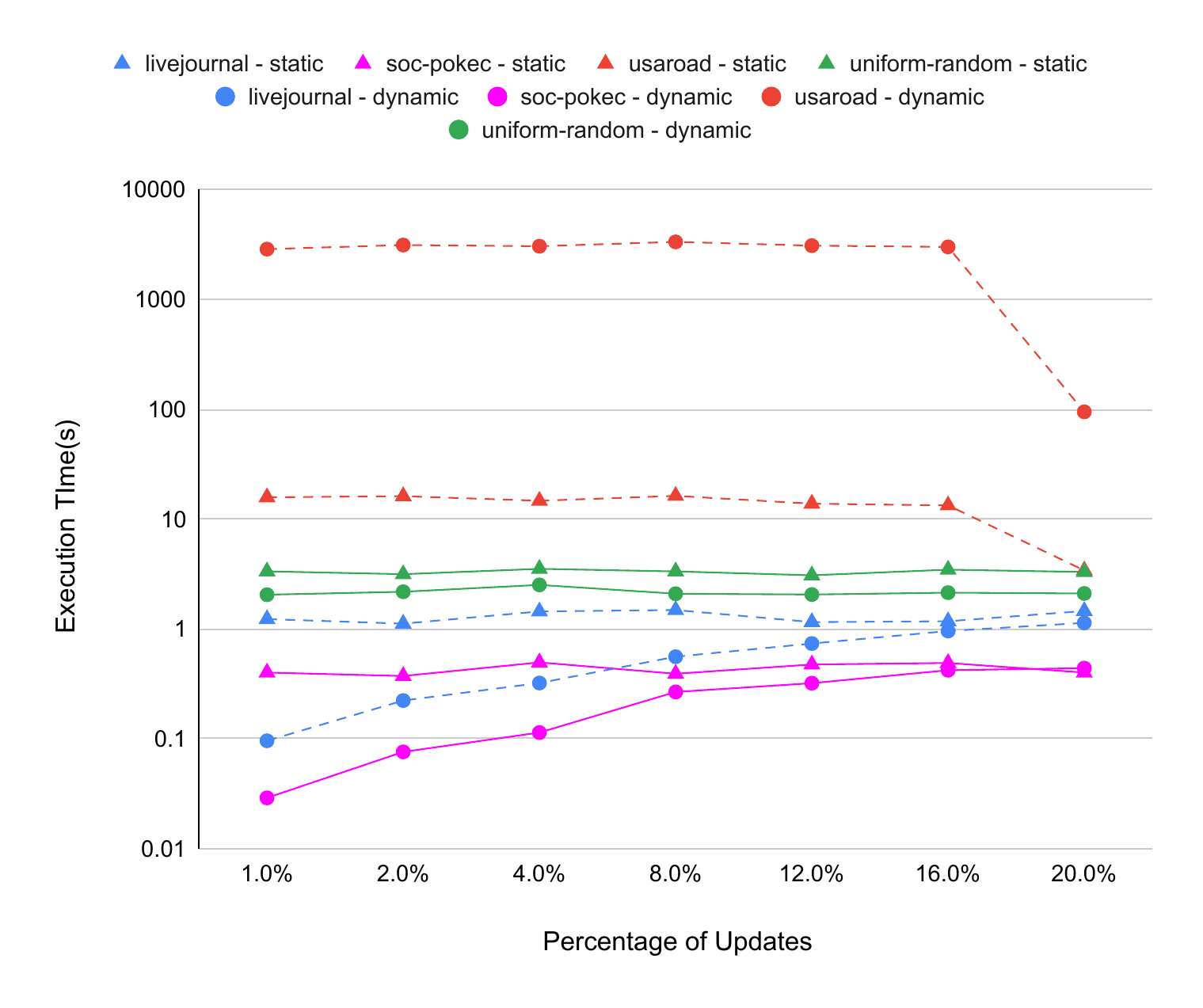}}
    \caption{Dynamic and Static Algorithms Runtime Comparison for Single Source Shortest Path (SSSP) for OpenMP. (Note different y-axis scales)}
    \label{fig:ssspUpdates}
    \end{center}
\end{figure}

\textbf{Single-Source Shortest Paths.} 
The plot in Fig~\ref{fig:ssspUpdates} illustrates that the dynamic variant outperforms the static variant in three out of four graphs up to a certain percentage of updates. Specifically, the dynamic variant is nearly five times faster than the static variant for lower percentages of updates. However, as the percentage of updates continues to increase, the runtime of the dynamic algorithm also increases, while the runtime of the static algorithm remains constant. Notably, for the USA road network (US), the dynamic version performs worse than the static version. This can be attributed to slower convergence resulting from a pull-based decremental implementation. Therefore, it is worth considering exploring a push-based version (each node actively sends or "pushes" updates or information to its neighbors), as it has the potential to be more efficient.

\begin{figure}[htpb]
    \begin{center}
    \resizebox{100mm}{!}{\includegraphics *{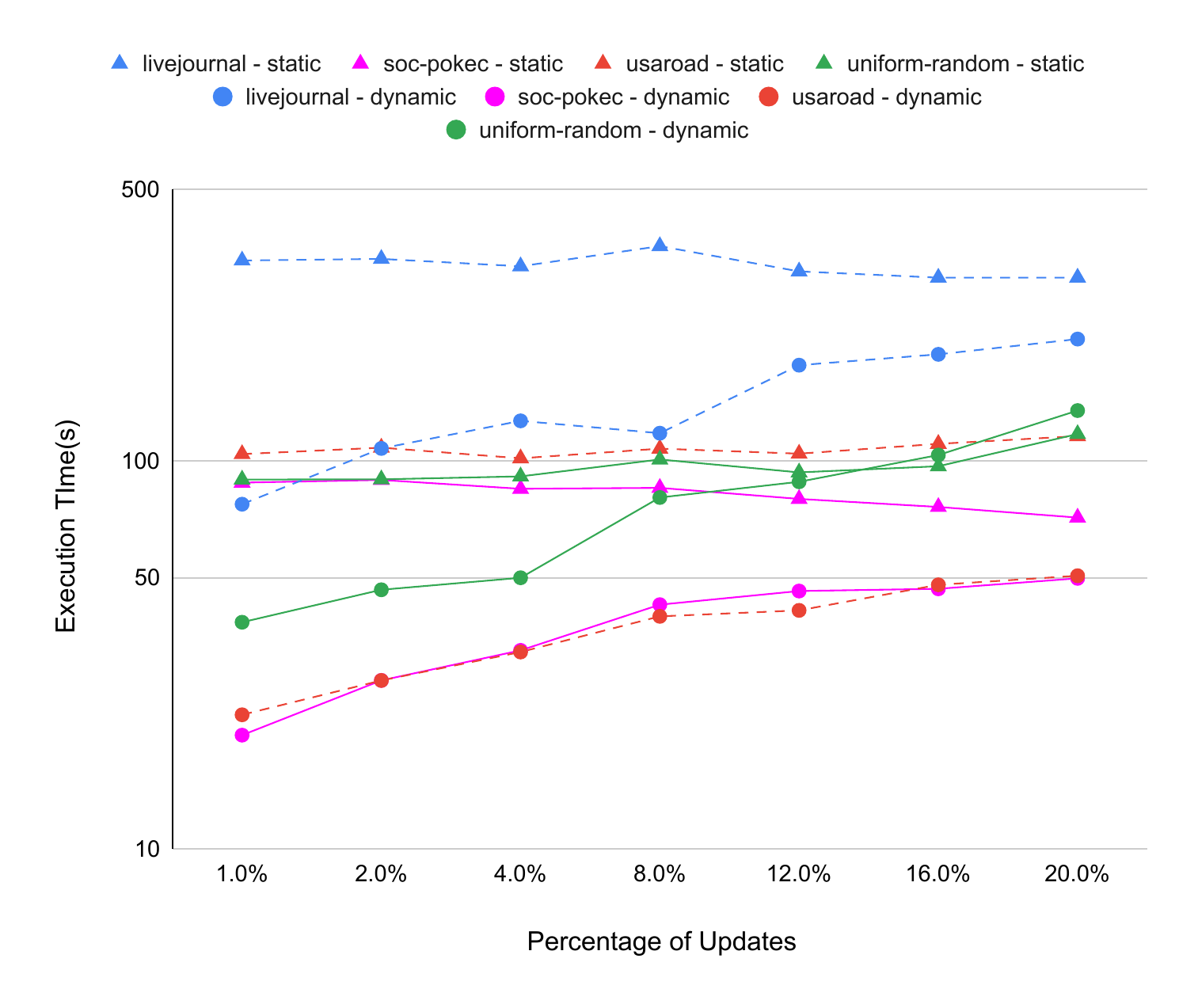}}
    \caption{Dynamic and Static Algorithms Runtime Comparison for Page Rank (PR) for OpenMP. (Note different y-axis scales)}
    \label{fig:prUpdates}
    \end{center}
\end{figure}

\textbf{Page Rank.} The plot in Fig~\ref{fig:prUpdates} provides evidence that the dynamic Page Rank algorithm outperforms the static variant in terms of performance across different graphs, even when a significant percentage of updates is introduced. Interestingly, the runtime of the static implementation remains constant irrespective of the percentage of updates, which could be attributed to the similar iteration counts required for convergence across different update percentages. On the other hand, the dynamic runtime exhibits an increase as the update percentage rises.

\arrayrulecolor{gray}
\begin{table*}
\caption{\name's OpenMP code performance comparison against Galois, Ligra, and Green-Marl (20 Threads). All times are in seconds. The minimum times for each graph are \textbf{boldfaced}. Galois framework failed to load the largest graph TW (exited with a segfault).}
\label{table:openmp-table}
\scalebox{0.9}{
\setlength{\tabcolsep}{2pt}
\begin{tabular}{|rr|r||r|r|r|r|r|r|r|r|r|r||r|}
\hline
\multicolumn{2}{|c|}{Algo.} & Framework & TW  & SW  & OK  & WK & LJ & PK & US & GR & RM & UR & Total\\ \hline

\multicolumn{2}{|l|}{PR}                             & \begin{tabular}[c]{@{}l@{}}Galois\\ {{Ligra}}\\ Green-Marl\\GRAFS\\\name\end{tabular} & \begin{tabular}[c]{@{}l@{}}---\\ 25.600\\ \textbf{0.585}\\ 132.000\\1.752\end{tabular}          & \begin{tabular}[c]{@{}l@{}}\textbf{0.510}\\ 162.660\\ 7.211\\ 843.000\\9.002\end{tabular}      & \begin{tabular}[c]{@{}l@{}}\textbf{0.647}\\ 5.050\\ 1.437\\ 18.800\\1.213\end{tabular}     & \begin{tabular}[c]{@{}l@{}}\textbf{0.371}\\ 3.930\\ 0.512\\ 11.100\\0.473\end{tabular}         & \begin{tabular}[c]{@{}l@{}}\textbf{0.474}\\ 3.623\\ 0.585\\ 9.830\\0.509\end{tabular}   & \begin{tabular}[c]{@{}l@{}}\textbf{0.156}\\ 0.836\\ 0.263\\ 2.640\\0.236\end{tabular}   & \begin{tabular}[c]{@{}l@{}}\textbf{0.607}\\ 2.050\\ 1.235\\ 2.950\\1.600\end{tabular}      & \begin{tabular}[c]{@{}l@{}} \textbf{0.224}\\ 0.822\\ 0.525\\ 1.840\\0.667\end{tabular}     & \begin{tabular}[c]{@{}l@{}}\textbf{0.324}\\ 5.880\\ 0.821\\ 33.800\\0.883\end{tabular} & \begin{tabular}[c]{@{}l@{}}\textbf{0.443}\\ 0.942\\ 0.688\\ 10.000\\0.619\end{tabular}  & \begin{tabular}[c]{@{}l@{}} --- \\ 211.393 \\ 13.862 \\ 1065.960\\16.954 \end{tabular} 
\\ \hline
\multicolumn{2}{|l|}{SSSP}                           & \begin{tabular}[c]{@{}l@{}}Galois\\ Ligra\\ Green-Marl\\GRAFS\\ \name\end{tabular} & \begin{tabular}[c]{@{}l@{}}{---}\\ 10.700\\ 2.182\\ \textbf{0.014}\\5.831\end{tabular}          & \begin{tabular}[c]{@{}l@{}} {0.132}\\ 0.148\\ 0.891\\\textbf{0.037} \\1.437\end{tabular}   & \begin{tabular}[c]{@{}l@{}}{0.404}\\ 5.136\\ 1.048\\\textbf{0.253}\\1.850\end{tabular} & \begin{tabular}[c]{@{}l@{}}\textbf{0.203}\\ 1.846\\ 1.160\\ 0.262\\1.759\end{tabular}   & \begin{tabular}[c]{@{}l@{}}\textbf{0.205}\\ 3.890\\ 0.761\\ 0.208\\2.412\end{tabular}  & \begin{tabular}[c]{@{}l@{}}{0.099}\\ 1.683\\ 0.292\\\textbf{0.071}\\0.846\end{tabular}   & \begin{tabular}[c]{@{}l@{}}{19.387}\\ 283.000\\ 193.548\\ \textbf{0.534}\\{{294.303}}\end{tabular}   & \begin{tabular}[c]{@{}l@{}}{6.798}\\ 9.043\\ 48.349\\ \textbf{0.376}\\{{53.395}}\end{tabular}    & \begin{tabular}[c]{@{}l@{}}\textbf{0.133}\\ 2.740\\0.464\\0.462 \\1.369\end{tabular}       & \begin{tabular}[c]{@{}l@{}}{0.480}\\ 10.400\\ 1.361\\ \textbf{0.181}\\5.237\end{tabular}  & \begin{tabular}[c]{@{}l@{}} ---\\ 328.586\\ 250.056 \\ 2.398\\368.439 \end{tabular} 
\\ \hline
\multicolumn{2}{|l|}{TC}                             & \begin{tabular}[c]{@{}l@{}}Galois\\ Ligra\\ {{Green-Marl}}\\ \name\end{tabular} & \begin{tabular}[c]{@{}l@{}}---\\ 2103.333\\  11611.029\\ 1414.323\end{tabular} & \begin{tabular}[c]{@{}l@{}}\textbf{56.432}\\ 188.660\\ 4257.498\\ 59.925\end{tabular} & \begin{tabular}[c]{@{}l@{}}33.110\\ \textbf{22.800}\\ 137.559\\ 23.420\end{tabular}   & \begin{tabular}[c]{@{}l@{}}\textbf{46.168}\\ 97.360\\ 4564.568\\ 111.430\end{tabular} & \begin{tabular}[c]{@{}l@{}}9.811\\ 10.460\\ 29.426\\ \textbf{7.544}\end{tabular} & \begin{tabular}[c]{@{}l@{}}3.008\\ 1.926\\ 12.705\\ \textbf{1.559}\end{tabular}  & \begin{tabular}[c]{@{}l@{}}0.061\\ 0.147\\ 0.065\\ \textbf{0.059}\end{tabular}   & \begin{tabular}[c]{@{}l@{}}\textbf{0.020}\\ 0.0698\\ 0.021\\ 0.024\end{tabular}     & \begin{tabular}[c]{@{}l@{}}184.260\\ \textbf{130.330}\\ 5647.156\\ 158.760\end{tabular} & \begin{tabular}[c]{@{}l@{}}2.350\\ 1.706\\ 1.435\\ \textbf{1.176}\end{tabular}  & \begin{tabular}[c]{@{}l@{}}---\\ 2556.792\\ 14650.430\\ 1778.220\end{tabular}     
\\ \hline

\end{tabular}
}

\label{openmp-table}
\end{table*}

\begin{table*}
\caption{\name's SSSP OpenMP code running times (seconds) with \texttt{static} scheduling}
\begin{tabular}{|l|l|l|l|l|l|l|l|l|l|l|}
\hline
\multirow{2}{*}{SSSP} & TW   & SW    & OK    & WK    & LJ    & PK    & US   & GR    & RM    & UR    \\ \cline{2-11} 
                      & 4.127 & 0.127 & 1.503 & 0.633 & 2.315 & 0.822 & 72.654 & 9.641 & 1.319 & 4.477 \\ \hline
\end{tabular}

\label{openmp-dyn-table}
\end{table*}

\textbf{Static algorithm runtime comparison for OpenMP with
other frameworks} Table~\ref{table:openmp-table} presents the running times of the codes in various frameworks for the three algorithms. The execution time
refers to only the algorithmic processing and excludes the graph reading time, and is an average over three runs (for
each backend). GRAFS uses the Ligra framework underneath.

\textit{Page Rank} For all inputs, the PR code generated by \name is competitive with that of Green-Marl. It's interesting to note that Ligra takes much longer and has slower implementation. The main cause of this is the loop separation that occurs between the difference of successive PR values for each vertex and the PR value computation. The slowest code is that synthesized by GRAFS caused by a change in the termination criterion. In every other framework, the termination is determined by the Page Rank values' convergence as well as a maximum number of iterations. However, GRAFS solely considers the number of iterations for determining convergence. Overall, Galois performs better than all other benchmarks, being roughly 3$\times$ faster than \name-generated code. This is due to the in-place update of the PR values for vertices, which leads to faster convergence. \name, Green-Marl, Ligra, and GRAFS follow a similar processing of updating the PR values using double buffering. 

\textit{Single-Source Shortest Paths.} We discover that GRAFS SSSP outperforms other benchmarks and \name's code. In terms of performance, Galois comes in second place and is orders of magnitude faster than Ligra, Green-Marl, and \name. The Galois framework~\cite{galois-scheduling} uses application-specific prioritized scheduling, which is the cause of this. For instance, processing tasks in the ascending distance order reduces the total amount of extra work done. When it comes to SSSP computation, Green-Marl and StarPlat have very comparable implementations. Both follow a dense push configuration for vertex processing which needs iterating over all the vertices to determine if they are active. For road networks, where a smaller frontier is active in each cycle, this is usually expensive. Furthermore, because road networks have a big diameter, this additional cost builds up over a number of iterations. However, for almost all graphs, Green-Marl outperforms StarPlat. Green-Marl reduces superfluous CPU cycles by implementing spin-lock with the back-off technique. Although StarPlat likewise employs a lock-free atomic-based implementation, unnecessary updates causes performance deterioration due to false sharing. Compared to other benchmarks, Ligra's performance is not very competitive. Depending on the frontier size, Ligra alternates between sparse and dense edge processing. However, with the exception of the road networks, this direction optimization does not significantly enhance performance for the provided graph suite.

\name creates OpenMP code with \texttt{dynamic} scheduling by default. For the most part, this is effective with a variety of graph types and methods. Table~\ref{openmp-dyn-table} indicates that, on the whole, SSSP code appears to perform better with static scheduling. The execution times of the big diameter graphs US and GR drop significantly from over a minute to a few seconds, indicating a significant difference.

\textit{Triangle Count} In TC, Galois, \name, and Green-Marl all adhere to a node-iterator paradigm. Ligra, on the other hand, uses an edge-iterator-based version, which has better load balance and is, therefore, expected to perform better for skewed degree graphs. We find that different frameworks perform better for different graphs. We observe that for different graphs, different frameworks outperform. It's interesting to note that the code created by Green-Marl performs much worse.

\subsection{MPI}\label{subsec exp mpi}
Table~\ref{table:all:mpi} presents an in-depth evaluation of static versus dynamic graph algorithms across ten diverse graph datasets. It outlines the runtime performance of SSSP, TC, and PageRank under different levels of update intensity, thereby shedding light on the efficiency and scalability characteristics of each method.
Now, we will discuss the trends for each algorithm one by one.

\begin{figure}[htpb]
    \begin{center}
    \resizebox{100mm}{!}{\includegraphics *{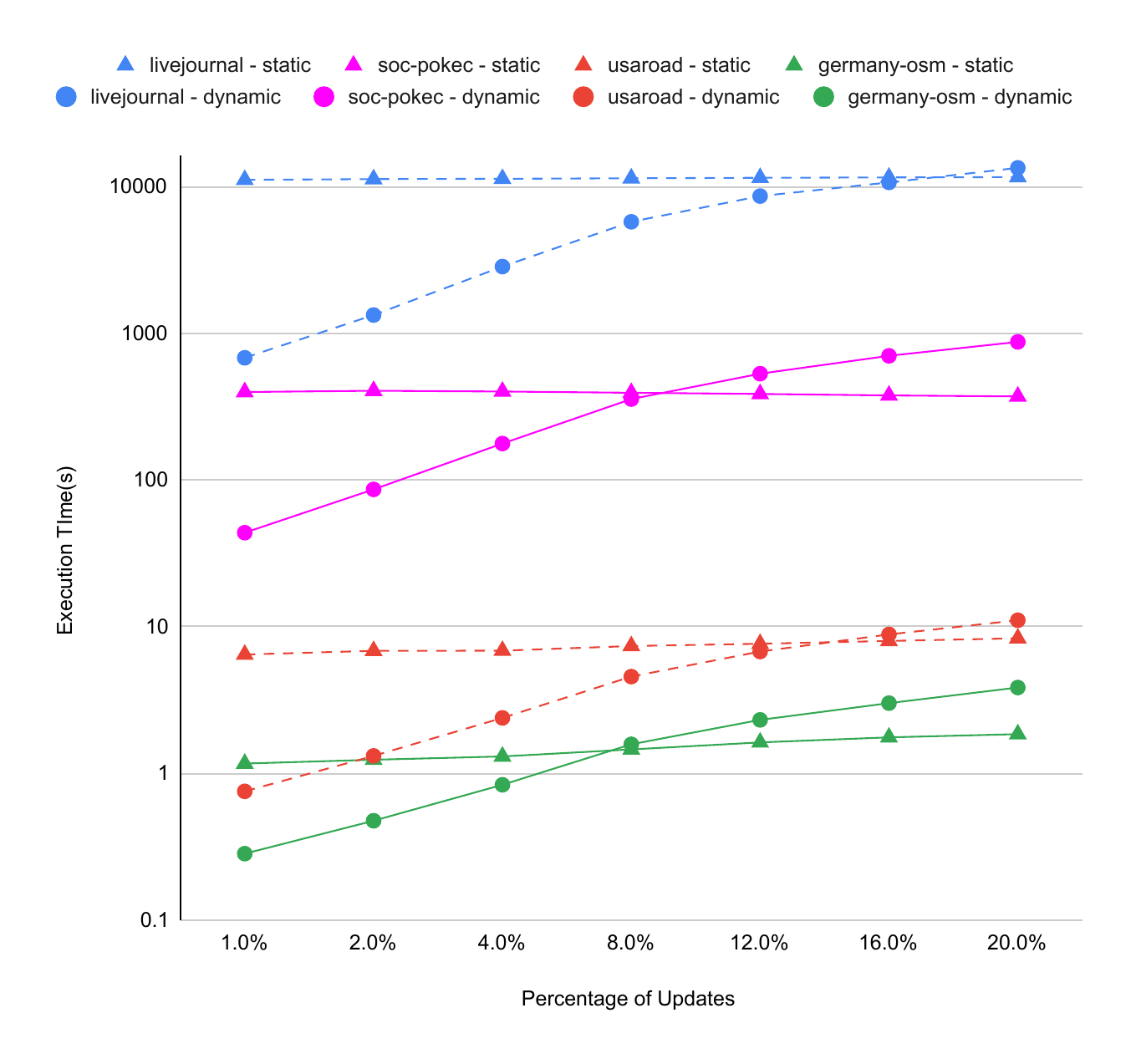}}
    \caption{Dynamic and Static Algorithms Runtime Comparison for Triangle Counting (TC) for MPI. (Note different y-axis scales)}
    \label{fig:tc_chart}
    \end{center}
\end{figure}

\textbf{Triangle Counting.} Figure \ref{fig:tc_chart} shows that the dynamic triangle counting algorithm outperforms its static counterpart for four of the nine graphs. The dynamic TC algorithm is faster than the static TC for these four graphs even when the percentage of updates is as high as 8\%. The dynamic algorithm's runtime follows an increasing trend with the increase in the percentage of updates and crosses the static runtime at a certain point, which itself remains relatively constant across different percentages of updates. On average, for the triangle counting algorithm, the dynamic variant outperformed the static variant by nearly 10 times for lower percentages of updates. For the remaining five graphs, most of which are social network graphs, the code did not terminate within 3 hrs. The reason for this is that the DSL TC algorithm uses a triple nested for loop to get neighbors (code \ref{TCdyn-dsl}), and because of the nature of the social network graphs, some of the nodes have a huge degree; getting neighbors of such a node from another process makes the communication between the two processes the bottleneck for the distributed backend.

\begin{figure}[htpb]
    \begin{center}
\resizebox{100mm}{!}{\includegraphics *{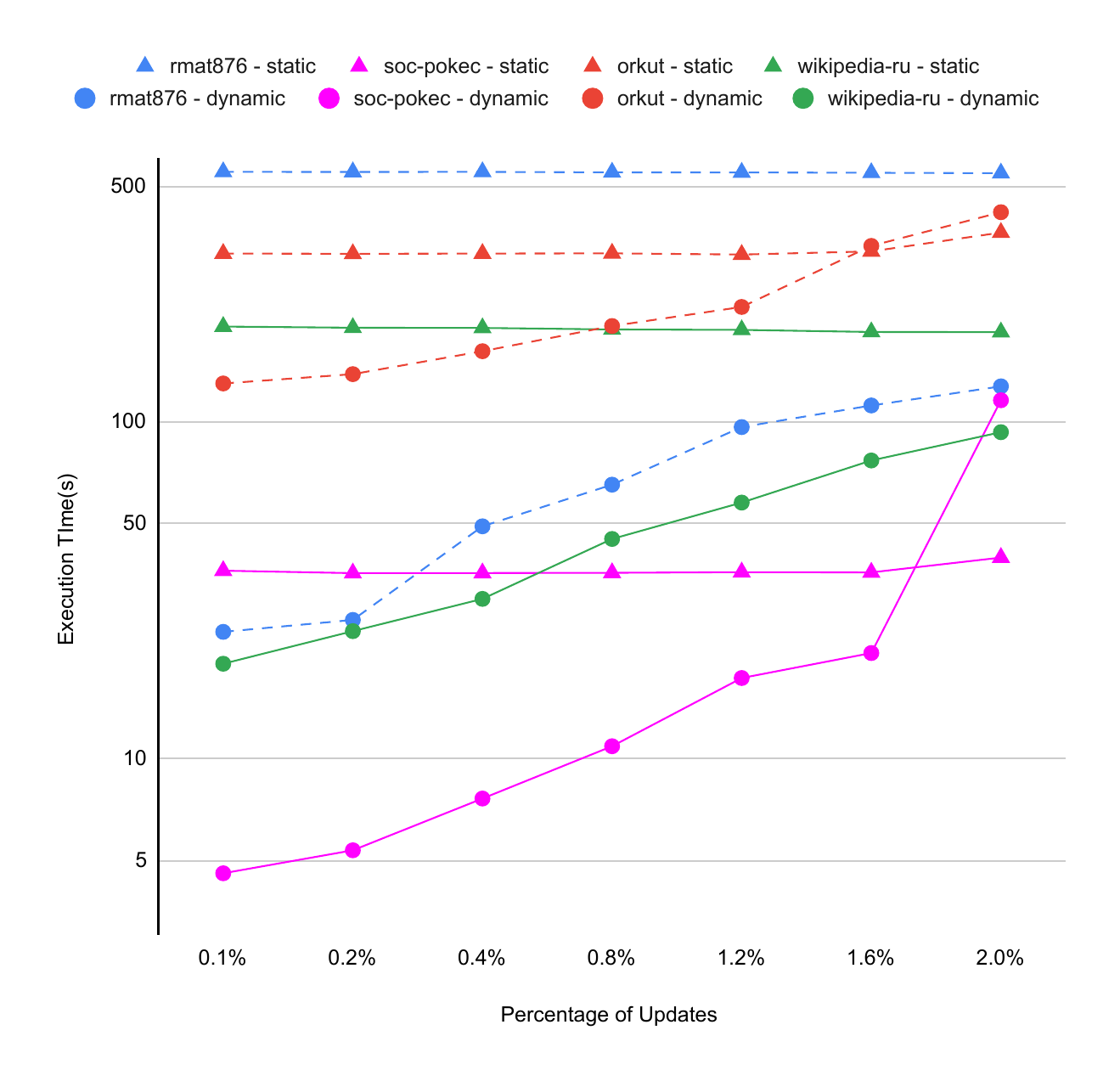}}
    \caption{Dynamic and Static Algorithms Runtime Comparison for Single Source Shortest Path (SSSP) for MPI. (Note different y-axis scales)}
    \label{fig:sssp_chart}
    \end{center}
\end{figure}

\textbf{Single-Source Shortest Paths.} Fig~\ref{fig:sssp_chart} shows the dynamic SSSP algorithm in comparison with its static counterpart. Here, for SSSP, we have used the percentage of updates ranging from 0.1\% to 2\%, unlike previous analyses where we have used updates from 1\% to 20\%. This is because we found this range provided a better comparison with its static counterpart as seen in Figure \ref{fig:sssp_chart} and a further higher percentage of updates led to a much steeper increase in the runtimes. We see that the dynamic variant performs better for the four graphs up to a certain percentage of updates, which is consistent with other previous observations. The dynamic algorithm is almost 10 times faster as compared to static for certain graphs for a lower percentage of updates. As the percentage of updates keeps on increasing, the runtime of the dynamic algorithm increases while the runtime of the static algorithm remains almost the same. At a higher percentage of updates, the dynamic algorithm becomes slower as compared to the static algorithm, and there is a steep rise in the runtime thereafter.  

\begin{figure}[htpb]
    \begin{center}
\resizebox{100mm}{!}{\includegraphics *{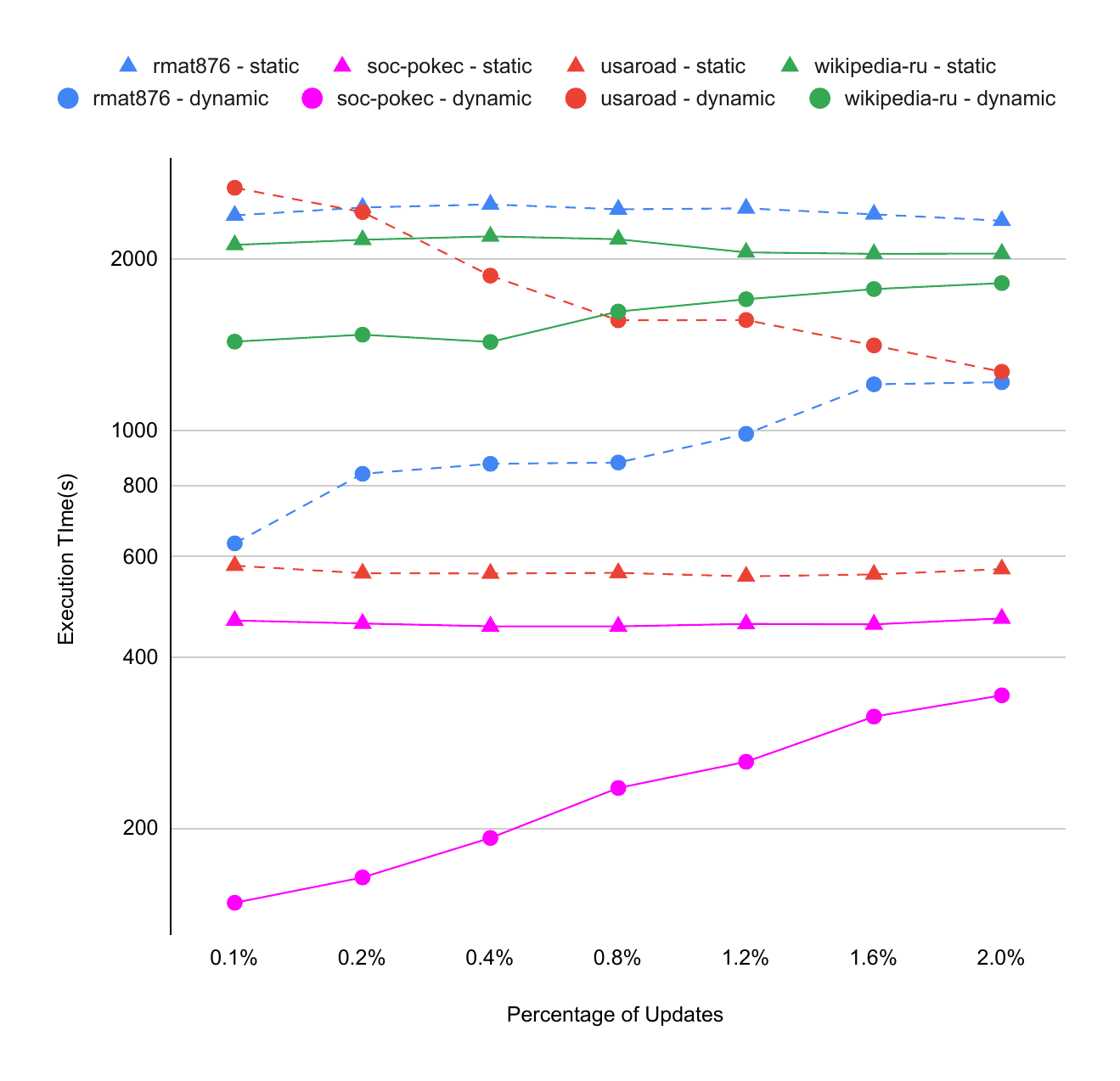}}
    \caption{Dynamic and Static Algorithms Runtime Comparison for Page Rank (PR) for MPI. (Note different y-axis scales)}
    \label{fig:pr_chart}
    \end{center}
\end{figure}

\textbf{Page Rank.} From Fig~\ref{fig:pr_chart} we see that the dynamic PR algorithm follows a similar trend. Like SSSP for PR, we also use a range of percentages of updates varying from 0.1\% to 2\%. For three out of four graphs, the dynamic PR algorithm performs better than the static PR algorithm up to a certain percentage of updates. For rmat876 (RM), soc-pokec (PK), and Wikipedia (WK), the dynamic algorithm performs better than its static counterpart even when the percentage of updates is as high as 2\%. Though for Graph USA Road (US) the dynamic algorithm's runtime decreases as the percentage of updates increases. The reason for this anomalous behavior is that the dynamic PR algorithm uses an in-built StarPlat method called propagateNodeFlags (Code \ref{PRdyn-dsl}). What this method does is propagate a particular boolean node property, which might be set for some nodes, to all the nodes in its connected component. Thus, the underlying implementation does a BFS traversal of the graph for propagateNodeFlags. As a result, when the percentage of updates (which are random) is small the propagateNodeFlags method requires a higher number of BFS iterations to propagate the flags to other nodes in its connected components as compared to when the number of percentage updates is large. As USA Road is a road network with a large diameter, the runtime for the propagateNodeFlags method is very high as compared to other graphs because of higher BFS iterations that would be required to traverse it. Hence the propagateNodeFlags method dominates the runtime for the USA road (US) network because of its higher diameter, but this does not happen for other graphs. And as the runtime for propagateNodeFlags decreases with increasing updates we see a decreasing trend in runtime for the dynamic algorithm for the two graphs but not for the other graphs. We observe the similar anomalous behaviour for the Germany Road network graph too for the MPI backend in the Table~\ref{table:all}.

\textbf{Static algorithm runtime comparison for MPI with other frameworks}
Table \ref{table:mpi-table} shows the comparison of execution times of StarPlat’s MPI code against the distributed Galois framework, Gemini framework, and Gemini target code generated by the GRAFS framework (PR and SSSP). For Page Rank StarPlat-generated PR code outperforms the Galois, Gemini, and GRAFS-synthesized codes for four out of the 10 graphs. From the comparison of execution times of SSSP we can observe that, Galois performs better for five graphs LJ, US, PK, RM and WK while GRAFS-generated Gemini code performs better for OK. StarPlat performs better for the remaining four graphs. The TC code generated by StarPlat is less efficient as it iterates through neighbors of each node multiple times and checks for neighborhood among these nodes. Due to this, StarPlat’s TC code takes much more time than Galois. For road graphs US and GR, StarPlat’s TC takes less time than the Galois code. For some graphs such as TW, SW, WK and RM StarPlat’s code did not give results even after 24 hours. TC implementation of Gemini is not available and hence it is not compared.

\arrayrulecolor{gray}
\begin{table*}[htb!]
\caption{\name's MPI code performance comparison (3 nodes - 32 processes per node). All times are in seconds. Note that PR for GRAFS(Gemini) doesn't set the value of beta = 0.001 and runs for max-iteration that is 100. The minimum times for each graph are \textbf{boldfaced}.}
\scalebox{0.93}{
\setlength{\tabcolsep}{3pt}
\begin{tabular}{|ll|l||r|r|r|r|r|r|r|r|r|r||r|}
\hline
\multicolumn{2}{|l|}{Algo.}                                      & Framework & \multicolumn{1}{c|}{TW} & \multicolumn{1}{c|}{SW} & \multicolumn{1}{c|}{OK} & \multicolumn{1}{c|}{WK} & \multicolumn{1}{c|}{LJ} & \multicolumn{1}{c|}{PK} & \multicolumn{1}{c|}{US} & \multicolumn{1}{c|}{GR} & \multicolumn{1}{c|}{RM} & \multicolumn{1}{c||}{UR} & \multicolumn{1}{c|}{Total} \\ \hline

\multicolumn{2}{|l|}{\multirow{2}{*}{PR}}                        & Galois    & 398                 & \textbf{761}                 & 314                  & \textbf{376}                  & 551                 & 479                 & \textbf{257}                 & 296                 & \textbf{244}                 & 268                 & 3944                 \\

\multicolumn{2}{|l|}{}                                           & Gemini  & 3027         & 3080              & 555         & 572         & 2998          & 341          & 3025          & 3026          & 3024         & 3051          & \multicolumn{1}{c|}{22699}   \\

\multicolumn{2}{|l|}{}                                           & GRAFS  & 3027         & 3361              & 464         & 474         & \textbf{441}          & \textbf{295}          & 3096          & 3014          & 3035         & 3079          & \multicolumn{1}{c|}{20286}   \\

\multicolumn{2}{|l|}{}                                           & StarPlat  & \textbf{262}         & 1917              & \textbf{81}         & 2027         & 1626          & 463          & 579          & \textbf{65}          & 2173        & \textbf{259}          & \multicolumn{1}{c|}{9452}   \\ \hline

\multicolumn{2}{|l|}{\multirow{2}{*}{SSSP}}                      & Galois    & 33         & 11                  & 30         & \textbf{63}                  & \textbf{27}         & \textbf{24}                  & \textbf{1933}       & 82                  & \textbf{20}         & 49                  & 2272                 \\

\multicolumn{2}{|l|}{}                                           & Gemini  & 255                & 58          & 56                & 502         & 79                  & 58         & 34522                & 8754         & 58                  & 128          &  44470               \\ 

\multicolumn{2}{|l|}{}                                           & GRAFS  & \textbf{3}                & 1          & \textbf{24}                & 503         & 34                  & 34         & 15956                & 8699         & 25                  & 74          &            25353     \\ 

\multicolumn{2}{|l|}{}                                           & StarPlat  & \textbf{0.4}                 & \textbf{0.1}          & 312                 & 191         & 166                  & 36         & 4895                & \textbf{76}         & 549                  & \textbf{42}          & 6367                \\ \hline

\multicolumn{2}{|l|}{\multirow{2}{*}{TC}}                        & Galois    & \textbf{71275}      & \textbf{436}        & \textbf{327}        & \textbf{884}        & \textbf{174}        & \textbf{19}          & 7                   & 9                   & \textbf{1238}       & \textbf{10}         & 74379                \\
\multicolumn{2}{|l|}{}                                           & StarPlat  & \textgreater 24 Hrs     & \textgreater 24 Hrs     & \textgreater 24 Hrs                & \textgreater 24 Hrs     & 11504                & 390                & \textbf{5}          & \textbf{1}          & \textgreater 24 Hrs     & 32                  & \multicolumn{1}{c|}{---}   \\ \hline
\end{tabular}
}
\label {table:mpi-table}
\end{table*}

\subsection{CUDA}\label{subsec exp CUDA}
Table~\ref{table:all:cuda} offers a detailed assessment of static and dynamic graph algorithms across a collection of ten varied graph datasets. It reports the execution times for SSSP, TC, and PageRank under varying degrees of graph updates, thereby providing insights into the efficiency and scalability of the respective approaches.
Now, we will discuss the trends for each algorithm one by one.

\begin{figure}[htpb]
    \begin{center}
    \resizebox{100mm}{!}{\includegraphics *{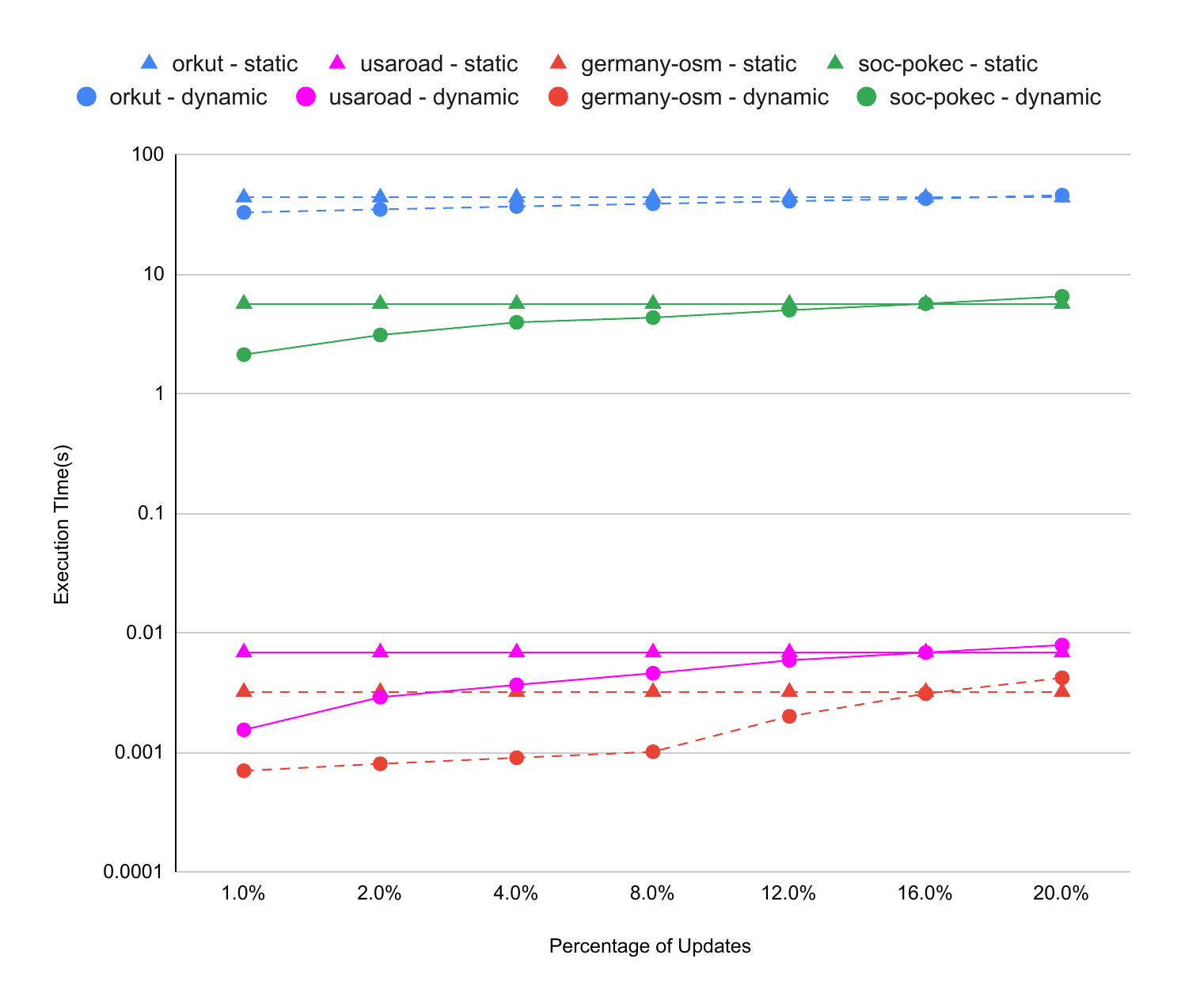}}
    \caption{Dynamic and Static Algorithms Runtime Comparison for Triangle Counting (TC) for CUDA. (Note different y-axis scales)}
    \label{fig:tc_chart_cuda}
    \end{center}
\end{figure}

\textbf{Triangle Counting.} Figure \ref{fig:tc_chart_cuda} demonstrates that the dynamic triangle counting approach yields a noteworthy improvement in performance compared to the static variant,
regardless of the percentage of updates. The execution times for the static implementation remained relatively consistent across different percentages of updates for a given graph. On average, for the triangle counting algorithm, the dynamic variant outperformed the static variant by nearly 5 times for
almost all percentages of updates except 20\%.

\begin{figure}[htpb]
    \begin{center}
\resizebox{100mm}{!}{\includegraphics *{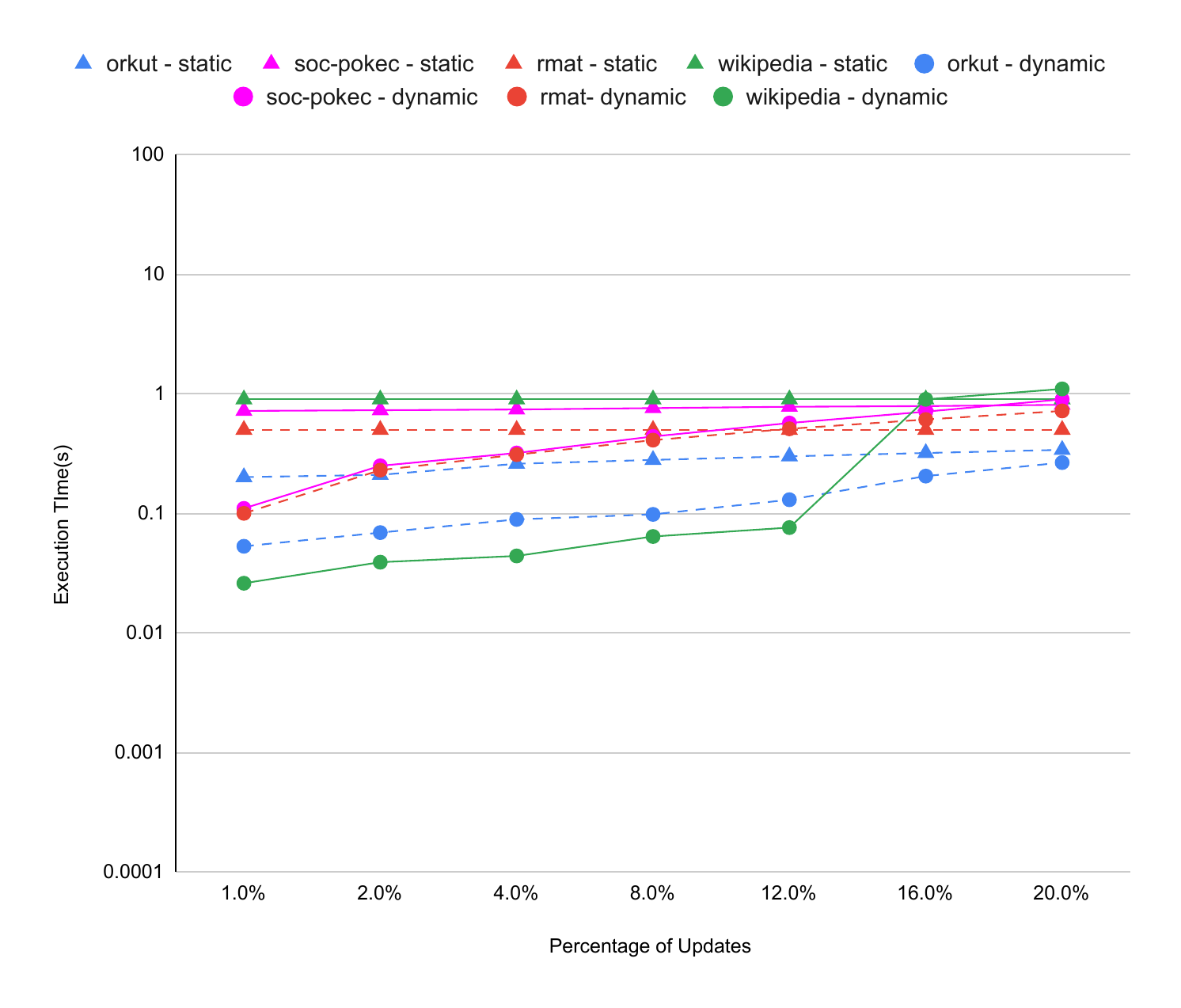}}
    \caption{Dynamic and Static Algorithms Runtime Comparison for Single Source Shortest Path (SSSP) for CUDA. (Note different y-axis scales)}
    \label{fig:sssp_chart_cuda}
    \end{center}
\end{figure}

\textbf{Single-Source Shortest Paths.} Fig.~\ref{fig:sssp_chart_cuda} vividly underscores the exceptional efficiency of the dynamic SSSP approach in comparison to its static counterpart, irrespective of the percentage of updates. The static implementation demonstrates relatively stable execution times across different update ratios, reflecting its inflexibility in adapting to changes. In contrast, the dynamic variant exhibits remarkable adaptability and responsiveness, consistently outperforming the static method by nearly a factor of four in most cases. Particularly noteworthy is its performance on the Wikipedia-ru (WK) graph, where at lower update percentages, the dynamic algorithm operates up to 34 times faster than the static version, highlighting its superior scalability and effectiveness in dynamic graph processing.

\begin{figure}[htpb]
    \begin{center}
\resizebox{100mm}{!}{\includegraphics *{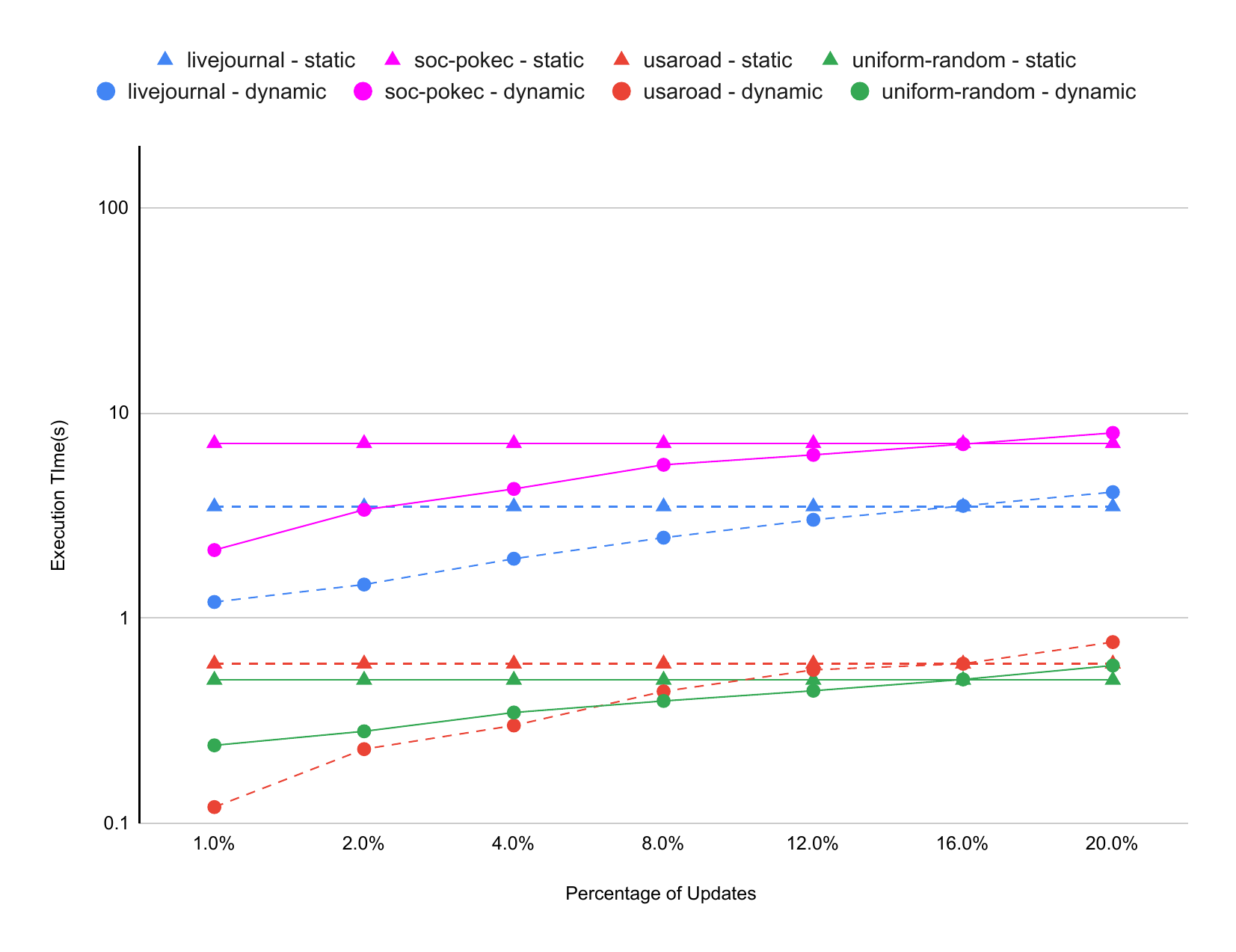}}
    \caption{Dynamic and Static Algorithms Runtime Comparison for Page Rank (PR) for CUDA. (Note different y-axis scales)}
    \label{fig:pr_chart_cuda}
    \end{center}
\end{figure}

\textbf{Page Rank.} From Fig.~\ref{fig:pr_chart_cuda} showcases the pronounced performance benefits of the dynamic Page Rank algorithm over its static counterpart. Across a broad range of graphs and update percentages, the dynamic approach consistently demonstrates superior efficiency and adaptability. In several instances, particularly at lower update rates, the dynamic variant achieves up to a 3 times acceleration in execution time. Most notably, on the usaroad (US) graph, the dynamic algorithm delivers an impressive 5 times speedup, highlighting its remarkable scalability and effectiveness in handling evolving graph structures.

\textbf{Static algorithm runtime comparison for CUDA with other frameworks}
Table \ref{table:cuda-table} shows the comparison of execution times of StarPlat’s CUDA code against LonestarGPU and Gunrock v1.2.

\textit{Triangle Counting} is not a propagation-based algorithm. In addition, it is
characterized by a doubly-nested loop inside the kernel (to iterate through neighbors of neighbors). Another iteration is required for checking edge, which can be implemented linearly or using binary search if the neighbors are sorted in the
CSR representation. Due to this variation in the innermost loop, the time difference across various implementations can
be pronounced, which we observe across the three frameworks. Their performances are mixed across the ten graphs, and clearly, StarPlat consumes considerably more time.

For \textit{Single-Source Shortest Paths}, LonestarGPU, and StarPlat outperform Gunrock on all ten graphs. Between
LonestarGPU and StarPlat, there is no clear winner. They, in fact, have competitive execution times across graphs.

For the \textit{Page Rank} algorithm, we observed that the three frameworks have consistent relative performance, with hand-crafted LonestarGPU codes outperforming the other two and StarPlat outperforming Gunrock. StarPlat exploits the double buffering
approach to read the current PR values and generate those for the next iteration (see Listing 7). This separation reduces synchronization requirements during the update but necessitates a barrier across iterations. LonestarGPU uses an in-place update of the PR values and converges faster.

\arrayrulecolor{gray}
\begin{table*}

\caption{\name's CUDA code performance comparison against LonestarGPU and Gunrock v1.2. All times are in seconds. LonestarGPU fails to load the largest graph TW ({\MLE} == out of memory). The minimum times for each graph are \textbf{boldfaced}.
}
\scalebox{0.9}{
\setlength{\tabcolsep}{3pt}
    \begin{tabular}{|lr|c|r|r|r|r|r|r|r|r|r|r||r|}
\hline
\multicolumn{2}{|l|}{Algo.} & Framework & TW & SW & OK & WK & LJ & PK & US & GR & RM & UR & Total \\ \hline
 
\multicolumn{2}{|l|}{\multirow{3}{*}{PR}} & LonestarGPU & - & \textbf{0.240} & 0.363 & \textbf{0.104} & \textbf{0.225} & \textbf{0.240} & \textbf{0.832} & \textbf{0.294} & \textbf{0.240} & \textbf{0.240} & --- \\
\multicolumn{2}{|l|}{} & Gunrock v1.2 & 15.230 & 36.910 & 2.430 & 2.460 & 2.952 & 1.085 & 13.345 & 6.499 & 9.170 & 5.487 & 95.568 \\
\multicolumn{2}{|l|}{} & \name & \textbf{4.081} & 7.112 & \textbf{0.256} & 1.780 & 1.300 & 0.257 & 3.420 & 0.679 & 0.891 & 0.257 & 20.033 \\ \hline
\multicolumn{2}{|l|}{\multirow{3}{*}{SSSP}} & LonestarGPU & - & 0.077 & 0.217 & 0.058 & 0.084 & 0.037 & \textbf{0.162} & \textbf{0.091} & 0.129 & 0.183 & --- \\
\multicolumn{2}{|l|}{} & Gunrock v1.2 & 2.272 & 4.057 & 0.616 & 0.556 & 0.562 & 0.311 & 1.283 & 1.140 & 1.034 & 0.915 & 12.746 \\
\multicolumn{2}{|l|}{} & \name & \textbf{0.001} & \textbf{0.002} & \textbf{0.078} & \textbf{0.044} & \textbf{0.027} & \textbf{0.012} & 1.667 & 0.695 & \textbf{0.120} & \textbf{0.028} & 2.674 \\ \hline
\multicolumn{2}{|l|}{\multirow{3}{*}{TC}} & LonestarGPU & - & 31.990 & 2.998 & 2.771 & \textbf{0.110} & \textbf{0.039} & 11.874 & 5.695 & \textbf{1.270} & 0.499 & --- \\
\multicolumn{2}{|l|}{} & Gunrock v1.2 & \textbf{67.718} & 7.369 & \textbf{0.843} & \textbf{0.997} & 0.850 & 0.404 & 1.490 & 0.712 & 3.200 & 1.040 & 84.623 \\
\multicolumn{2}{|l|}{} & \name & 10540.002 & \textbf{1.410} & 46.700 & 4.009 & 3.006 & 0.655 & \textbf{0.001} & \textbf{0.001} & 824.620 & \textbf{0.034} & 11420.430 \\ \hline
\end{tabular}
}


\label {table:cuda-table}
\end{table*}

\section{Related Work} \label{sec relatedwork}
In the subsequent two subsections, we systematically categorize the pertinent body of related work based on the nature of the underlying algorithms. The discourse primarily encompasses prior research on static graph algorithms, wherein the graph topology remains invariant, and dynamic graph algorithms, which are designed to efficiently manage and adapt to evolving graph structures. A comprehensive and meticulously curated summary of existing literature on dynamic graph algorithms is presented in Table~\ref{related-work-dynamic}. It shows the chronological order of different data structures or frameworks which evolved over time for the dynamic graph algorithms.


\begin{table}
\caption{Related Work Summary For Dynamic Graph Algorithms. F: Framework D: Data Structure YOP: Year of Publication
Note that the table is sorted in chronological order of YOP} 
\centering
\small
\setlength{\tabcolsep}{4pt}
\begin{tabular}{l|c|c|c}

 \hline
  \textbf{Related Work} & \textbf{F or D}  & \textbf{Backend Support} & \textbf{YOP} 
  \\

 \hline

  \textsf{STINGER~\cite{STINGER}} & D  &  CPU (OpenMP) & 2012
 \\

 \textsf{cuSTINGER~\cite{cutstinger}} & D  & GPU  & 2016
 \\
\textsf{GPMA~\cite{Sha2017Accelerating}} & D  & GPU & 2017
 \\
 \textsf{aimGraph~\cite{autonomous}} & D  & GPU  & 2017
 \\

\textsf{SlabGraph~\cite{slabgraph, slabgraph1}} & D  & GPU & 2018
 \\
 \textsf{Hornet~\cite{hornet}} & D  & GPU & 2018
 \\
 \textsf{faimGraph~\cite{famigraph}} & D &  GPU & 2018
 \\
 \textsf{Aspen~\cite{aspen}} & F & CPU Multithreading & 2019
 \\
 
  \textsf{GraphOne~\cite{graphone}} & F & CPU & 2020
 \\

 \textsf{Terrace~\cite{terrace}} & F  &  CPU (OpenMP) & 2021
 \\

  \textsf{Teseo~\cite{tesco}} & D & CPU (PThreads) & 2021
 \\
  \textsf{Sortledon~\cite{Sortledton}} & D  &  CPU Multithreading & 2022
 \\

  \textsf{GraphFly~\cite{graphfly}} & F & MPI & 2022
 \\
 
 \textsf{EGraph~\cite{Egraph}} & F  & GPU  & 2023
 \\

\textsf{LPMA~\cite{Zoudynamic}} & D & GPU & 2023
 \\

 \textsf{CommonGraph~\cite{commongraph}} & F &  CPU (OpenMP) & 2023
 \\

\textsf{Meerkat~\cite{concessao2024meerkat}} & F &  GPU & 2024
 \\
 
 \hline
\end{tabular}

\label{related-work-dynamic}
\end{table}

\subsection{Dynamic Graph Algorithms}

\subsubsection{Dynamic Data Structures}

cuSTINGER~\cite{cutstinger}is a graph data structure intended for NVIDIA GPUs and is used for streaming graphs that change over time. It offers users the option to update the graph in batches or one edge at a time, depending on the requirements of the application. To regulate how much memory is allotted to each vertex, it provides a number of built-in memory allocators.  In order to boost performance for dynamic graph algorithms, it can employ several allocators that trade memory usage.
In contrast to GT-STINGER, which can do both insertion and deletion simultaneously, cuSTINGER isolates these functions.
This is partly because of a GPU's enhanced parallelization and more intricate memory management.

faimGraph~\cite{famigraph} is a 
Graphics Processing Unit (GPU) graph data structure that is entirely dynamic. It uses autonomous memory management right on the GPU to provide rapid update rates with a small memory footprint. The data structure is completely dynamic, enabling modifications to both edges and vertices.

aimGraph~\cite{autonomous} provides a novel dynamic graph data structure that uses autonomous memory management directly on the GPU to provide rapid update speeds with a small memory footprint. Fast initialization times and effective graph structure updating are made possible by moving memory management to the GPU. This is because the device handles memory allocation and reallocation directly, eliminating the need for extra calls.

While many of the current formats for sparse data representations on parallel architectures are only applicable to static data problems, those for dynamic data are inefficient in terms of memory footprint and speed. Hornet\cite{hornet} is a unique data format designed to address dynamic data challenges. Throughout the data evolution process, Hornet does not require any re-allocation or re-initialization of data and is scalable with respect to the input size.

STINGER~\cite{STINGER} is a scalable, high performance graph data structure which enables fast insertions, deletions, and updates on semantic graphs with skewed degree distributions.
STINGER is in C program that uses the Cray MTA and OpenMP pragmas for parallelization. It can run and compile on the Cray XMT supercomputing platform, as well as the Intel and AMD x86 platforms. Python and Java on x86 systems are experimentally supported.

LPMA~\cite{Zoudynamic} represents a significant advancement in dynamic graph processing on GPUs by addressing fundamental limitations in existing approaches. The leveled structure effectively reduces unnecessary re-balancing operations during expansions, while the hybrid strategy adaptively selects the most efficient processing approach based on data characteristics.
The comprehensive experimental evaluation validates the theoretical benefits of LPMA and demonstrates substantial performance improvements across various graph operations and datasets. This work contributes valuable insights into efficient data structure design for high-throughput dynamic graph processing on GPUs, with potential applications in network analysis, social networks, and communication log processing.

GraphTau~\cite{iyerdynamic} represents an important step forward in processing time-evolving graph data at scale. By providing efficient abstractions for handling graph dynamics, fault tolerance, and incremental computation, it enables developers to implement sophisticated analytics on dynamic graph-structured data with less complexity and better performance. The system's unification of data streaming and graph processing paradigms makes it particularly valuable for real-world applications where both models are needed simultaneously.

SlabGraph~\cite{slabgraph} introduces a fully concurrent dynamic hash table for GPUs that achieves performance comparable to state-of-the-art static hash tables. The research addresses a significant limitation in the GPU ecosystem: the lack of data structures that efficiently support incremental updates. It proposes several novel techniques to enable high-performance dynamic operations on GPUs like
Warp-cooperative work sharing strategy: A technique that reduces branch divergence and provides an efficient alternative to traditional per-thread/per-warp work assignments
Slab list: A dynamic non-blocking concurrent linked list that supports asynchronous, concurrent updates (insertions and deletions) as well as search queries
Slab hash: A dynamic hash table implemented using the slab list with chaining for collision resolution
SlabAlloc: A warp-synchronous dynamic memory allocator specifically designed for the high-performance needs of the slab hash.
The hash table implementation uses a linked-list approach for collision resolution, which conveniently allows for concurrent operations using Compare-And-Swap (CAS) operations. The slab hash maintains high performance even at high load factors, making it suitable for dynamic workloads in applications such as computational geometry and bioinformatics.

GPMA~\cite{Sha2017Accelerating} introduces innovative approaches to address a significant challenge in graph processing: efficiently handling dynamic, frequently evolving graphs on Graphics Processing Units (GPUs). It is a lock-based approach that performs effectively in scenarios where few concurrent updates conflict, such as small-sized update batches with randomly distributed edge updates.

Teseo~\cite{tesco} is an innovative system designed for storing and analyzing dynamic structural graphs in main memory with full transactional support. The research presents a significant advancement over existing systems by addressing key limitations in graph data management.
It introduces a novel architecture based on sparse arrays (large arrays interleaved with gaps) and a specialized data structure called "fat trees." Unlike conventional B+ trees, fat trees feature exceptionally large leaves (approximately 2MB each), which serve as the foundation for storing graph data. This design dramatically differs from earlier systems that typically lack transactional support and rely on vertex tables as primary indices. A distinguishing feature of Teseo is its comprehensive transactional support, guaranteeing snapshot isolation for all operations. The system employs an adaptive strategy where sparse array segments are transparently transformed into layouts better suited for either reads or writes based on the observed workload patterns.

Sortledton~\cite{Sortledton} is a universal, transactional graph data structure designed to efficiently handle dynamic graphs with diverse workloads, including analytics, traversals, and pattern matching. It addresses the challenge of supporting high update throughput while maintaining performance comparable to static graph representations like Compressed Sparse Row (CSR). It uses sorted adjacency lists stored in memory-friendly blocks, combining fast scans (similar to contiguous memory) with efficient edge insertions/deletions. For hub vertices with large neighborhoods, it employs an unrolled skip list to balance access speed and update efficiency. Further it provides a unified solution for dynamic graph processing, bridging the gap between update-heavy and compute-intensive workloads with a simpler architecture than specialized alternatives.

\subsubsection{Dynamic Frameworks}

Meerkat~\cite{concessao2024meerkat} represents a significant advancement in dynamic graph processing, demonstrating that GPUs can efficiently handle both computational and update workloads. Key innovations include its hashtable-based graph representation, warp-cooperative execution, and work-efficient iterative patterns. Meerkat implements dynamic versions of Breadth-First Search (BFS), Single-Source Shortest Paths (SSSP), Page Rank and Weakly Connected Components (WCC).

GraphOne~\cite{graphone} represents a significant advancement in graph data store design by successfully bridging the gap between high-performance analytics and rapid data ingestion. Its hybrid storage architecture, dual versioning mechanism, and novel GraphView abstraction provide a unified solution for real-time graph analytics that outperforms specialized systems. GraphOne introduces a new data abstraction called GraphView that provides application-level flexibility:
Static View - Offers real-time versioning of the latest data for batch analytics.
Stream View - Supports stream analytics with the most recent updates.
These views provide visibility of data updates at two granularity levels:
Edge-level granularity via the edge log.
Coarse granularity via the adjacency store.
This abstraction allows applications to make performance tradeoffs based on their specific visibility requirements, enabling diverse analytical workloads on the same underlying data store.

GraphFly~\cite{graphfly} is a high-performance streaming graph processing system that introduces novel techniques to significantly improve efficiency through dependency-aware processing. It addresses a critical inefficiency found in typical streaming graph processing systems. Conventional systems employ a two-phase approach-refinement and recomputation-to ensure correctness in incremental computation. However, this approach creates substantial redundant memory accesses due to unnecessary synchronization among independent edge updates.

Terrace~\cite{terrace} is an innovative system for processing streaming graphs that employs a hierarchical data structure designed to address the challenges of skewed degree distributions in real-world dynamic graphs. It is implemented as a C++ library parallelized using Cilk Plus and the Tapir branch of the LLVM compiler. The code is available as an open-source project, allowing contributions from the community. The system's ability to adapt to skewness makes it particularly valuable for real-world graph applications that exhibit power-law degree distributions.

Aspen~\cite{aspen} enables high-performance, memory-efficient graph streaming on a scale previously unattainable with existing technologies. At the core of the authors' contribution is the C-tree, a compressed purely-functional search tree data structure designed to overcome the limitations of standard purely-functional trees. The fundamental insight behind C-trees is the application of a chunking technique that stores multiple entries per tree node instead of the traditional one-element-per-node approach.

CommonGraph~\cite{commongraph} is an innovative approach for efficiently processing analytics queries on evolving graphs. It is incorporated by extending the KickStarter streaming framework and implementing optimizations to efficiently handle edge additions without expensive in-place graph mutations. Further, it avoids the need to mutate graphs and enables efficient reuse of edges by snapshots that share them. It achieves impressive performance gains of 1.38×-8.17× improvement over the KickStarter baseline across multiple benchmarks.

EGraph~\cite{Egraph} is a groundbreaking system that addresses significant challenges in processing large-scale dynamic graphs using Graphics Processing Units (GPUs). It introduce a novel approach for efficiently executing concurrent Timing iterative Graph Processing (TGP) jobs on dynamic graphs, particularly when dealing with graphs too large to fit entirely in GPU memory.
It is presented as the first GPU-based dynamic graph processing system specifically designed for efficient concurrent execution of TGP jobs on dynamic graphs. A distinctive feature of EGraph is its ability to be integrated into existing out-of-GPU-memory static graph processing systems, enhancing their capabilities to handle dynamic graphs efficiently. At the core of EGraph lies the innovative Loading-Processing-Switching (LPS) execution model. This model introduces a sophisticated approach to dynamic graph processing that effectively reduces CPU-GPU data transfer overhead by intelligently managing what graph data needs to be transferred.

\subsection{Static Graph Algorithms}
The lightweight graph processing framework Ligra~\cite{Ligra} is designed specifically for shared-memory processors, making the writing of graph traversal algorithms simple. The two very basic routines in the framework are one for mapping over vertices and one for mapping over edges.
The framework is promising for different graph traversal algorithms operating on subsets of the vertices because the routines can be applied to any subset of the vertices.
An integer label set for the vertices that are included in the vertex subset is preserved in abstraction, and each integer is assigned the user-supplied function via the mapping over vertices procedure.

Galois~\cite{Galois} is a simple infrastructure for graph analytics that facilitates a more expressive programming style that is more broad in nature. A library for scalable data structures and autonomous scheduling of fine-grained jobs with application-specific priorities are supported by the infrastructure. High throughput can be achieved by implementing existing DSLs on top of the Galois system.
Parallel execution of iterations allows the system to take use of parallelism.  Programmers must employ a library of pre-built concurrent data structures like graphs, worklists, etc. in order to guarantee serializability of iterations.

For shared memory graph processing, there is a DSL called Green-Marl~\cite{Green-Marl}. With the language's matching structures, programs that rely on BFS/DFS traversals can be written succinctly. Some graph algorithms, however, require the user to explicitly define the iteration across vertices or edges. As a result, the built-in data-level parallelism can be shown while writing the algorithmic description in a comprehensible manner utilizing the language's components.
By taking use of data-level parallelism, the Green-Marl compiler converts high-level algorithmic descriptions expressed in Green-Marl into an effective C++ implementation. Moreover, a collection of optimizations that leverage the high-level semantic knowledge embedded in the Green-Marl DSL are applied by the Green-Marl compiler.

GRAFS~\cite{grafs-ligra} is a synthesizer for declarative graph analytics. Ligra~\cite{Ligra}, PowerGraph~\cite{PowerGraph}, GraphIt~\cite{GraphIt}, GridGraph~\cite{gridgraph}, and Gemini~\cite{GeminiGraph} are the five frameworks used to illustrate it. PowerGraph and Gemini support MPI, while Ligra, GridGraph, and GraphIt support the OpenMP implementation. While GRAFS's architecture can support many backend frameworks, it differs from StarPlat in terms of its ultimate objectives. Cross-API optimizations (like fusion) are carried out by GRAFS to facilitate the synthesis of efficient backend code. In contrast, StarPlat's philosophy is to ease out the programming burden of the domain experts. We contrast the GRAFS Ligra backend with StarPlat using SSSP and Page Rank. Combining StarPlat's productivity increases with GRAFS's synthesis capability would be intriguing.

Gluon, a communication optimization tool, was interfaced with the Galois shared memory graph processing system~\cite{Gluon-DGalois} to build D-Galois, a distributed graph processing system. In our experiments, we compare against Gluon.

The BSP programming model~\cite{Pregel,GiraphPaper} served as an inspiration for the Pregel and Giraph graph processing frameworks.

Pregel breaks down the computing involved in graph processing into a series of iterations, wherein the vertices receive and process data from the previous iteration, modify their state and transmit messages to their neighboring vertices, which they then get from the subsequent iteration. To the Pregel framework~\cite{GPS}, GPS provides features including repartitioning of massive adjacency lists, dynamic graph repartitioning, and global computation.

A compiler automatically generates Pregel code from a subset of programs called Pregel canonical written using Green-Marl DSL~\cite{Pregel_GreenMarl}. It also applies certain transformations such as edge flipping, dissecting nested loops, translating random access, transforming BFS and DFS traversals to convert non-Pregel canonical pattern into Pregel canonical. Another compiler named DisGCo can translate all the Green-Marl programs to equivalent MPI RMA programs~\cite{DisGCo}. It handled the challenges such as syntactic differences, differences in memory view, intermixed serial code with parallel code, and graph representation while translating a Green-Marl program to MPI RMA code. 

The DH-Falcon DSL and compiler~\cite{DH-Falcon} allowed the specification and implementation of non-vertex centric graph algorithms and morph graph algorithms for heterogeneous distributed systems. A distributed graph analytics technique called Gemini~\cite{GeminiGraph} offered a hybrid solution that combines both sparse and dense mode processing by using push and pull-based compute approaches. It also made use of partitioning techniques and optimized graph representation for both intra and inter-node load balancing.

GraphLab provides a fault-tolerant distributed graph processing abstraction that can perform dynamic, asynchronous graph processing for machine learning algorithms~\cite{GraphLab}. Another parallel distributed graph processing abstraction for processing power law graphs with a caching mechanism and a distributed graph placement mechanism was introduced by PowerGraph~\cite{PowerGraph}.
GraphChi supports asynchronous processing of dynamic graphs using a sliding approach on the disk-based systems~\cite{GraphChi}.

Gunrock~\cite{Gunrock} leverages the parallel processing power of GPUs to speed up graph algorithms, making it particularly useful for tasks in social network analysis, recommendation systems, and scientific research where graph-based data is common. Gunrock uses a frontier-based approach, which focuses on manipulating subsets of graph vertices or edges, called frontiers, to efficiently handle tasks such as breadth-first search (BFS), single-source shortest path (SSSP), Page Rank, and others. Its primary advantage lies in its ability to exploit the parallelism in GPU hardware, making it one of the faster frameworks for complex graph analytics.

LonestarGPU~\cite{lonestargpu} optimizes graph algorithms for GPU architectures, allowing for faster processing of large-scale graphs by utilizing the parallel computing power of GPUs. LonestarGPU focuses on asynchronous, data-driven execution models for irregular workloads, which are common in graph processing. 

Medusa~\cite{medusa} Medusa is a GPU-based framework specifically designed for parallel graph processing, which simplifies the complexities associated with programming on GPUs. By enabling users to write sequential C/C++ code using Medusa’s APIs, it hides much of the intricate GPU architecture details and instead translates code to CUDA for execution on GPUs.

nvGRAPH~\cite{nvidia1} represents a paradigm shift in graph analytics, leveraging GPU parallelism to address large-scale graph problems with unprecedented efficiency. Introduced in 2016 as part of the CUDA Toolkit, nvGRAPH targets graphs with up to 2 billion edges on a single GPU (e.g., M40 with 24GB memory). Its architecture bridges graph theory and sparse linear algebra, modeling graph operations as semi-ring sparse matrix-vector products (SPMV). This approach transforms graph traversal and computation into linear algebra primitives optimized for GPU execution, enabling efficient handling of power-law distributed networks common in real-world datasets.
It comprises of three algorithms Page Rank, SSSP, and Single Source Widest Path.

\section{Conclusion} \label{sec conclusion}

This work delineates a sophisticated framework that amalgamates high-level programming abstractions with an intelligent code generation pipeline, purpose-built to facilitate the efficient execution of dynamic graph algorithms across a spectrum of parallel computing backends. The proposed domain-specific constructs are designed to elegantly capture batched incremental and decremental updates, enabling the specification of dynamic behaviors through succinct and semantically rich logic.
By abstracting away low-level complexities, this approach empowers developers to effortlessly synthesize highly optimized parallel code for prominent graph algorithms such as Single Source Shortest Path (SSSP), PageRank (PR), and Triangle Counting (TC). Experimental evaluations conducted over a diverse suite of real-world and synthetic graph datasets reveal that the performance of the generated dynamic code substantially surpasses that of re-executing the equivalent static algorithms—particularly under moderate update workloads. We posit that this contribution marks a significant stride toward productive, scalable, and architecture-aware dynamic graph processing. As part of our future trajectory, we intend to extend the expressiveness of the StarPlat Dynamic DSL to encompass more intricate graph algorithms and to rigorously assess its efficacy across an expanded corpus of graph benchmarks and computational scenarios.
\begin{acks}
We gratefully acknowledge the use of the computing resources at HPCE, IIT Madras. This work is supported by India's National Supercomputing Mission grant CS1920/1123/MEIT/008606.
\end{acks}

\bibliographystyle{plain}
\bibliography{9REF}

\appendix
\input{91codes}
\end{document}

%% file: 10expt-openmp.tex
\begin{table*}

\caption{StarPlat’s OpenMP dynamic code performance comparison against static code for various percentages of updates. All times
are in seconds.
}

\scalebox{0.85}{
\setlength{\tabcolsep}{3pt}
\begin{tabular}{|lr|c|r|r|r|r|r|r|r|r|r|r|}
\hline
\multicolumn{2}{|l|}{Algo.} & Framework & TW & SW & OK & WK & LJ & PK & US & GR & RM & UR  \\ \hline
\multicolumn{1}{|l|}{} & 1 & Static & >3hrs & 0.851 & 1.002 & 0.909 & 1.23 & 0.401 & \textbf{15.808} & \textbf{5.883} & 1.88 & 2.05  \\
\multicolumn{1}{|l|}{} & 1 & Dynamic & >3hrs & \textbf{0.83} & \textbf{0.06} & \textbf{0.09} & \textbf{0.09} & \textbf{0.03} & {2869.03} & {1462.09} & \textbf{0.24} & \textbf{0.35}  \\

 \cline{2-13}

\multicolumn{1}{|l|}{} & 2 & {Static} & >3hrs & \textbf{0.826} & 1.019 & 0.775 & 1.118 & 0.373 & \textbf{16.179} & \textbf{6.084} & 1.931 & 2.181  \\
\multicolumn{1}{|l|}{} & 2 & Dynamic & >3hrs & 0.93 & \textbf{0.09} & \textbf{0.14} & \textbf{0.22} & \textbf{0.07} & 3126.66 & 1842.36 & \textbf{0.21} & \textbf{0.51}   \\

 \cline{2-13}

\multicolumn{1}{|l|}{} & 4 & Static & >3hrs & \textbf{0.857} & 0.979 & 0.759 & 1.444 & 0.496 & \textbf{14.687} & \textbf{7.307} & 1.959 & 2.518  \\ 
\multicolumn{1}{|l|}{} & 4 & Dynamic & >3hrs & 1.7 & \textbf{0.19} &\textbf{0.46} & \textbf{0.32} & \textbf{0.11} & 3049.52 & 2660.53 & 0.37 & 0.77    \\

 \cline{2-13} 

 \multicolumn{1}{|l|}{SSSP} & 8 & Static & >3hrs & \textbf{0.915} & 0.998 & \textbf{0.763} & 1.487 & 0.391 & \textbf{16.279} & \textbf{2.377} & 1.932 & 2.093  \\ 
\multicolumn{1}{|l|}{} & 8 & Dynamic & >3hrs & 2.34 & \textbf{0.46} & 1.10 & \textbf{0.56} & \textbf{0.26} & 3341.04 & 233.23 & \textbf{0.59} & \textbf{1.57}  \\

 \cline{2-13}

\multicolumn{1}{|l|}{} & 12 & Static & >3hrs & \textbf{0.939} & 0.95 & \textbf{0.897} & 1.155 & 0.475 & \textbf{13.814} & \textbf{1.435} & 2.079 & 2.058  \\
\multicolumn{1}{|l|}{} & 12 & Dynamic & >3hrs & 3.95 & \textbf{0.58} & 1.45 & \textbf{0.73} & \textbf{0.32} & 3086.56 & 2.09 & \textbf{0.73} & \textbf{1.83} \\

\cline{2-13} 

\multicolumn{1}{|l|}{} & 16 & static & >3hrs & \textbf{0.882} & 0.978 & \textbf{0.798} & 1.175 & 0.49 & \textbf{13.343} & \textbf{1.127} & 2.248 & \textbf{2.14}  \\
\multicolumn{1}{|l|}{} & 16 & Dynamic & {>3hrs} & {5.91} & \textbf{0.71} & {1.52} & \textbf{0.95} & \textbf{0.42} & 3006.14 & 1.70 & \textbf{1.04} & {2.40}  \\

\cline{2-13} 

\multicolumn{1}{|l|}{} & 20 & Static & >3hrs & \textbf{0.835} & 1 & \textbf{0.893} & 1.458 & \textbf{0.401} & \textbf{3.412} & \textbf{1.392} & 2.197 & \textbf{2.104}  \\
\multicolumn{1}{|l|}{} & 20 & Dynamic & {>3hrs} & {6.92} & \textbf{0.85} & {1.67} & \textbf{1.13} & {0.44} & 94.77 & 2.09 & \textbf{1.18} & {2.77}  \\

\hline

\multicolumn{1}{|l|}{} & 1 & Static & >3hrs & >3hrs & >3hrs & >3hrs & >3hrs & 1282.685 & 5.695 & 2.626 & >3hrs & 65.193  \\
\multicolumn{1}{|l|}{} & 1 & Dynamic & {>3hrs} & {>3hrs} & {>3hrs} & {>3hr} & {>3hrs} & \textbf{12.95} & \textbf{0.10} & \textbf{0.04} & {>3hrs} & \textbf{1.83} \\

 \cline{2-13}

\multicolumn{1}{|l|}{} & 2 & {Static} & >3hrs & >3hrs & >3hrs & >3hrs & >3hrs & 1293.803 & 6.068 & 2.605 & >3hrs & 62.162 \\
\multicolumn{1}{|l|}{} & 2 & Dynamic & >3hrs & >3hrs & >3hrs & >3hrs & >3hrs &  \textbf{24.26} & \textbf{0.21} & \textbf{0.07}& >3hrs & \textbf{3.34}  \\

 \cline{2-13}

\multicolumn{1}{|l|}{} & 4 & Static & >3hrs & >3hrs & >3hrs & >3hrs & >3hrs & 1200.045 & 5.435 & 2.666 & >3hrs & 65.093  \\ 
\multicolumn{1}{|l|}{} & 4 & Dynamic & >3hrs & >3hrs & >3hrs & >3hrs & >3hrs &  \textbf{51.78} & \textbf{0.44} & \textbf{0.15} & >3hrs & \textbf{7.57}  \\

 \cline{2-13} 

 \multicolumn{1}{|l|}{TC} & 8 & Static & >3hrs & >3hrs & >3hrs & >3hrs & >3hrs & 1120.378 & 5.664 & 2.623 & >3hrs & 61.152  \\ 
\multicolumn{1}{|l|}{} & 8 & Dynamic & >3hrs & >3hrs & >3hrs & >3hrs & >3hrs &  \textbf{101.57} & \textbf{0.83} & \textbf{0.31} & >3hrs & \textbf{14.22}  \\

 \cline{2-13}

 \multicolumn{1}{|l|}{} & 12 & Static & >3hrs & >3hrs & >3hrs & >3hrs & >3hrs & 1045.173 & 5.324 & 2.587 & >3hrs & 58.446  \\
\multicolumn{1}{|l|}{} & 12 & Dynamic & {>3hrs} & {>3hrs} & {>3hrs} & {>3hrs} & {>3hrs} & \textbf{148.06} & \textbf{1.33} & \textbf{0.43} & {>3hrs} & \textbf{21.54}  \\

\cline{2-13}

 \multicolumn{1}{|l|}{} & 16 & Static & >3hrs & >3hrs & >3hrs & >3hrs & >3hrs & 971.916 & 5.561 & 2.576 & >3hrs & 59.175   \\
\multicolumn{1}{|l|}{} & 16 & Dynamic & {>3hrs} & {>3hrs} & {>3hrs} & {>3hrs} & {>3hrs} & \textbf{203.63} & \textbf{1.67} & \textbf{0.61} & {>3hrs} & \textbf{28.64}  \\

\cline{2-13}

\multicolumn{1}{|l|}{} & 20 & Static & >3hrs & >3hrs & >3hrs & >3hrs & >3hrs & 890.531 & 5.451 & 2.484 & >3hrs & 57.305  \\
\multicolumn{1}{|l|}{} & 20 & Dynamic & {>3hrs} & {>3hrs} & {>3hrs} & {>3hrs} & {>3hrs} & \textbf{269.03} & \textbf{2.21} & \textbf{0.71} & {>3hrs} & \textbf{36.88} \\

 \hline

\multicolumn{1}{|l|}{} & 1 & Static & 2748.916 & >3hrs & 468.521 & 325.241 & 328.201 & 87.808 & 104.022 & 47.122 & 617.283 & 89.427  \\
\multicolumn{1}{|l|}{} & 1 & Dynamic & \textbf{827.74} & {>3hrs} & \textbf{187.22} & \textbf{119.88} & \textbf{77.26} & \textbf{19.61} & \textbf{22.13} & \textbf{9.07} & \textbf{140.14} & \textbf{38.34}  \\

 \cline{2-13}

\multicolumn{1}{|l|}{} & 2 & {Static} & 2976.898 & >3hrs & 467.722 & 318.246 & 331.331 & 89.21 & 108.08 & 50.934 & 601.218 & 89.536  \\
\multicolumn{1}{|l|}{} & 2 & Dynamic & \textbf{1008.99} & >3hrs & \textbf{247.15} & \textbf{147.65} & \textbf{107.57} & \textbf{27.147} & \textbf{32.13} & \textbf{11.85} & \textbf{164.26} & \textbf{46.5} \\

 \cline{2-13}

\multicolumn{1}{|l|}{} & 4 & Static & 2927.867 & >3hrs & 470.689 & 331.404 & 317.112 & 84.646 & 101.44 & 49.248 & 612.808 & 91.12  \\ 
\multicolumn{1}{|l|}{} & 4 & Dynamic & \textbf{1268.87} & >3hrs & \textbf{285.13} & \textbf{188.45} & \textbf{126.70} & \textbf{32.44} & \textbf{32.13} & \textbf{14.18} & \textbf{159.32} & \textbf{49.92}   \\

 \cline{2-13} 

 \multicolumn{1}{|l|}{PR} & 8 & Static & >3hrs & >3hrs & 460.412 & 317.83 & 357.419 & 85.119 & 107.444 & 51.741 & 593.664 & 100.742  \\ 
\multicolumn{1}{|l|}{} & 8 &  Dynamic &>3hrs & >3hrs & \textbf{355.14} & \textbf{219.65} & \textbf{117.74} & \textbf{42.55}& \textbf{39.70} & \textbf{17.46} & \textbf{207.75} & \textbf{80.39}   \\

\cline{2-13} 

 \multicolumn{1}{|l|}{} & 12 & Static & >3hrs & >3hrs & 448.886 & 294.982 & 307.606 & 79.716 & 104.276 & 58.992 & 630.93 & 93.309  \\
\multicolumn{1}{|l|}{} & 12 & Dynamic & {>3hrs} & {>3hrs} & \textbf{322.50} & \textbf{250.13} & \textbf{176.32} & \textbf{46.14} & \textbf{41.13} & \textbf{21.7} & \textbf{264} & \textbf{88.24}   \\

\cline{2-13} 

 \multicolumn{1}{|l|}{} & 16 & Static & >3hrs & >3hrs & 454.456 & 293.993 & 296.384 & 76.003 & 110.544 & 62.497 & 599.825 & \textbf{96.781}  \\
\multicolumn{1}{|l|}{} & 16 & Dynamic & {>3hrs} & {>3hrs} & \textbf{338.78} & \textbf{259.33} & \textbf{188.05} & \textbf{46.75} & \textbf{47.88} & \textbf{25.08} & \textbf{278.06} & 103.43  \\

\cline{2-13}

\multicolumn{1}{|l|}{} & 20 & Static &  >3hrs & >3hrs & 446.175 & 285.543 & 296.53 & 71.352 & 115.712 & 65.965 & 626.272 & \textbf{117.155} \\
\multicolumn{1}{|l|}{} & 20 & Dynamic & {>3hrs} & {>3hrs} & \textbf{319.70} & \textbf{255.40} & \textbf{205.88} & \textbf{49.75} & \textbf{50.52} & \textbf{27.40} & \textbf{324.8} & 134.68 \\

\hline

\end{tabular}
}
\label{table:all:openmp}
\end{table*}

%% file: 11expt-mpi.tex
\begin{table*}

\caption{StarPlat’s MPI dynamic code performance comparison against static code for various percentages of updates. All times
are in seconds.
}

\scalebox{0.85}{
\setlength{\tabcolsep}{3pt}
\begin{tabular}{|lr|c|r|r|r|r|r|r|r|r|r|r|}
\hline
\multicolumn{2}{|l|}{Algo.} & Framework & TW & SW & OK & WK & LJ & PK & US & GR & RM & UR  \\ \hline

\multicolumn{1}{|l|}{} & 0.1 & Static & \textbf{0.042} & \textbf{0.112} & {314.764} & {191.149} & {166.86} & {36.085} & \textbf{4895.808} & \textbf{132.953} & {549.731} & {44.437}  \\

\multicolumn{1}{|l|}{} & 0.1 & Dynamic & 1.174 & 2.08 & \textbf{129.591} & \textbf{19.113} & \textbf{18.024} & \textbf{4.569} & >3hrs & >3hrs & \textbf{23.761} & \textbf{20.75} \\

 \cline{2-13}

\multicolumn{1}{|l|}{} & 0.2 & Static & \textbf{0.043} & \textbf{0.152} & 314.14 & 189.639 & 165.423 & 35.485 & \textbf{4921.769} & \textbf{129.24} & 549.443 & 44.313 \\

\multicolumn{1}{|l|}{} & 0.2 & Dynamic & 2.081 & 3.436 & \textbf{138.117} & \textbf{23.877} & \textbf{19.468} & \textbf{5.346} & >3hrs & >3hrs & \textbf{25.778} & \textbf{29.211} \\

 \cline{2-13}

\multicolumn{1}{|l|}{} & 0.4 & Static & \textbf{0.041} & \textbf{0.115} & 314.67 & 189.354 & 165.427 & 35.511 & \textbf{4929.9} & \textbf{129.272} & 549.716 & 44.58 \\
 
\multicolumn{1}{|l|}{} & 0.4 & Dynamic & 3.879 & 6.132 & \textbf{161.531} & \textbf{29.763} & \textbf{29.841} & \textbf{7.612} & >3hrs & >3hrs & \textbf{48.858} & \textbf{34.063} \\

 \cline{2-13} 

\multicolumn{1}{|l|}{SSSP} & 0.8 & Static & \textbf{0.043} & \textbf{0.11} & 315.152 & 187.412 & 165.192 & 35.564 & \textbf{4922.257} & \textbf{129.36} & 548.051 & 45.12 \\

\multicolumn{1}{|l|}{} & 0.8 & Dynamic & 7.482 & 11.564 & \textbf{191.848} & \textbf{44.817} & \textbf{42.761} & \textbf{10.88} & >3hrs & >3hrs & \textbf{64.931} & \textbf{44.16} \\

 \cline{2-13}

\multicolumn{1}{|l|}{} & 1.2 & Static & \textbf{0.044} & \textbf{0.117} & 312.575 & 186.854 & 166.843 & 35.724 & \textbf{4927.46} & \textbf{129.326} & 547.625 & \textbf{45.708} \\

\multicolumn{1}{|l|}{} & 1.2 & Dynamic & 11.111 & 17.33 & \textbf{218.548} & \textbf{57.458} & \textbf{53.395} & \textbf{17.336} & >3hrs & >3hrs & \textbf{96.217} & 58.618 \\

\cline{2-13}

\multicolumn{1}{|l|}{} & 1.6 & Static & \textbf{0.045} & \textbf{0.123} & \textbf{319.414} & 184.212 & 168.667 & 35.664 & \textbf{4925.509} & \textbf{129.227} & 546.39 & \textbf{45.971} \\

\multicolumn{1}{|l|}{} & 1.6 & Dynamic & 14.688 & 22.62 & 331.623 & \textbf{76.596} & \textbf{96.238} & \textbf{20.563} & >3hrs & >3hrs & \textbf{111.521} & 71.752 \\

\cline{2-13}

\multicolumn{1}{|l|}{} & 2.0 & Static & \textbf{0.043} & \textbf{0.149} & \textbf{363.247} & 184.035 & 167.871 & \textbf{39.463} & \textbf{4919.144} & \textbf{129.261} & 544.799 & \textbf{46.522} \\

\multicolumn{1}{|l|}{} & 2.0 & Dynamic & 18.254 & 28.283 & 417.376 & \textbf{92.917} & \textbf{118.75} & 115.573 & >3hrs & >3hrs & \textbf{127.072} & 78.154 \\
 
\hline


\multicolumn{1}{|l|}{} & 1 & Static & >3hrs & >3hrs & >3hrs & >3hrs & 11110.08 & 397.726 & 6.461 & 1.168 & >3hrs & 63.874 \\

\multicolumn{1}{|l|}{} & 1 & Dynamic & >3hrs & >3hrs & >3hrs & >3hrs & \textbf{680.367} & \textbf{43.737} & \textbf{0.755} & \textbf{0.284} & >3hrs & \textbf{2.852} \\

 \cline{2-13}

\multicolumn{1}{|l|}{} & 2 & Static & >3hrs & >3hrs & >3hrs & >3hrs & 11234.186 & 405.665 & 6.853 & 1.24 & >3hrs & 66.352 \\

\multicolumn{1}{|l|}{} & 2 & Dynamic & >3hrs & >3hrs & >3hrs & >3hrs & \textbf{1330.833} & \textbf{86.28} & \textbf{1.317} & \textbf{0.476} & >3hrs & \textbf{5.749} \\

 \cline{2-13}

\multicolumn{1}{|l|}{} & 4 & Static & >3hrs & >3hrs & >3hrs & >3hrs & 11291.851 & 400.822 & 6.866 & 1.306 & >3hrs & 68.432 \\
 
\multicolumn{1}{|l|}{} & 4 & Dynamic & >3hrs & >3hrs & >3hrs & >3hrs & \textbf{2847.051} & \textbf{177.115} & \textbf{2.391} & \textbf{0.837} & >3hrs & \textbf{11.21} \\

 \cline{2-13} 

\multicolumn{1}{|l|}{TC} & 8 & Static & >3hrs & >3hrs & >3hrs & >3hrs & 11383.513 & 393.23 & 7.388 & \textbf{1.458} & >3hrs & 70.825 \\

\multicolumn{1}{|l|}{} & 8 & Dynamic & >3hrs & >3hrs & >3hrs & >3hrs & \textbf{5749.978} & \textbf{355.97} & \textbf{4.572} & 1.58 & >3hrs & \textbf{22.299} \\

 \cline{2-13}

\multicolumn{1}{|l|}{} & 12 & Static & >3hrs & >3hrs & >3hrs & >3hrs & 11467.071 & \textbf{385.704} & 7.654 & \textbf{1.628} & >3hrs & 74.494 \\

\multicolumn{1}{|l|}{} & 12 & Dynamic & >3hrs & >3hrs & >3hrs & >3hrs & \textbf{8604.504} & 530.086 & \textbf{6.775} & 2.318 & >3hrs & \textbf{33.52} \\

\cline{2-13}

\multicolumn{1}{|l|}{} & 16 & Static & >3hrs & >3hrs & >3hrs & >3hrs & 11531.5 & \textbf{376.968} & \textbf{7.997} & \textbf{1.762} & >3hrs & 76.162 \\

\multicolumn{1}{|l|}{} & 16 & Dynamic & >3hrs & >3hrs & >3hrs & >3hrs & \textbf{10665.043} & 701.857 & 8.863 & 3.014 & >3hrs & \textbf{44.567} \\

\cline{2-13}

\multicolumn{1}{|l|}{} & 20 & Static & >3hrs & >3hrs & >3hrs & >3hrs & \textbf{11597.472} & \textbf{371.144} & \textbf{8.336} & \textbf{1.853} & >3hrs & 78.886 \\

\multicolumn{1}{|l|}{} & 20 & Dynamic & >3hrs & >3hrs & >3hrs & >3hrs & 13401.138 & 873.172 & 11.067 & 3.851 & >3hrs & \textbf{55.821} \\

 \hline

\multicolumn{1}{|l|}{} & 0.1 & Static & >3hrs & >3hrs & >3hrs & 2120.135 & 1626.326 & 463.796 & \textbf{579.088} & \textbf{65.517} & \textbf{2388.215} & 279.683 \\

\multicolumn{1}{|l|}{} & 0.1 & Dynamic & >3hrs & >3hrs & >3hrs & \textbf{1433.332} & \textbf{563.655} & \textbf{148.044} & 2671.001 & 1007.987 & 633.7 & \textbf{239.453} \\

 \cline{2-13}

\multicolumn{1}{|l|}{} & 0.2 & Static & >3hrs & >3hrs & >3hrs & 2164.517 & 1693.333 & 458.204 & \textbf{561.518} & \textbf{68.43} & \textbf{2466.32} & \textbf{275.65} \\

\multicolumn{1}{|l|}{} & 0.2 & Dynamic & >3hrs & >3hrs & >3hrs & \textbf{1474.223} & \textbf{558.286} & \textbf{164.022} & 2420.55 & 788.033 & 839.633 & 304.406 \\

 \cline{2-13}

\multicolumn{1}{|l|}{} & 0.4 & Static & >3hrs & >3hrs & >3hrs & 2194.122 & 1684.515 & 453.074 & \textbf{560.971} & \textbf{68.734}& \textbf{2497.341} & \textbf{276.671} \\
 
\multicolumn{1}{|l|}{} & 0.4 & Dynamic & >3hrs & >3hrs & >3hrs & \textbf{1431.401} & \textbf{716.817} & \textbf{192.442} & 1871.904 & 769.33 & 874.286 & 379.408 \\

 \cline{2-13} 

\multicolumn{1}{|l|}{PR} & 0.8 & Static & >3hrs & >3hrs & >3hrs & 2168.976 & 1670.088 & 452.873 & \textbf{562.492} & \textbf{69.508} & \textbf{2447.768} & \textbf{281.393} \\

\multicolumn{1}{|l|}{} & 0.8 & Dynamic & >3hrs & >3hrs & >3hrs & \textbf{1618.92} & \textbf{811.724} & \textbf{235.548} & 1562.551 & 744.992 & 878.84 & 385.268 \\

 \cline{2-13}

\multicolumn{1}{|l|}{} & 1.2 & Static & >3hrs & >3hrs & >3hrs & 2056.642 & 1686.707 & 457.507 & \textbf{554.478} & \textbf{70.172} & \textbf{2457.735} & \textbf{279.02} \\

\multicolumn{1}{|l|}{} & 1.2 & Dynamic & >3hrs & >3hrs & >3hrs & \textbf{1701.479} & \textbf{905.924} & \textbf{261.873} & 1564.432 & 515.388 & 986.911 & 386.572 \\

\cline{2-13}

\multicolumn{1}{|l|}{} & 1.6 & Static & >3hrs & >3hrs & >3hrs & 2044.639 & 1667.447 & 456.643 & \textbf{558.895} & \textbf{70.929} & \textbf{2398.797} & \textbf{283.299} \\

\multicolumn{1}{|l|}{} & 1.6 & Dynamic & >3hrs & >3hrs & >3hrs & \textbf{1772.943} & \textbf{1065.067} & \textbf{314.412} & 1411.546 & 483.162 & 1206.162 & 389.735 \\

\cline{2-13}

\multicolumn{1}{|l|}{} & 2.0 & Static & >3hrs & >3hrs & >3hrs & 2047.229 & 1679.913 & 467.717 & \textbf{571.007} & \textbf{71.22} & \textbf{2337.766} & \textbf{278.516} \\

\multicolumn{1}{|l|}{} & 2.0 & Dynamic & >3hrs & >3hrs & >3hrs & \textbf{1816.382} & \textbf{1167.336} & \textbf{342.631} & 1268.661 & 397.587 & 1216.225 & 420.235 \\
\hline

\end{tabular}
}
\label{table:all:mpi}
\end{table*}

%% file: 12expt-cuda.tex
\begin{table*}

\caption{StarPlat’s CUDA dynamic code performance comparison against static code for various percentages of updates. All times are in seconds.
}

\scalebox{0.85}{
\setlength{\tabcolsep}{3pt}
\begin{tabular}{|lr|c|r|r|r|r|r|r|r|r|r|r|}
\hline
\multicolumn{2}{|l|}{Algo.} & Framework & TW & SW & OK & WK & LJ & PK & US & GR & RM & UR  \\ \hline

\multicolumn{1}{|l|}{} & 1 & Static & 27.21 & {13.95} & {0.21} & {0.90} & {1.87} & {0.72} & {8.01} & {8.23} & {0.50} & {2.01}  \\
\multicolumn{1}{|l|}{} & 1 & Dynamic & {\textbf{15.91}} & {\textbf{10.41}} & {\textbf{0.05}} & \textbf{0.02} & \textbf{0.91} & {\textbf{0.11}} & \textbf{7.05} & \textbf{4.85} & {\textbf{0.10}} & {\textbf{1.11}}  \\

 \cline{2-13}

\multicolumn{1}{|l|}{} & 2 & Static & 27.23 & 13.98 & 0.21 & 0.91 & 1.91 & 0.73 & 8.55 & 8.25 & 0.52 &  2.05  \\

\multicolumn{1}{|l|}{} & 2 & {Dynamic} & \textbf{15.01} & \textbf{10.65} & \textbf{0.06} & \textbf{0.03} & \textbf{1.14} & \textbf{0.25} & \textbf{6.32} & \textbf{5.10} & \textbf{0.23} &  \textbf{1.21} \\

 \cline{2-13} 

\multicolumn{1}{|l|}{} & 4 & Static & 27.22 & 14.01 & {0.26} & 0.93 & 1.94  & 0.74 & 8.97 & 8.28 & 0.54 & 2.07  \\

\multicolumn{1}{|l|}{} & 4 & Dynamic & \textbf{16.12} & \textbf{11.20} & \textbf{0.08} & \textbf{0.04} & \textbf{1.37} & \textbf{0.32}  & \textbf{6.79} & \textbf{6.31} & {\textbf{0.31}} &  \textbf{1.27} \\

 \cline{2-13} 

\multicolumn{1}{|l|}{} & 8 & Static & 27.25 & 14.03 & {0.28} & 0.94  & 1.97  & 0.76 & 9.10 & 8.29 & 0.57 & 2.10  \\
 \multicolumn{1}{|l|}{SSSP} & 8 & Dynamic & \textbf{19.10} & \textbf{12.79} & \textbf{0.09} & \textbf{0.06} & \textbf{1.58} & \textbf{0.44} & \textbf{7.35} & \textbf{0.41} & {\textbf{0.41}} & \textbf{1.49}  \\

 \cline{2-13}

\multicolumn{1}{|l|}{} & 12 & Static & 27.26 & 14.05 & {0.30} & 0.95 & 1.87 & 0.78 & 9.27 & 8.35 & 0.58 & 2.12 \\

\multicolumn{1}{|l|}{} & 12 & Dynamic & {\textbf{22.74}} & {\textbf{13.72}} & {\textbf{0.13}} & \textbf{0.07}  & \textbf{1.75}  & {\textbf{0.57}} & \textbf{8.72} & \textbf{8.19} & {\textbf{0.51}} & {\textbf{1.72}}  \\

\cline{2-13} 

\multicolumn{1}{|l|}{} & 16 & Static & {27.28} & {\textbf{14.07}} & {0.32} & {0.97} & {\textbf{1.90}} & {0.79} & 9.55 & \textbf{8.67} & {\textbf{0.50}} & {2.13}  \\
\multicolumn{1}{|l|}{} & 16 & Dynamic & {\textbf{24.92}} & {16.01} & {\textbf{0.20}} & \textbf{0.90} & 1.93 & {\textbf{0.71}} & \textbf{8.81} & 9.22 & {0.61} & {\textbf{2.11}}  \\

\cline{2-13} 

\multicolumn{1}{|l|}{} & 20 & Static & {\textbf{27.33}} & {\textbf{14.08}} & {0.34} & {\textbf{0.98}} & {\textbf{1.91}} & {\textbf{0.81}} & 9.78 & \textbf{9.01} & {\textbf{0.54}} & {\textbf{2.14}}  \\
\multicolumn{1}{|l|}{} & 20 & Dynamic & {32.42} & {16.10} & {\textbf{0.26}} & 1.10 & 2.14 & {0.90} & \textbf{2.61}  & 10.34 & {0.72} & {2.30}  \\

\hline

\multicolumn{1}{|l|}{} & 1 & Static & {>3hrs} & {>3hrs} & {44.01} & {>3hrs} & {>3hrs} & {5.67} & 0.00 & 0.003 & {>3hrs} & {>3hrs} \\

\multicolumn{1}{|l|}{} & 1 & Dynamic & {>3hrs} & {>3hrs} & {33.12} & >3hrs & >3hrs & {2.13} & 0.001  & 0.0007 & {>3hrs} & {>3hrs}  \\

 \cline{2-13}

\multicolumn{1}{|l|}{} & 2 & Static & >3hrs & >3hrs & 44.04 & >3hrs & >3hrs & 5.69 & 0.006 & 0.003 & >3hrs & >3hrs  \\

\multicolumn{1}{|l|}{} & 2 & {Dynamic} & >3hrs & >3hrs & 35.35 & >3hrs & >3hrs & 3.11 & 0.002 & 0.0008 & >3hrs & >3hrs \\

 \cline{2-13} 

\multicolumn{1}{|l|}{} & 4 & Static & >3hrs & >3hrs & 44.10 & >3hrs & >3hrs & 5.73 & 0.006 & 0.003 & >3hrs & >3hrs  \\

\multicolumn{1}{|l|}{} & 4 & Dynamic & >3hrs & >3hrs & 37.23 & >3hrs & >3hrs & 3.97 & 0.003 & 0.0009 & {>3hrs} & >3hrs  \\

 \cline{2-13} 

\multicolumn{1}{|l|}{} & 8 & Static & >3hrs & >3hrs & 44.15  & >3hrs & >3hrs & 5.75 & 0.006 & 0.003  & >3hrs & >3hrs  \\

 \multicolumn{1}{|l|}{TC} & 8 & Dynamic & >3hrs & >3hrs & 39.14  & >3hrs & >3hrs & 4.35  & 0.004 & 0.001 & {>3hrs} & >3hrs  \\

 \cline{2-13}

\multicolumn{1}{|l|}{} & 12 & Static & {>3hrs} & {>3hrs} & {44.18} & {>3hrs} & {>3hrs} & {5.79} & 0.006  & 0.004 & {>3hrs} & {>3hrs}  \\

 \multicolumn{1}{|l|}{} & 12 & Dynamic & {>3hrs} & {>3hrs} & {41.17} & >3hrs & >3hrs & {5.02} & 0.002 & 0.002  & {>3hrs} & {>3hrs}  \\

\cline{2-13}

\multicolumn{1}{|l|}{} & 16 & Static & {>3hrs} & {>3hrs} & {44.21} & {>3hrs} & {>3hrs} & {5.83} & 0.006 & 0.004 & {>3hrs} & {>3hrs}  \\

 \multicolumn{1}{|l|}{} & 16 & Dynamic & {>3hrs} & {>3hrs} & {43.24} & >3hrs & >3hrs & {5.68} & 0.006 & 0.003 & {>3hrs} & {>3hrs}  \\

\cline{2-13}

\multicolumn{1}{|l|}{} & 20 & Static & {>3hrs} & {>3hrs} & {44.25} & {>3hrs} & {>3hrs} & {5.86} & 0.006 & 0.004 & {>3hrs} & {>3hrs} \\

\multicolumn{1}{|l|}{} & 20 & Dynamic & {>3hrs} & {>3hrs} & {46.54} & >3hrs & >3hrs & {6.55} & 0.007 & 0.004 & {>3hrs} & {>3hrs}  \\

 \hline

\multicolumn{1}{|l|}{} & 1 & Static & {4.08} & {7.11} & {0.58} & {1.78} & {3.50} & {7.12} & 0.61 & 0.67 & {2.20} & {0.50}  \\

\multicolumn{1}{|l|}{} & 1 & Dynamic & {\textbf{3.85}} & {\textbf{4.96}} & {\textbf{0.11}} & \textbf{1.69} & \textbf{1.20} & {\textbf{2.15}} & \textbf{0.12} & \textbf{0.55} & {\textbf{0.77}} & {\textbf{0.24}}  \\

 \cline{2-13}

\multicolumn{1}{|l|}{} & 2 & Static & 4.97  & 8.23 &  0.59 & 1.98 & 3.52 & 7.14 & 0.63 & 0.77 & 2.25 & 0.51 \\

\multicolumn{1}{|l|}{} & 2 & {Dynamic} & \textbf{4.38} & \textbf{5.45} & \textbf{0.23}  & \textbf{1.80} & \textbf{1.46} & \textbf{3.38} & \textbf{0.23} & \textbf{0.59} & \textbf{0.88}  &  \textbf{0.28} \\

 \cline{2-13}

\multicolumn{1}{|l|}{} & 4 & Static & 5.79 & 9.16 & {0.61} & 2.20  & 3.54 & 7.15 & 0.64 & \textbf{0.86} & 2.47 &  0.52  \\

\multicolumn{1}{|l|}{} & 4 & Dynamic & \textbf{5.20} & \textbf{6.74} & \textbf{0.35} & \textbf{0.35} & \textbf{1.95}  & \textbf{1.95} & \textbf{0.30} & 0.88 & {\textbf{1.29}} &  \textbf{0.34} \\

 \cline{2-13}

\multicolumn{1}{|l|}{} & 8 &  Static & 6.46 & 10.23 & {0.63} & 2.70 & 3.55 & 7.16 & 0.66 & \textbf{0.92} & 2.53 &  0.54  \\

 \multicolumn{1}{|l|}{PR} & 8 & Dynamic & \textbf{6.20} & \textbf{7.25} & \textbf{0.46} & \textbf{2.24} & \textbf{2.47} & \textbf{5.60} & \textbf{0.44} & 1.06  & {\textbf{1.57}} & \textbf{0.39}  \\

\cline{2-13} 

\multicolumn{1}{|l|}{} & 12 & Static & {7.80} & {11.45} & {0.67} & {3.14} & {3.57} & {7.18} & 0.69 & \textbf{1.04} & {2.58} & {0.57}   \\

 \multicolumn{1}{|l|}{} & 12 & Dynamic & {\textbf{7.17}} & {\textbf{8.03}} & {\textbf{0.56}} & \textbf{2.98} & \textbf{3.02} & {\textbf{6.26}} & \textbf{0.56} & 1.47 & {\textbf{1.95}} & {\textbf{0.44}}  \\

\cline{2-13}

\multicolumn{1}{|l|}{} & 16 & Static & {8.33 } & {12.20} & {0.69} & {3.99} & {3.59} & {7.20} & 0.71 & \textbf{1.21}  & {2.60} & {0.59}  \\

 \multicolumn{1}{|l|}{} & 16 & Dynamic & {\textbf{8.06}} & {\textbf{9.14}} & {\textbf{0.64}} & \textbf{3.46} & \textbf{3.53} & {\textbf{7.06}} & \textbf{0.60} & 1.69 & {\textbf{2.35}} & {\textbf{0.50}} \\

\cline{2-13}

\multicolumn{1}{|l|}{} & 20 & Static & {9.84} & {13.10} & {\textbf{0.71}} & {4.41} & {\textbf{3.50}} & {\textbf{7.22}} & \textbf{0.73} & \textbf{1.32} & {\textbf{2.62}} & {0.62} \\

\multicolumn{1}{|l|}{} & 20 &  Dynamic & {\textbf{9.14}} & {\textbf{10.34}} & {0.75} & \textbf{4.05} & 4.12 & {8.01} & 0.76 & 1.95 & {2.97} & {\textbf{0.58}}  \\

\hline

\end{tabular}
}
\label{table:all:cuda}
\end{table*}

%% file: 91codes.tex
\section{Graph Algorithms in Dynamic \name}\label{starplat:codes}
StarPlat specifications for Dynamic Triangle Counting (TC), PageRank (PR) and Single Source Shortest Algorithm (SSSP) shown below in Figures~\ref{TCdyn-dsl}, \ref{PRdyn-dsl}, and \ref{ssspdyn-dsl} respectively.

\begin{lstlisting}[language=NEAR, style=mystyle, label=TCdyn-dsl, caption = Dynamic TC coded in StarPlat DSL, %xleftmargin = 0.01\textwidth
]
Static staticTC(Graph g) {
  long triangle_count = 0;
  
  forall(v in g.nodes()) {
      forall(u in g.neighbors(v).filter(u < v)) {
           forall(w in g.neighbors(v).filter(w > v)) {
                 if(g.is_an_edge(u,w)) 
                    triangle_count+=1;
  }   }    }     

   return triangle_count ;
}

Incremental(Graph g, int triangle_countSent, propEdge<bool> modified, updates<g> addBatch) {
 long triangle_count = triangle_countSent;
 int count1 = 0;
 int count2 = 0;
 int count3 = 0;
 
  forall(update in addBatch) {
      int v1 = update.source;
      int v2 = update.destination;
      
      forall(v3 in g.neighbors(v1).filter(v3!=v2 && v3!=v1 && v1!=v2)) {
        edge e1 = g.get_edge(v1, v3); 
        int newEdge = 1;
        bool isTriangle = False;
            
        if (e1.modified)
            newEdge = newEdge + 1;

        if (g.is_an_edge(v2, v3)) {
            edge e2 = g.get_edge(v2, v3); 
            isTriangle = True;
            if(e2.modified)
                newEdge = newEdge + 1;
        }
        if (isTriangle) {
            if (newEdge == 1) count1 += 1;
            else if(newEdge == 2) count2 += 1;   
            else count3 += 1;
        }              
      } 
  }
  triangle_count = triangle_count + (count1/2 + count2/4 + count3/6);

   return triangle_count ;
}

Decremental(Graph g, int triangle_countSent, propEdge<bool> modified, updates<g> deleteBatch) {
  long triangle_count = triangle_countSent;
  int count1 = 0;
  int count2 = 0;
  int count3 = 0;

   forall(update in deleteBatch) {
      int v1 = update.source;
      int v2 = update.destination;
      
      forall(v3 in g.neighbors(v1).filter(v3!=v2 && v3!=v1 && v1!=v2)) {
           edge e1 = g.get_edge(v1, v3); 
           int newEdge = 1;
           bool isTriangle = False;
          
           if (e1.modified)
              newEdge = newEdge + 1;

           if (g.is_an_edge(v2, v3)) {
              edge e2 = g.get_edge(v2, v3); 
              isTriangle = True;
              if(e2.modified)
                  newEdge = newEdge + 1;
           }

          if(isTriangle) {
            if(newEdge == 1) count1 += 1;
            else if(newEdge == 2) count2 += 1;   
            else count3 += 1;
         } 
      }       
   }

   triangle_count = triangle_count - (count1/2 + count2/4 + count3/6);
   
   return  triangle_count ;
}

Dynamic DynTC(Graph g, updates<g> updateBatch, int batchSize) {
  int triangleCount = staticTC(g);

  Batch(updateBatch:batchSize) {  
     propEdge<bool> modified_del; 
     g.attachEdgeProperty(modified_del = false);

     updates<g> deleteBatch = updateBatch.currentBatch(0);
     updates<g> addBatch = updateBatch.currentBatch(1);

     OnDelete(u in updateBatch.currentBatch()) { 
            int src = u.source;
            int dest = u.destination;

            for(nbr in g.neighbors(src)) {
                 edge e = g.get_edge(src,nbr);
                 if(nbr == dest)
                   e.modified_del = True;
            }
    }
   triangleCount = Decremental(g, triangleCount, modified_del, deleteBatch);
     
   g.updateCSRDel(updateBatch); 
   g.updateCSRAdd(updateBatch);
   
   propEdge<bool> modified_add; 
   g.attachEdgeProperty(modified_add = false);

   OnAdd(u in updateBatch.currentBatch()) {
        int src = u.source;
        int dest = u.destination;

        for(nbr in g.neighbors(src)) {
            edge e = g.get_edge(src,nbr);
            if(nbr == dest)
                e.modified_add = True;
        }             
    }  
    triangleCount = Incremental(g, triangleCount, modified_add, addBatch);    
  }
}
\end{lstlisting}

\begin{lstlisting}[language=NEAR, style=mystyle, label=PRdyn-dsl, caption = Dynamic PR coded in StarPlat DSL, %xleftmargin = 0.01\textwidth
]
Static staticPR(Graph g, float beta, float delta, int maxIter, propNode<float> pageRank) {
 float num_nodes = g.num_nodes();
 propNode<float> pageRank_nxt;
 g.attachNodeProperty(pageRank = 1/num_nodes, pageRank_nxt = 0);
 int iterCount=0;
 
 float diff;
 do {
    diff=0.0;

    forall(v in g.nodes()) {    
        float sum=0.0;
 
        for(nbr in g.nodes_to(v)) {
            sum=sum+nbr.pageRank / g.count_outNbrs(nbr);
        }
        float val = (1 - delta) / num_nodes + delta * sum;
        diff += val - v.pageRank; 
        v.pageRank_nxt = val;
    }
    pageRank = pageRank_nxt ; 
    iterCount++;
  } while((diff > beta) && (iterCount < maxIter));
}

Incremental(Graph g, float beta, float delta, int maxIter, propNode<float> pageRank, propNode<bool> modified) {
 propNode<float> pageRank_nxt;
 g.attachNodeProperty(pageRank_nxt = 0);
 int iterCount = 0;
 float diff;

 do {
    diff=0.0;

    forall(v in g.nodes().filter(modified == True)) {    
        float sum=0.0;
 
        for(nbr in g.nodes_to(v)) {
            sum = sum + nbr.pageRank / g.count_outNbrs(nbr);
        }

        float val = (1 - delta) / g.num_nodes() + delta * sum;
        diff += val - v.pageRank; 
        v.pageRank_nxt = val;
    }
     
    forall(v in g.nodes().filter(modified == True)) {
          v.pageRank = v.pageRank_nxt ; 
    }
    iterCount++;
  } while((diff > beta) && (iterCount < maxIter));
} 

Decremental(Graph g, float beta, float delta, int maxIter, propNode<float> pageRank, propNode<bool> modified) {
 propNode<float> pageRank_nxt;
 g.attachNodeProperty(pageRank_nxt = 0);
 int iterCount=0;
 float diff;

 do {
    diff=0.0;

    forall(v in g.nodes().filter(modified == True)) {    
        float sum = 0.0;
 
        for(nbr in g.nodes_to(v)) {
            sum = sum + nbr.pageRank / g.count_outNbrs(nbr);
        }

        float val = (1 - delta) / g.num_nodes() + delta * sum;
        diff += val - v.pageRank; 
        v.pageRank_nxt = val;
    }
     
    forall(v in g.nodes().filter(modified == True)) {
          v.pageRank = v.pageRank_nxt ; 
    }
    iterCount++;
  } while((diff > beta) && (iterCount < maxIter));
} 

 Dynamic DynPR(Graph g, float beta, float delta, int maxIter, propNode<float> pageRank, updates<g> updateBatch, int batchSize) {  
   staticPR(g, beta, delta, maxIter, pageRank);

   Batch(updateBatch:batchSize) {
     propNode<bool> modified;
     propNode<bool> modified_add;

     g.attachNodeProperty(modified = False, modified_add = False);

     updates<g> deleteBatch = updateBatch.currentBatch(0);
     updates<g> addBatch = updateBatch.currentBatch(1);

     OnDelete(u in updateBatch.currentBatch()) { 
            node src = u.source;
            node dest = u.destination;
            dest.modified = True;
     }
     g.propagateNodeFlags(modified);
     g.updateCSRDel(updateBatch); 
     Decremental(g, beta, delta, maxIter, pageRank, modified);

     OnAdd(u in updateBatch.currentBatch()) {
           node src = u.source;
           node dest = u.destination;    
           dest.modified_add = True;      
     }  
     g.propagateNodeFlags(modified_add);
     g.updateCSRAdd(updateBatch);      
     Incremental(g, beta, delta, maxIter, pageRank, modified_add);
  }
}
\end{lstlisting}

\begin{lstlisting}[language=NEAR, style=mystyle, label=ssspdyn-dsl, caption = Dynamic SSSP coded in StarPlat DSL, %xleftmargin = 0.01\textwidth
]
Static staticSSSP(Graph g, propNode<int> dist, propNode<int> parent, propEdge<int> weight, int src) {
  propNode <bool> modified;
  propNode <bool> modified_nxt;
  g.attachNodeProperty(dist=INF, modified = False, modified_nxt = False, parent = -1);
  src.modified = True; 
  src.dist=0;
  bool finished =False;
  fixedPoint until (finished:!modified) {
    forall (v in g.nodes().filter(modified == True) ) {
      forall (nbr in g.neighbors(v)) {          
        edge e = g.get_edge(v, nbr);
        <nbr.dist,nbr.modified_nxt,nbr.parent> = <Min(nbr.dist, v.dist + e.weight), True, v>;
      }
    }
    modified = modified_nxt;
    g.attachNodeProperty(modified_nxt = False); 
  }
}

Incremental(Graph g, propNode<int> dist, propNode<int> parent, propEdge<int> weight, propNode<bool> modified) {
  propNode <bool> modified_nxt;
  g.attachNodeProperty(modified_nxt = False);
  bool finished =False;

  fixedPoint until (finished:!modified) {
    forall (v in g.nodes().filter(modified == True) ) {
      forall (nbr in g.neighbors(v)) {          
        edge e = g.get_edge(v, nbr);
        <nbr.dist,nbr.modified_nxt,nbr.parent> = <Min (nbr.dist, v.dist + e.weight), True,v>;
      }
    }
    modified = modified_nxt;
    g.attachNodeProperty(modified_nxt = False);
  }
}

Decremental(Graph g, propNode<int> dist, propNode<int> parent, propEdge<int> weight, propNode<bool> modified) {
  bool finished = False;
  while(!finished) {
    finished = true;
    forall (v in g.nodes().filter(modified == False) ) {
      node parent_v = v.parent;
      if(parent_v>-1 && parent_v.modified ) {
        v.dist = INT_MAX/2;
        v.modified = True;
        v.parent= -1;
        finished = false;
      }    
    }
  } 
  finished = False;
  while(!finished) {
    finished = true;
    forall (v in g.nodes().filter(modified == True) ) {
      forall (nbr in g.nodes_to(v)) {          
        edge e = g.get_edge(nbr, v);
        if(v.dist > nbr.dist + e.weight) {
          v.dist = nbr.dist + e.weight;
          v.parent = nbr;
          finished = false;
        }
      }
    }
  }  
}

Dynamic DynSSSP(Graph g, propNode<int> dist, propNode<int> parent, propEdge<int> weight, updates<g> updateBatch, int batchSize, int src) {
  staticSSSP(g, dist, parent, weight, src);
  Batch(updateBatch:batchSize) {
    propNode<bool> modified;
    propNode<bool> modified_add;
    g.attachNodeProperty(modified = false, modified_add = false);
    OnDelete(u in updateBatch.currentBatch()): { 
      int src = u.source;
      int dest = u.destination;
      if(dest.parent == src) {
        dest.dist = INT_MAX/2;
        dest.modified = True;
        dest.parent = -1;
      }
    }
    g.updateCSRDel(updateBatch); 
    Decremental(g, dist, parent, weight, modified);   
    OnAdd(u in updateBatch.currentBatch()):{
      int src = u.source;
      int dest = u.destination;
      if(dest.dist > src.dist + 1) {
        dest.modified_add = True;
        src.modified_add = True;
      }
    }          
    g.updateCSRAdd(updateBatch);      
    Incremental(g, dist, parent, weight, modified_add);
  }
}
\end{lstlisting}

%% file: 0main.bbl
\begin{thebibliography}{10}

\bibitem{tweet}
D. Sayce. 10 billions tweets, number of tweets per day. http://www.dsayce.com/
  social-media/10-billions-tweets/., 2016-10-18, {Online; accessed
  12-May-2017}.

\bibitem{nvidia1}
Nvidia, "nvgraph".
\newblock https://developer.nvidia.com/nvgraph, 2016, {Online; accessed
  12-May-2017}.

\bibitem{commongraph}
Mahbod Afarin, Chao Gao, Shafiur Rahman, Nael Abu-Ghazaleh, and Rajiv Gupta.
\newblock Commongraph: Graph analytics on evolving data (abstract).
\newblock In {\em Proceedings of the 2023 ACM Workshop on Highlights of
  Parallel Computing}, HOPC '23, page 1–2, New York, NY, USA, 2023.
  Association for Computing Machinery.

\bibitem{bladyg}
Sabeur Aridhi, Alberto Montresor, and Yannis Velegrakis.
\newblock Bladyg: A graph processing framework for large dynamic graphs.
\newblock {\em Big Data Research}, 9:9--17, 2017.

\bibitem{slabgraph}
Saman Ashkiani, Martin Farach-Colton, and John~D. Owens.
\newblock A dynamic hash table for the gpu.
\newblock In {\em 2018 IEEE International Parallel and Distributed Processing
  Symposium (IPDPS)}, pages 419--429, 2018.

\bibitem{slabgraph1}
Muhammad~A. Awad, Saman Ashkiani, Serban~D. Porumbescu, and John~D. Owens.
\newblock Dynamic graphs on the gpu.
\newblock In {\em 2020 IEEE International Parallel and Distributed Processing
  Symposium (IPDPS)}, pages 739--748, 2020.

\bibitem{starplat}
Nibedita Behera, Ashwina Kumar, Ebenezer {Rajadurai T}, Sai Nitish,
  Rajesh~Pandian M, and Rupesh Nasre.
\newblock Starplat: A versatile dsl for graph analytics.
\newblock {\em Journal of Parallel and Distributed Computing}, 194:104967,
  2024.

\bibitem{braundynamic}
Lucas Braun, Thomas Etter, Georgios Gasparis, Martin Kaufmann, Donald Kossmann,
  Daniel Widmer, Aharon Avitzur, Anthony Iliopoulos, Eliezer Levy, and Ning
  Liang.
\newblock Analytics in motion: High performance event-processing and real-time
  analytics in the same database.
\newblock In {\em Proceedings of the 2015 ACM SIGMOD International Conference
  on Management of Data}, SIGMOD '15, page 251–264, New York, NY, USA, 2015.
  Association for Computing Machinery.

\bibitem{lonestargpu}
Martin Burtscher, Rupesh Nasre, and Keshav Pingali.
\newblock {A quantitative study of irregular programs on GPUs}.
\newblock In {\em Proceedings of the 2012 {IEEE} International Symposium on
  Workload Characterization, {IISWC} 2012, La Jolla, CA, USA, November 4-6,
  2012}, pages 141--151, New York, NY, USA, 2012. {IEEE} Computer Society.

\bibitem{hornet}
Federico Busato, Oded Green, Nicola Bombieri, and David~A. Bader.
\newblock Hornet: An efficient data structure for dynamic sparse graphs and
  matrices on gpus.
\newblock In {\em 2018 IEEE High Performance extreme Computing Conference
  (HPEC)}, pages 1--7, 2018.

\bibitem{graphfly}
D.~Chen et~al.
\newblock {Graphfly: Efficient Asynchronous Streaming Graphs Processing via
  Dependency-Flow}.
\newblock In {\em Proceedings of the International Conference on High
  Performance Computing, Networking, Storage and Analysis (SC '22)}. IEEE
  Press, 2022.

\bibitem{DH-Falcon}
Unnikrishnan Cheramangalath, Rupesh Nasre, and Y.~N. Srikant.
\newblock Dh-falcon: {A} language for large-scale graph processing on
  distributed heterogeneous systems.
\newblock In {\em 2017 {IEEE} International Conference on Cluster Computing,
  {CLUSTER} 2017, Honolulu, HI, USA, September 5-8, 2017}, pages 439--450, New
  York, NY, USA, 2017. {IEEE} Computer Society.

\bibitem{concessao2024meerkat}
Kevin~Jude Concessao, Unnikrishnan Cheramangalath, MJ~Ricky Dev, and Rupesh
  Nasre.
\newblock Meerkat: A framework for dynamic graph algorithms on gpus.
\newblock {\em International Journal of Parallel Programming},
  52(5-6):400--453, 2024.

\bibitem{Gluon-DGalois}
Roshan Dathathri, Gurbinder Gill, Loc Hoang, Hoang-Vu Dang, Alex Brooks, Nikoli
  Dryden, Marc Snir, and Keshav Pingali.
\newblock Gluon: A communication-optimizing substrate for distributed
  heterogeneous graph analytics.
\newblock In {\em Proceedings of the 39th ACM SIGPLAN Conference on Programming
  Language Design and Implementation}, PLDI 2018, page 752–768, New York, NY,
  USA, 2018. Association for Computing Machinery.

\bibitem{tesco}
Dean De~Leo and Peter Boncz.
\newblock Teseo and the analysis of structural dynamic graphs.
\newblock {\em Proc. VLDB Endow.}, 14(6):1053–1066, February 2021.

\bibitem{aspen}
Laxman Dhulipala, Julian Shun, and Guy~E. Blelloch.
\newblock Low-latency graph streaming using compressed purely-functional trees.
\newblock {\em CoRR}, abs/1904.08380, 2019.

\bibitem{Dinan2016AnIA}
James Dinan, Pavan Balaji, Darius Buntinas, David Goodell, William Gropp, and
  Rajeev Thakur.
\newblock An implementation and evaluation of the mpi 3.0 one‐sided
  communication interface.
\newblock {\em Concurrency and Computation: Practice and Experience}, 28:4385
  -- 4404, 2016.

\bibitem{STINGER}
David Ediger, Robert McColl, E.~Jason Riedy, and David~A. Bader.
\newblock Stinger: High performance data structure for streaming graphs.
\newblock {\em 2012 IEEE Conference on High Performance Extreme Computing},
  pages 1--5, 2012.

\bibitem{Sortledton}
Per Fuchs, Domagoj Margan, and Jana Giceva.
\newblock Sortledton: a universal, transactional graph data structure.
\newblock {\em Proc. VLDB Endow.}, 15(6):1173–1186, February 2022.

\bibitem{PowerGraph}
Joseph~E. Gonzalez, Yucheng Low, Haijie Gu, Danny Bickson, and Carlos Guestrin.
\newblock Powergraph: Distributed graph-parallel computation on natural graphs.
\newblock In {\em 10th {USENIX} Symposium on Operating Systems Design and
  Implementation ({OSDI} 12)}, pages 17--30, Hollywood, CA, October 2012.
  {USENIX} Association.

\bibitem{dynamicgraphRECENT}
Kathrin Hanauer, Monika Henzinger, and Christian Schulz.
\newblock Recent advances in fully dynamic graph algorithms – a quick
  reference guide.
\newblock {\em ACM J. Exp. Algorithmics}, 27, December 2022.

\bibitem{GreenMarl}
Sungpack Hong, Hassan Chafi, Eric Sedlar, and Kunle Olukotun.
\newblock Green-marl: a {DSL} for easy and efficient graph analysis.
\newblock In Tim Harris and Michael~L. Scott, editors, {\em Proceedings of the
  17th International Conference on Architectural Support for Programming
  Languages and Operating Systems, {ASPLOS} 2012, London, UK, March 3-7, 2012},
  pages 349--362, London, UK, 2012. {ACM}.

\bibitem{Pregel_GreenMarl}
Sungpack Hong, Semih Salihoglu, Jennifer Widom, and Kunle Olukotun.
\newblock Simplifying scalable graph processing with a domain-specific
  language.
\newblock In {\em Proceedings of Annual IEEE/ACM International Symposium on
  Code Generation and Optimization}, CGO '14, page 208–218, New York, NY,
  USA, 2014. Association for Computing Machinery.

\bibitem{grafs-ligra}
Farzin Houshmand, Mohsen Lesani, and Keval Vora.
\newblock Grafs: Declarative graph analytics.
\newblock {\em Proc. ACM Program. Lang.}, 5(ICFP), aug 2021.

\bibitem{iyerdynamic}
Anand~Padmanabha Iyer, Li~Erran Li, Tathagata Das, and Ion Stoica.
\newblock Time-evolving graph processing at scale.
\newblock GRADES '16, New York, NY, USA, 2016. Association for Computing
  Machinery.

\bibitem{graphone}
P.~Kumar and H.~H. Huang.
\newblock {Graphone: A Data Store for Real-Time Analytics on Evolving Graphs}.
\newblock {\em ACM Transactions on Storage (TOS)}, 15(4):1--40, 2020.

\bibitem{GraphChi}
Aapo Kyrola, Guy Blelloch, and Carlos Guestrin.
\newblock Graphchi: Large-scale graph computation on just a {PC}.
\newblock In {\em 10th {USENIX} Symposium on Operating Systems Design and
  Implementation ({OSDI} 12)}, pages 31--46, Hollywood, CA, October 2012.
  {USENIX} Association.

\bibitem{GraphLab}
Yucheng Low, Joseph Gonzalez, Aapo Kyrola, Danny Bickson, Carlos Guestrin, and
  Joseph~M. Hellerstein.
\newblock Distributed graphlab: A framework for machine learning in the cloud,
  2012.

\bibitem{Pregel}
Grzegorz Malewicz, Matthew~H. Austern, Aart~J.C Bik, James~C. Dehnert, Ilan
  Horn, Naty Leiser, and Grzegorz Czajkowski.
\newblock Pregel: A system for large-scale graph processing.
\newblock In {\em Proceedings of the 2010 ACM SIGMOD International Conference
  on Management of Data}, SIGMOD '10, page 135–146, New York, NY, USA, 2010.
  Association for Computing Machinery.

\bibitem{diffcsr}
Gaurav Malhotra, Hitish Chappidi, and Rupesh Nasre.
\newblock Fast dynamic graph algorithms.
\newblock In Lawrence Rauchwerger, editor, {\em Languages and Compilers for
  Parallel Computing}, pages 262--277, Cham, 2019. Springer International
  Publishing.

\bibitem{10.1007/978-3-030-35225-7_17}
Gaurav Malhotra, Hitish Chappidi, and Rupesh Nasre.
\newblock Fast dynamic graph algorithms.
\newblock In Lawrence Rauchwerger, editor, {\em Languages and Compilers for
  Parallel Computing}, pages 262--277, Cham, 2019. Springer International
  Publishing.

\bibitem{andrewdynamic}
Andrew McGregor.
\newblock Graph stream algorithms: a survey.
\newblock {\em SIGMOD Rec.}, 43(1):9–20, May 2014.

\bibitem{McGregordynamic}
Andrew McGregor.
\newblock Graph stream algorithms: a survey.
\newblock {\em SIGMOD Rec.}, 43(1):9–20, May 2014.

\bibitem{Galois}
Donald Nguyen, Andrew Lenharth, and Keshav Pingali.
\newblock A lightweight infrastructure for graph analytics.
\newblock In Michael Kaminsky and Mike Dahlin, editors, {\em {ACM} {SIGOPS}
  24th Symposium on Operating Systems Principles, {SOSP} '13, Farmington, PA,
  USA, November 3-6, 2013}, pages 456--471, London, UK, 2013. {ACM}.

\bibitem{galois-scheduling}
Donald Nguyen and Keshav Pingali.
\newblock Synthesizing concurrent schedulers for irregular algorithms.
\newblock In Rajiv Gupta and Todd~C. Mowry, editors, {\em Proceedings of the
  16th International Conference on Architectural Support for Programming
  Languages and Operating Systems, {ASPLOS} 2011, Newport Beach, CA, USA, March
  5-11, 2011}, pages 333--344, New York, NY, USA, 2011. {ACM}.

\bibitem{cutstinger}
David A.~Bader Oded~Green.
\newblock custinger: Supporting dynamic graph algorithms for gpus.
\newblock In {\em IEEE High Performance Extreme Computing Conference (HPEC)},
  2016.

\bibitem{terrace}
Prashant Pandey, Brian Wheatman, Helen Xu, and Aydin Buluc.
\newblock Terrace: A hierarchical graph container for skewed dynamic graphs.
\newblock In {\em Proceedings of the 2021 International Conference on
  Management of Data}, SIGMOD '21, page 1372–1385, New York, NY, USA, 2021.
  Association for Computing Machinery.

\bibitem{alibaba}
Xiafei Qiu, Wubin Cen, Zhengping Qian, You Peng, Ying Zhang, Xuemin Lin, and
  Jingren Zhou.
\newblock Real-time constrained cycle detection in large dynamic graphs.
\newblock {\em Proc. VLDB Endow.}, 11(12):1876–1888, August 2018.

\bibitem{DisGCo}
Anchu Rajendran and V.~Krishna Nandivada.
\newblock Disgco: A compiler for distributed graph analytics.
\newblock {\em ACM Trans. Archit. Code Optim.}, 17(4), September 2020.

\bibitem{GiraphPaper}
Sherif Sakr, Faisal~Moeen Orakzai, Ibrahim Abdelaziz, and Zuhair Khayyat.
\newblock {\em Large-Scale Graph Processing Using Apache Giraph}.
\newblock Springer, New York, NY, USA, 2016.

\bibitem{GPS}
Semih Salihoglu and Jennifer Widom.
\newblock Gps: A graph processing system.
\newblock In {\em Proceedings of the 25th International Conference on
  Scientific and Statistical Database Management}, SSDBM, New York, NY, USA,
  2013. Association for Computing Machinery.

\bibitem{Sha2017Accelerating}
Mo~Sha, Yuchen Li, Bingsheng He, and Kian-Lee Tan.
\newblock Accelerating dynamic graph analytics on gpus.
\newblock {\em Proceedings of the VLDB Endowment}, 11(1):107--120, 2017.

\bibitem{Ligra}
Julian Shun and Guy~E. Blelloch.
\newblock Ligra: a lightweight graph processing framework for shared memory.
\newblock In Alex Nicolau, Xiaowei Shen, Saman~P. Amarasinghe, and Richard~W.
  Vuduc, editors, {\em {ACM} {SIGPLAN} Symposium on Principles and Practice of
  Parallel Programming, PPoPP '13, Shenzhen, China, February 23-27, 2013},
  pages 135--146, Shenzhen, China, 2013. {ACM}.

\bibitem{Gunrock}
Yangzihao Wang, Andrew~A. Davidson, Yuechao Pan, Yuduo Wu, Andy Riffel, and
  John~D. Owens.
\newblock Gunrock: a high-performance graph processing library on the {GPU}.
\newblock In Rafael Asenjo and Tim Harris, editors, {\em Proceedings of the
  21st {ACM} {SIGPLAN} Symposium on Principles and Practice of Parallel
  Programming, PPoPP 2016, Barcelona, Spain, March 12-16, 2016}, pages
  11:1--11:12, Barcelona, Spain, 2016. {ACM}.

\bibitem{famigraph}
Martin Winter, Daniel Mlakar, Rhaleb Zayer, Hans-Peter Seidel, and Markus
  Steinberger.
\newblock faimgraph: high performance management of fully-dynamic graphs under
  tight memory constraints on the gpu.
\newblock In {\em Proceedings of the International Conference for High
  Performance Computing, Networking, Storage, and Analysis}, SC '18. IEEE
  Press, 2018.

\bibitem{autonomous}
Martin Winter, Rhaleb Zayer, and Markus Steinberger.
\newblock Autonomous, independent management of dynamic graphs on gpus.
\newblock In {\em 2017 IEEE High Performance Extreme Computing Conference
  (HPEC)}, pages 1--8, 2017.
\newblock 2017 IEEE High Performance Extreme Computing Conference : HPEC 2017,
  HPEC {\textquoteleft}17 ; Conference date: 12-09-2017 Through 14-09-2017.

\bibitem{HPECGraph}
Martin Winter, Rhaleb Zayer, and Markus Steinberger.
\newblock Autonomous, independent management of dynamic graphs on gpus.
\newblock In {\em 2017 IEEE High Performance Extreme Computing Conference
  (HPEC)}, pages 1--7, 2017.

\bibitem{Egraph}
Yu~Zhang, Yuxuan Liang, Jin Zhao, Fubing Mao, Lin Gu, Xiaofei Liao, Hai Jin,
  Haikun Liu, Song Guo, Yangqing Zeng, Hang Hu, Chen Li, Ji~Zhang, and Biao
  Wang.
\newblock Egraph: Efficient concurrent gpu-based dynamic graph processing.
\newblock 35(6):5823–5836, June 2023.

\bibitem{GraphIt}
Yunming Zhang, Mengjiao Yang, Riyadh Baghdadi, Shoaib Kamil, Julian Shun, and
  Saman~P. Amarasinghe.
\newblock Graphit: a high-performance graph {DSL}.
\newblock {\em Proc. {ACM} Program. Lang.}, 2({OOPSLA}):121:1--121:30, 2018.

\bibitem{medusa}
Jianlong Zhong and Bingsheng He.
\newblock Medusa: Simplified graph processing on gpus.
\newblock {\em IEEE Transactions on Parallel and Distributed Systems},
  25(6):1543--1552, 2014.

\bibitem{GeminiGraph}
Xiaowei Zhu, Wenguang Chen, Weimin Zheng, and Xiaosong Ma.
\newblock Gemini: A computation-centric distributed graph processing system.
\newblock In {\em 12th {USENIX} Symposium on Operating Systems Design and
  Implementation ({OSDI} 16)}, pages 301--316, Savannah, GA, November 2016.
  {USENIX} Association.

\bibitem{gridgraph}
Xiaowei Zhu, Wentao Han, and Wenguang Chen.
\newblock {GridGraph: Large-Scale Graph Processing on a Single Machine Using
  2-Level Hierarchical Partitioning}.
\newblock In {\em Proceedings of the 2015 USENIX Conference on Usenix Annual
  Technical Conference}, USENIX ATC '15, page 375–386, USA, 2015. USENIX
  Association.

\bibitem{Zoudynamic}
Lei Zou, Fan Zhang, Yinnian Lin, and Yanpeng Yu.
\newblock An efficient data structure for dynamic graph on gpus.
\newblock {\em IEEE Transactions on Knowledge and Data Engineering},
  35(11):11051--11066, 2023.

\end{thebibliography}
